\documentclass[]{article}
\usepackage{authblk}

\usepackage{url}
\usepackage[T1]{fontenc}    % use 8-bit T1 fonts
\usepackage{fullpage}
\usepackage{caption}
\usepackage[noend]{algorithmic}
\usepackage[vlined,ruled]{algorithm2e}
\usepackage{amsmath}
\usepackage{amssymb} 
\usepackage{mathrsfs}
\usepackage{mathtools}
\usepackage{bm} 
\usepackage{color}
\usepackage{enumitem}
\usepackage{subfig}
\usepackage{adjustbox}
\usepackage{cancel} % for strikethough the text
\usepackage{Definitions}

\usepackage{color}

\newcommand{\redqueen}{\textsc{RedQueen}\xspace}
\newcommand{\cheshire}{\textsc{Cheshire}\xspace}

\newcommand{\xhdr}[1]{\vspace{0.1mm}\noindent{{\bf #1. }}}

\newcommand{\qb}{\boldsymbol{\omega}}

\newcommand\blfootnote[1]{%
  \begingroup
  \renewcommand\thefootnote{}\footnote{#1}%
  \addtocounter{footnote}{-1}%
  \endgroup
}

 \graphicspath{{./figures/}}

\SetKwProg{Fn}{Function}{}{end}
\title{Steering Social Activity: A Stochastic \\ Optimal Control Point Of View$^{*}$}

\author[1]{Ali Zarezade}
\author[2]{Abir De}
\author[2]{Utkarsh Upadhyay}
\author[1]{Hamid R. Rabiee}
\author[2]{Manuel Gomez-Rodriguez}

\affil[1]{Sharif University of Technology, zarezade@ce.sharif.edu, rabiee@sharif.edu}
\affil[2]{Max Planck Institute for Software Systems, ade@mpi-sws.org, utkarshu@mpi-sws.org, manuelgr@mpi-sws.org}

\date{}

\begin{document}
% Set the space before and after equations
%\setlength{\abovedisplayskip}{2pt}
%\setlength{\belowdisplayskip}{2pt}


\maketitle

\begin{abstract}%
\noindent User engagement in online social networking depends critically on the level of \emph{social activity} in the corresponding platform---the number 
of online \emph{actions}, such as posts, shares or replies, taken by their users.
Can we design data-driven algorithms to increase social activity?
At a user level, such algorithms may increase activity by helping users decide when to take an action to be more likely to be \emph{noticed} 
by their peers. At a network level, they may increase activity by \emph{incentivizing} a few influential users to take more actions, which in turn 
will trigger additional actions by other users.

In this paper, we model social activity using the framework of marked temporal point processes, derive an alternate representation of these processes 
using stochastic differential equations (SDEs) with jumps and, exploiting this alternate representation, develop two efficient online algorithms with 
provable guarantees to steer social activity both at a user and at a network level.
In doing so, we establish a previously unexplored connection between optimal control of jump SDEs and doubly stochastic marked 
temporal point processes, which is of independent interest.
Finally, we experiment both with synthetic and real data gathered from Twitter and show that our algorithms consistently steer social activity more effectively
than the state of the art.
\end{abstract}

%\begin{keywords}
%marked temporal point processes, stochastic optimal control, stochastic diffe\-rential equa\-tions with jumps, social networks, 
%information networks
%\end{keywords}

\blfootnote{$^{*}$Preliminary version of this work appeared in~\cite{redqueen17wsdm}.}

%-------------------------------------------------------
%-------------------------------------------------------
\section{Introduction}
\label{sec:intro}
People perform a wide variety of online actions in a growing number of online social networking sites.
% social media platforms and
%
In most cases, these online actions can be broadly categorized into \emph{exogenous} actions, taken by users at their
own initiative, and \emph{endogenous} actions, taken by users as a \emph{response} to previous actions by other users.
For example,
users often post small pieces of information at their own initiative, which can then trigger additional posts, shares or replies
by other users. % ;
In this paper, our goal is designing online algorithms that, by steering exo\-ge\-nous actions, are able to increase the number
of endogenous actions and thus the overall user activity.

We first address the above problem from the perspective of an individual user whose posts compete for \emph{attention} with dozens, if not
hundreds, of posts \emph{simultaneously} shared by other users that her \emph{followers} follow~\cite{backstrom2011center,gomez14icwsm}.
In this context, recent empirical studies have shown that stories at the top of their followers'{} feed are more likely to be \emph{noticed} and
consequently liked, shared or replied to~\cite{hodas2012visibility,kang2015vip,lerman2014leveraging}.
Therefore, we design an online algorithm, \redqueen, to help a user decide when-to-post to increase her chances to stay at the top, increasing
her \emph{visibility}.
Then, we tackle the problem from the perspective of an entire online social networking site, where an endless stream of stories posted by its users
are constantly eliciting a variable number of likes, shares or replies by other users~\cite{goel2012structure, shaping14nips, rizoiu2017expecting}.
Here, we design another online algorithm, \cheshire, to find how much should we incentivize a small number of influential users to post more over
time to increase the overall number of additional posts, shares or replies in the site.

More specifically, we represent users'{} actions using the framework of marked temporal point processes, which characterizes the continuous time interval between actions
using conditional intensity functions~\cite{AalBorGje08}, and model endogenous and exogenous actions using multidimensional Hawkes processes~\cite{hawkes1971spectra}.
Then, we derive an alternate representation of these processes using stochastic differential equations (SDEs) with jumps and, exploiting this alternate representation, address
the design of both algorithms from the perspective of optimal control for SDEs with jumps~\cite{hanson2007}.
Our approach differs from previous literature in two key technical aspects, which are of independent interest:
\begin{itemize}% [noitemsep,nolistsep,leftmargin=1cm]
\item[I.] The control signals are conditional (posting) intensities, which are used to sample stochastic events (\ie, stories to post). In contrast, previous work considered the control signal
to be a time-varying real vector. As a consequence, our approach requires another layer of stochasticity.
\item[II.] The (posting) intensities are stochastic Markov processes and thus the dynamics are doubly stochastic.
This requires us to redefine the cost-to-go to incorporate the instantaneous value of these intensities as additional arguments. Previous work has typically considered constant intensities
until very recently~\cite{wang2016stochastic, wang2017variational}.
\end{itemize}
These technical aspects have implications beyond steering social activity since they enable us to establish a previously unexplored connection between optimal control of jump SDEs and
doubly stochastic temporal point processes (\eg, Hawkes processes), which have been increasingly used to model social activity~\cite{shaping14nips,coevolve15nips,zhao2015seismic}.

In both cases, we find that the solution to the corresponding optimal control problem is surprisingly simple. At a user level, the optimal posting intensity for a user to achieve high visibility depends
linearly on the position of her most recent post on each of her follower'{}s feed and, at a network level, the optimal level of incentivized actions depends linearly on the current level of overall
actions.
Moreover, at a user level, the coefficients of the linear relationship only depend on tunable parameters, which are predefined, and, at a network level, the coefficients can be found by solving a matrix
Riccati differential equation~\cite{garrett2013} and a first order differential equation, which has a closed form solution.
As a consequence, both \redqueen and \cheshire are simple and highly efficient online procedures, which can be implemented in a few lines of code (refer to Algorithms~\ref{alg:sampling-redqueen}
and~\ref{alg:sampling-cheshire}).
Finally, we perform experiments on both synthetic and real-data gathered from Twitter and show that our algorithms can consistently steer social activity at a
user and at a network level more effectively than the state of the art.

\subsection{Related Work}
%
% summary
Our work relates to previous work on the when-to-post problem, the activity shaping problem, stochastic optimal control, and temporal point processes.

% when-to-post problem.
%
The ``when-to-post'' problem was first studied by~\cite{spasojevic2015post}, who performed a large empirical study on the best times for a user to post in Twitter and Facebook,
measuring attention a user elicits by means of the number of responses to her posts. Moreover, they designed several heuristics to pinpoint at the times that elicit the greatest attention in a training
set and showed that these times also lead to more responses in a held-out set.
Since then, algorithmic approaches to the ``when-to-post'' problem with provable guarantees have been largely lacking. Only very recently,~\cite{karimi2016smart} introduced a convex
optimization framework to find optimal broadcasting strategies, measuring attention a user elicits as the time that at least one of her posts is among the $k$ most recent stories received in her
followers'{} feed.
However, their algorithm requires expensive data pre-processing, it does not adapt to changes in the users'{} feeds dynamics, and in practice, it is less effective than our algorithm \redqueen,
as shown in Section~\ref{sec:experiments}.

% activity shaping
The ``activity shaping" problem was first studied by~\cite{Farajtabar2015}, who derived a time dependent linear relation between the \emph{intensity} of exogenous actions and the
overall intensity of actions in a social network under a model of actions based on multidimensional Hawkes processes and, exploiting this connection, developed a convex optimization framework
for activity shaping.
One of the main shortcomings of their framework is that it provides deterministic exogenous in\-ten\-si\-ties that do not adapt to changes in the users'{} intensities and, as a consequence, it is less
effective than our algorithm \cheshire, as shown in Section~\ref{sec:experiments}.
More recently,~\cite{farajtabar2016msc} developed a heuristic method that splits the time window of interest into stages and adapts to changes in the users'{} intensities at the beginning
of each stage. However, their method is suboptimal, it does not have \emph{provable} guarantees, and it achieves lower performance than our method.

% optimal control
In the traditional control literature~\cite{hanson2007}, two key aspects of our approach---intensities as control signals and stochastic intensities---have been largely understudied.
Only very recently,~\cite{wang2016stochastic} and~\cite{wang2017variational} have considered these aspects. However, in~\cite{wang2016stochastic}, the intensities are not stochastic and
the resulting algorithm needs to know the future actions, hindering its applicability, and, in~\cite{wang2017variational}, the solution is open-loop and the control policy depends on the expectation 
of the uncontrolled dynamics, which needs to be calculated approximately by a time consuming sampling process.
In contrast, our framework consider double stochastic intensities, our solution is closed-loop, our control policies only depend on the current state of the dynamics, and the feedback coefficients only 
need to be calculated once off-line.

% temporal point processes
Temporal point processes have been increasingly used for representation and modeling in a wide range of applications in social and information
systems, \eg, information propagation~\cite{Rodriguez2011, du13nips, zhao2015seismic}, opinion dynamics~\cite{de2016learning}, product
competition~\cite{Valera2015,zarezade2015correlated}, spatiotemporal social activity~\cite{jankowiak2017uncovering,zarezade2016spatio}, information reliability~\cite{reliability2017tabibian}, or human learning~\cite{hdhp2017learning}.  
However, in such context, algorithms based on stochastic optimal control of temporal point processes have been lacking.

%-------------------------------------------------------
%-------------------------------------------------------
\section{Preliminaries}
\label{sec:preliminaries}
In this section, we first revisit the framework of temporal point processes~\cite{AalBorGje08} and then describe how to use such framework to represent actions
and feeds~\cite{karimi2016smart} as well as to model endogenous and exogenous actions in social networks~\cite{shaping14nips}.

\subsection{Temporal Point Processes}
A univariate temporal point process is a stochastic process whose realization consists of a sequence of discrete events localized in time, $\Hcal = \{t_i\in \RR^{+} \,\vert\, i\in \mathbb{N}^+,\, t_i<t_{i+1} \}$.
It can also be represented as a counting process $N(t)$, which counts the number of events before time $t$, \ie,
\begin{equation*}
N(t) = \sum_{t_i \in \Hcal} \mathbb{I}(t - t_i \geq 0),
\end{equation*}
where $\mathbb{I}(\cdot)$ is the indicator function.
Then, we can characterize the counting process using the conditional intensity function $\lambda^{*}(t)$, which is the
conditional probability of observing an event in an infinitesimal window $[t, t + dt)$ given the history of event times up to time $t$, $\Hcal(t) = \{t_i \in \Hcal \,\vert\, t _i < t \}$, \ie,
\begin{equation*}
\lambda^{*}(t) \, dt = \PP\cbr{\text{event in $[t, t+dt) \,\vert\, \Hcal(t)$}} = \EE[dN(t) \,\vert\, \Hcal(t)],
\end{equation*}
where $dN(t) := N(t+dt) - N(t) \in \{0, 1\}$, the sign $^{*}$ means that the intensity may depend on the history $\Hcal(t)$.
Moreover, given a function $f(t)$, it will be useful to define the convolution with respect to $dN(t)$ as
\begin{equation*}
f(t) \star dN(t) := \int_{0}^{t} f(t - s) dN(s) = \sum_{t_i \in \Hcal(t)} f(t - t_i).
\end{equation*}

One can readily extend the above definitions to multivariate (or multidimensional) temporal point processes, which have been recently used to represent many different types of event data
produced in social networks, such as the times of tweets~\cite{shaping14nips}, retweets~\cite{zhao2015seismic}, or links~\cite{coevolve15nips}.
More specifically, a realization of an $m$-dimensional temporal point process, $\Hcal = \{(t_i,u_i) \,\vert\, i \in \mathbb{N}^+, t_i \in \RR_{+}, u_i \in [m], t_i < t_{i+1} \}$, consists of $m$ sequences of discrete events localized in time, $\Hcal = \cup_{u \in [m]} \Hcal_u$, where
$\Hcal_u = \{(t_i,u_i) \in \Hcal \,\vert\, u_i=u \}$.
Equivalently, it can be represented by an $m$-dimensional counting process $\mathbf{N}(t)$, where $N_u(t)$ counts the number of events in the $u$-th sequence before time $t$,
and this counting process can be characterized by $m$ intensity functions, \ie,
\begin{equation*}
\lambdab^{*}(t)dt = \EE[d\Nb(t)|\Hcal(t)],
\end{equation*}
where $\Hcal(t) = \{(t_i,u_i) \in \Hcal \,\vert\, t _i < t\}$, $d\Nb(t):=(dN_u(t))_{u\in [m]} := (N_u(t+dt) - N_u(t))_{u \in [m]}$, and $\lambdab^*(t):=(\lambda^*_u(t))_{u\in [m]}$
denotes the associated intensities, which may depend on history $\Hcal(t)$.
Finally, given a function $f(t)$, one can define the convolution with respect to $d\Nb(t)$ as
\begin{equation*}
f(t) \star d\Nb(t) := \int_{0}^{t} f(t - s) d\Nb(s) = \left( \sum_{t_i \in \Hcal_u(t)} f(t - t_i) \right)_{u \in [m]}.
\end{equation*}
In the remainder of the paper, to simplify the notation, we drop the sign $^{*}$ from the intensities.

\subsection{Representation of Actions and Feeds} \label{sec:representation}
Given a (directed) social network $\mathcal{G}=(\Vcal,\Ecal)$ with $|\mathcal{V}|=n$ users, we assume that each user $i$ can take a variety of online actions, \eg, posting,
sharing or replying, and she will be exposed to the online actions taken by the users she \emph{follows} through her feed. Here, we assume that user $i$ follows user
$j$ if and only if $(i, j) \in \Ecal$.

Then, we represent the times when users take online actions as a multidimensional coun\-ting process $\Nb(t)$, where the $i$-th dimension, $N_{i}(t)$, counts the
number of actions taken by user $i$ up to time $t$. Here, we denote the history of times of the actions taken by user $i$ by time $t$ as $\Hcal_i(t)$, the entire history of times as $\Hcal(t) = \cup_{i \in \Vcal} \Hcal_i(t)$,
and characterize this multidimensional process using $n$ intensity functions, \ie, $\mathbb{E}[d\bm{N}(t) | \Hcal(t)] = \lambdab(t) \, dt$.

Given the adjacency matrix $\Ab \in \{0,1\}^{n\times n}$, where $\Ab_{ij}=1$ indicates that user $j$ follows user $i$, we can represent the times of the actions users are exposed to
through their feeds as a sum of counting processes, $\Ab^T \bm{N}(t)$, and calculate the corresponding conditional intensities as $\gammab(t) = \Ab^T \lambdab(t)$.
Here, we denote the history of times of the actions user $j$ is exposed to by time $t$ as $\Fcal_j(t) := \cup_{i \in \Ncal(j)} \Hcal_{i}(t)$, where $\Ncal(j)$ is the set of users that $j$
follows.

Finally, from the perspective of a user $i$, it is useful to define the multidimensional counting process $\Mb_{{\scriptscriptstyle\setminus} i}(t) =  \Ab^T \bm{N}(t) - \Ab_i N_i(t)$, in
which the $j$-th dimension, $M_{j {\scriptscriptstyle\setminus} i}(t)$, counts the number of actions taken by other users that user $j$ follows up to time $i$, and $\Ab_i$ is the $i$-th
row of the adjacency matrix $\Ab$.
Moreover, for each dimension, the conditional intensity is given by $\gamma_{j {\scriptscriptstyle\setminus} i}(t) = \gamma_{j}(t) - \lambda_i(t)$ and the history is given by
$\Fcal_{j {\scriptscriptstyle\setminus} i}(t) := \Fcal_j(t) \backslash \Hcal_{i}(t)$.
When there is no ambiguity, we will omit the subscript $i$ and write $\Mb_{{\scriptscriptstyle\setminus} i}(t) = \Mb(t)$ and $\gammab_{{\scriptscriptstyle\setminus} i}(t) = \gammab(t)$.

\subsection{Modeling Endogenous and Exogenous Actions}
Following previous work~\cite{shaping14nips, coevolve15nips, zhao2015seismic, de2016learning, rizoiu2017expecting}, we model endogenous and exogenous actions using
multidimensional Hawkes processes~\cite{hawkes1971spectra}. More specifically, we proceed as follows.

From the perspective of an individual user $i$, we assume we observe the online actions her followers are exposed to through their feeds. Then, consider the following
functional form for the intensities $\gammab_{\backslash i}(t) = \gammab(t)$ of the followers'{} feeds from Section~\ref{sec:representation}:
\begin{equation} \label{eq:multi-hawkes-redqueen}
	\gammab(t) = \gammab_0(t) + \Db\,\kappa_{\omega}(t) \star d\Mb(t) = \gammab_0(t) +  \Db \int_0^t \kappa_{\omega}(t-s) \, d\Mb(s)
\end{equation}
where $\gammab_0(t)$ are time-varying functions that model exogenous actions, \ie, actions users take at their own initiative, the second term, with $\Db = \diag(\alpha_j), \alpha_j \geq 0$,
models endogenous actions, \ie, actions users take as \emph{response} to previous actions (their own as well as the ones taken by others), and $\kappa_{\omega}(t) = e^{-\omega t} \, \II(t \geq 0)$
is an exponential kernel modeling the decay on influence over time\footnote{Exponential kernels have been shown to provide relatively good predictive 
performance~\cite{shaping14nips, coevolve15nips, zhao2015seismic, de2016learning, rizoiu2017expecting}, however, we acknowledge that power-law may have higher predictive performance 
in some scenarios, as recently shown by \cite{Mishra2016}. Here, we opt for exponential kernels since, in such case, the dynamics of the corresponding intensity can be expressed by a linear SDE 
with jumps and this will be helpful in the derivation of our stochastic optimal control algorithms.}.

From the perspective of an entire online social networking site, we assume we observe the online actions taken by all the users\footnote{We acknowledge that there may be scenarios in which one 
only has access to a subset of the online actions taken by the users. However, in those cases, one could resort to an increasing number of methods to fit multidimensional Hawkes processes 
from incomplete observations, \eg, refer to \cite{xu2017learning}}. Thus, we consider the following functional form for the users'{} intensities $\lambdab(t)$ from 
Section~\ref{sec:representation}:
\begin{equation} \label{eq:multi-hawkes-cheshire}
	\lambdab(t) = \lambdab_0(t) + \Bb\,\kappa_{\omega}(t) \star d\bm{N}(t) = \lambdab_0(t) +  \Bb \int_0^t \kappa_{\omega}(t-s) \, d\bm{N}(s),
\end{equation}
where $\lambdab_0(t) \geq 0$ are time-varying functions that model exogenous actions and the second term models endogenous actions---actions users take as response
to the actions taken by the users they follow.
Here, we parameterize the strength of \emph{influence} between users using a sparse nonnegative \emph{influence matrix} $\Bb = (\beta_{u v{}}) \in \mathbb{R}_{+}^{m\times m}$, where $\beta_{u v{}}$
means user u'{}s actions directly triggers \emph{follow-ups} from user $v{}$ and we assume $\beta_{u v{}} = 0$ if $(u, v) \notin \Ecal$.

In both cases, the second term makes the intensity dependent on the history and a stochastic process itself. Moreover, the following alternative representation of multidimensional Hawkes processes
will be useful to design our stochastic optimal control algorithms (proven in Appendix~\ref{app:hawkes-dynamics}):
\begin{proposition} \label{prop:hawkes}
	Let $\bm{N}(t)$ be an $m$-dimensional Hawkes process with an associated intensity $\lambdab(t) = \lambdab_0(t) +  \Cb \int_0^t \kappa_{\omega}(t-s) \, d\bm{N}(s)$, where $\lambda_0(t) \geq 0$ is a differentiable
	time-varying function, $\kappa_{\omega}(t) = e^{-\omega t} \, \II(t \geq 0)$ is an exponential triggering kernel, and $\omega \geq 0$ and $\Cb \in \mathbb{R}_{+}^{m\times m}$ are given parameters. Then, the tuple $(\bm{N}(t), \lambdab(t))$
	is a doubly stochastic Markov process, whose dynamics can be defined by the following jump SDEs:
	\begin{align} \label{eq:hawkes-dyn}
		d\lambdab(t) = \left[\lambdab'_0(t) + \omega \lambdab_0(t) - \omega \lambdab(t) \right] dt + \Cb \, d\bm{N}(t),
	\end{align}
	with the initial condition $\lambdab(0)= \lambdab_0(0)$.
\end{proposition}

\xhdr{Remark}
From the perspective of an individual user, we assume the user has a local view, \ie, she only observes the online actions her followers are exposed to through their feeds. Alternatively,
if the user would have a global view, one could naturally opt for the following functional form for the intensities $\gammab(t)$ of the followers'{} feeds:
\begin{equation} \label{eq:multi-hawkes-redqueen-ideal}
	\gammab(t) = \Ab^T \lambdab(t) = \Ab^T \lambdab_0(t) +  \Ab^T \Bb \int_0^t \kappa_{\omega}(t-s) \, d\Nb(s).
\end{equation}
However, for simplicity, from the perspective of an individual user, we will assume the user has a local view and thus consider the functional form defined
by Eq.~\ref{eq:multi-hawkes-redqueen}. Considering an individual user has a global view is an interesting, albeit challenging venue for future work.

%-------------------------------------------------------
%-------------------------------------------------------
\section{A Local Problem: When-to-post}
\label{sec:redqueen}
In this section, we take the perspective of an individual user and design an online algorithm to help her decide when-to-post to be \emph{noticed}---achieve high \emph{visibility}.
To this aim, we first define our visibility measure and then derive a jump stochastic differential equation that links our measure to the user'{}s posting intensity and the feeds'{} intensities
due to other users her followers follow. Finally, we formally state the when-to-post problem for this visibility measure and address the problem from the perspective of stochastic optimal
control of SDEs with jumps.
For ease of exposition, we assume users can only take one type of exogenous and endogenous action---posting stories. Moreover, throughout the section, we refer to users who posts stories
as \emph{broadcasters}.

\subsection{Visibility Definition and Dynamics}
Given a broadcaster $i$ and one of her followers $j$, we define the visibility function $r_{ij}(t)$ as the position or \emph{rank} of the most recent story posted by $i$ in $j$'{}s feed by time $t$, which clearly
depends on the feed ranking mechanism in the corresponding social network.
Here, for simplicity, we assume each user'{}s feed ranks stories in reverse chro\-no\-lo\-gi\-cal order\footnote{\scriptsize At the time of writing, Weibo rank stories in
reverse chronological order by default and
Twitter and Facebook allow choosing such an ordering.}.
However, our framework can be easily extended to any feed ranking mechanisms, as long as its rank dynamics can be expressed as a jump SDE\footnote{\scriptsize This would require either having access to the
corresponding feed ranking mechanism or reverse engineering it, which is out of the scope of this work.}.

Under the reverse chronological ordering assumption, position at time $t$ is simply the number of stories that other broadcasters posted in $j$'{}s feed from the time of the most recent story
posted by $i$ until $t$.
Then, when a new story arrives to a user'{}s feed, it appears at the top of the feed and the other stories are shifted down by one.
If we denote the time of the most recent story posted by $i$ by time $t$ as $\tau_{i}(t) = \max\{t_k \in \Hcal_i(t)\}$, then the visibility is formally defined as 
\begin{align}\label{equ:rank}
	r_{ij}(t) = M_{j {\scriptscriptstyle\setminus} i}(t) - M_{j {\scriptscriptstyle\setminus} i}(\tau_{i}(t)).
\end{align}
Note that, if the last story posted by $i$ is at the top of $j$'{}s feed at time $t$,
then $r_{ij}(t) = 0$.

\begin{figure}[!t]
\centering
\includegraphics[width=0.55\textwidth]{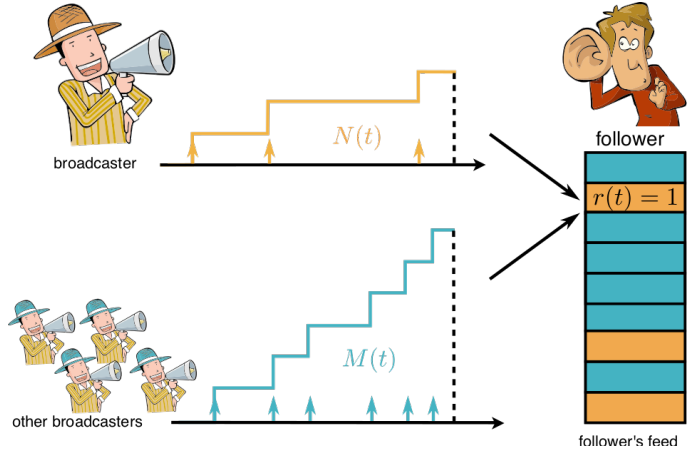}
\caption{The dynamics of visibility. A broadcaster $i$ posts $N_i(t) = N(t)$ stories with intensity $\lambda_{i}(t) = u(t)$. Her stories accumulate in her follower $j$'{}s
feed, competing for attention with $M_{j\backslash i}(t) = M(t)$ other stories posted by other broadcasters user $j$ follows with intensity $\gamma_{j \backslash i}(t) = \gamma(t)$.
The visibility function $r_{ij}(t) = r(t)$ is the position or rank of the most recent story posted by broadcaster $i$ in the follower $j$'{}s feed by time t.}
\label{fig:model}
\vspace{-2mm}
\end{figure}

\xhdr{Dynamics of visibility}
Given a broadcaster $i$ with broadcasting counting process $N_i(t)$ and one of her followers $j$ with feed counting process due to other broadcasters $M_{j {\scriptscriptstyle\setminus} i}(t)$, the rank of $i$ in $j$'s feed
$r_{ij}(t)$ satisfies the following equation:
\begin{equation}
	r_{ij}(t+dt) =
	\underset{\text{\scriptsize 1. Increases by one}}{\underbrace{(r_{ij}(t)+1) dM_{j {\scriptscriptstyle\setminus} i}(t) (1-dN_i(t))}}
	+\underset{\text{\scriptsize 2. Becomes zero}}{\underbrace{0}} \nonumber +
	\underset{\text{\scriptsize 3. Remains the same}}{\underbrace{r_{ij}(t)(1 - dM_{j {\scriptscriptstyle\setminus} i}(t))(1- dN_i(t))}}, \nonumber
\end{equation}
where each term models one of the three possible situations:
\begin{itemize}[noitemsep,nolistsep,leftmargin=1cm]
\item[1.] The other broadcasters post a story in $(t,t+dt]$, $dM_{j {\scriptscriptstyle\setminus}i}(t)=1$, and broadcaster $i$ does not post, $dN_i(t)=0$. The position of the last story posted by
$i$ in $j$'{}s feed steps down by one, \ie, $r_{ij}(t+dt)=r_{ij}(t)+1$.
\item[2.] Broadcaster $i$ posts a story in $(t,t+dt]$, $dN_i(t)=1$, and the other broadcasters do not, $dM_{j {\scriptscriptstyle\setminus}i}(t)=0$. No matter what the previous rank was, the new
rank is $r_{ij}(t+dt)=0$ since the newly posted story appears at the top of $j$'{}s feed.
\item[3.] No one posts any story in $(t,t+dt]$, $dN_i(t)=0$ and $dM_{j {\scriptscriptstyle\setminus}i}(t)=0$. The rank remains the same, \ie, $r_{ij}(t+dt)=r_{ij}(t)$.
\end{itemize}
We skip the case in which $M_{j {\scriptscriptstyle\setminus}i}(t)=1$ and $dN_i(t)=1$ in the same time interval $(t,t+dt]$ because, by the Blumenthal zero-one law~\cite{blumenthal1957extended},
it has zero probability. Now, by rearranging terms and using that $dN_i(t) dM_{j {\scriptscriptstyle\setminus}i}(t)=0$, we uncover the following jump SDE for the visibility
(or rank) dynamics:
\begin{align}\label{eq:rank-dynamics}
dr_{ij}(t) &= -r_{ij}(t)\,dN_i(t) + dM_{j {\scriptscriptstyle\setminus}i}(t),
\end{align}
where $dr_{ij}(t)=r_{ij}(t+dt)-r_{ij}(t)$. Figure~\ref{fig:model} illustrates the concept of visibility for one broadcaster and one follower.

\subsection{Problem Formulation}
Given a broadcaster $i$ and her followers $\Ncal(i)$, our goal is to find the optimal posting intensity $\lambda_i(t) = u(t)$ that minimizes the expected value of a 
particular nondecreasing convex loss function $\ell(\bm{r}(t), u(t))$ of the broadcaster'{}s visibility on each of her follower'{}s feed, $\bm{r}(t) = [r_{ij}(t)]_{j \in \Ncal(i)}$,
and the intensity itself, $u(t)$, over a time window $(t_0, t_f]$, \ie, 
\begin{align} \label{eq:average-loss}
\underset{u(t_0, t_f]}{\text{minimize}} & \quad \EE_{(N_i,\bm{M}_{{\scriptscriptstyle\setminus}i})(t_0, t_f]}\left[ \phi(\bm{r}(t_f)) + \int_{t_0}^{t_f} \ell(\bm{r}(s), u(s)) ds \right] \nonumber \\
\text{subject to} & \quad u(t) \geq 0 \quad \forall t \in (t_0, t_f],
\end{align}
where $u(t_0, t_f]$ denotes the broadcaster $i$'{}s intensity from $t_0$ to $t_f$, the expectation is taken over all possible realizations of the counting processes associated to the broadcaster
$i$ and all other broadcasters from $t_0$ to $t_f$, denoted as $(N_i,\bm{M}_{{\scriptscriptstyle\setminus}i})(t_0, t_f]$, and $\phi(\bm{r}(t_f))$ is an arbitrary penalty function\footnote{\scriptsize The
final penalty function $\phi(\bm{r}(t_f))$ is necessary to derive the optimal intensity $u^*(t)$ in Section~\ref{sec:optimal-control-redqueen}. However, the actual optimal intensity $u^*(t)$ will not
depend on the particular choice of terminal condition.}.
Here, by considering a loss which is nondecreasing in the rank $r(t)$ and the posting intensity $u(t)$, we penalize times when the position of the most recent story on each of the follower'{}s feeds 
is high (\ie, the most recent story does not stay \emph{at the top}) and we limit the posting intensity, which in turn limits the number of stories the broadcaster can post.
Finally, note that the optimal intensity $u(t)$ for broadcaster $i$ at time $t$ may depend on the visibility $\bm{r}(t)$ with respect to each
of her followers and, thus, the associated counting process $N_i(t)$ may be doubly stochastic.

\subsection{Stochastic Optimal Control Algorithm} \label{sec:optimal-control-redqueen}

In this section, we tackle the when-to-post problem defined by Eq.~\ref{eq:average-loss} from the perspective of stochastic optimal control of jump SDEs~\cite{hanson2007}. More specifically, we first derive a solution to the problem considering only one follower, provide an efficient practical implementation of the solution and then generalize it to the case of multiple followers.
We conclude this section by deriving a solution to the problem given an (idealized) oracle that knows the times of all stories in the
followers'{} feeds \emph{a priori}, which we will use as baseline in our experiments (refer to Section~\ref{sec:experiments-redqueen}).

\xhdr{Optimizing for one follower} Given a broadcaster $i$ with $N_i(t) = N(t)$ and $\lambda_i(t) = u(t)$ and only one of her followers $j$ with $M_{j {\scriptscriptstyle\setminus} i}(t) = M(t)$ and
$\gamma_{j {\scriptscriptstyle\setminus} i}(t)=\gamma(t)$, we can rewrite the when-to-post problem defined by Eq.~\ref{eq:average-loss} as
\begin{align}
\underset{u(t_0, t_f]}{\text{minimize}} & \quad \EE_{(N,M)(t_0, t_f]}\left[ \phi(r(t_f)) + 	\int_{t_0}^{t_f} \ell(r(s),u(s)) \, ds \right] \nonumber \\
\text{subject to} & \quad u(t) \geq 0 \quad \forall t \in (t_0, t_f], \label{eq:average-loss-one-follower}
\end{align}
where, using Eqs.~\ref{eq:multi-hawkes-redqueen},~\ref{eq:hawkes-dyn} and~\ref{eq:rank-dynamics}, the dynamics of $M(t)$ and $r(t)$ are given by the following two coupled jump SDEs:
\begin{align}
	d\gamma(t) &= \left[\gamma_0'(t)+\omega \gamma_0(t) -\omega \gamma(t)\right]dt + \alpha \, dM(t), \nonumber \\
	dr(t) &= -r(t)\,dN(t) + dM(t), \label{eq:dynamics-one-follower}
\end{align}
with initial conditions $r(t_0)=r_0$ and $\gamma(t_0)=\gamma_0$, and the dynamics of $N(t)$ are given by the intensity $u(t)$ that we aim to optimize.

Next, we will define a novel optimal cost-to-go function that accounts for the above unique aspects of our problem, showing that the Bellman'{}s principle of optimality still follows, and finally find
the optimal solution using the corresponding Hamilton-Jacobi-Bellman (HJB) equation.
\begin{definition}\label{thm:cost-def-redqueen}
The optimal cost-to-go $J(r(t),\gamma(t),t)$ is defined as the minimum of the expected value of the cost of going from state $r(t)$ with intensity $\gamma(t)$ at time $t$ to a final state at time $t_f$, \ie,
\begin{equation}\label{eq:cost-to-go-redqueen}
	J(r(t),\gamma(t),t) =
	\min_{u(t,t_f]} \mathbb{E}_{(N,M)(t,t_f]} \left[ \phi(r(t_f)) + \int_t^{t_f} \ell(r(s),u(s)) \, ds \right ],
	%% \big\vert r(t),\gamma(t),u(t)
\end{equation}
where the expectation is taken over all trajectories of the control and noise jump process, $N$ and $M$, in the $(t,t_f]$ interval, given the initial values of $r(t)$, $\gamma(t)$
and $u(t)$.
\end{definition}
To find the optimal control $u(t, t_f]$ and cost-to-go $J$, we break the problem into smaller subproblems, using the Bellman'{}s principle of optimality~\cite{bertsekas1995dynamic}, which the above definition
allows (proven in Appendix~\ref{app:bellman-opt-cond-redqueen}):
\begin{lemma}[Bellman'{}s Principle of Optimality] \label{lem:bellman-opt-cond-redqueen}
The optimal cost satisfies the follo\-wing recursive equation:
\begin{equation}
	J(r(t),\gamma(t),t) =
	\min_{u(t,t+dt]} \big\{ \mathbb{E} \left[ J(r(t+dt),\gamma(t+dt),t+dt)\right ] + \ell(r(t),u(t)) \, dt\big\}, \label{eq:bellman}
	%% \big\vert r(t),\gamma(t)
\end{equation}
\end{lemma}
where the expectation is taken over all trajectories of the control and noise jump processes, $N$ and $M$, in $(t, t+dt]$.
Then, we use the Bellman'{}s principle of optimality to derive a partial differential equation on $J$, often called the Hamilton-Jacobi-Bellman (HJB)
equation~\cite{hanson2007}.
To do so, we first assume $J$ is continuous and then rewrite Eq.~\ref{eq:bellman} as
\begin{align}
	J(r(t),\gamma(t),t) &= \min_{u(t,t+dt]} \big\{ \mathbb{E} \left[ J(r(t),\gamma(t),t) + dJ(r(t),\gamma(t),t) \right ] +\ell(r(t),u(t)) \, dt \big\} \nonumber \\ % _{(N,M)(t,t+dt]}
	0 &= \min_{u(t,t+dt]} \big\{ \mathbb{E} \left[dJ(r(t),\gamma(t),t) \right ] + \ell(r(t),u(t)) \, dt\big\}, \label{eq:diff-j-eq-short} % _{(N,M)(t,t+dt]}
\end{align}
where $dJ(r(t),\gamma(t),t) = J(r(t+dt),\gamma(t+dt),t+dt) - J(r(t),\gamma(t),t)$. Then, we differentiate $J$ with respect to time $t$, $r(t)$ and $\gamma(t)$ using to following Lemma
(proven in the Appendix~\ref{app:diff-cost-redqueen}):
\begin{lemma}\label{lem:diff-cost-redqueen}
	The differential $dJ(r(t),\gamma(t),t)$ of the cost-to-go function $J(r(t),\gamma(t),t)$, as defined by Eq.~\ref{eq:cost-to-go-redqueen}, is given by 
	\begin{align*}
	dJ(r(t),\gamma(t),t) &= J_t(r(t),\gamma(t),t)dt  + \left[\gamma_0'(t) + \omega \gamma_0(t) - \omega \gamma(t) \right]J_{\gamma}(r(t),\gamma(t),t)dt \\
	&\quad +[J(0,\gamma(t),t)-J(r(t),\gamma(t),t)] dN(t) \\
	&\quad + [J(r(t)+1,\gamma(t)+\alpha,t)-J(r(t),\gamma(t),t)] dM(t),
	\end{align*}
	where $J_t$ and $J_{\gamma}$ are derivatives of $J$ with respect to $t$ and $\gamma$, respectively.
\end{lemma}
Next, if we plug in the above equation in Eq.~\ref{eq:diff-j-eq-short}, it follows that
%J(r(t),\gamma(t),t)
\begin{align*}
	0 = \min_{u(t,t+dt]} & \Big\{J_t(r(t),\gamma(t),t) dt +\left[\gamma_0'(t)+ \omega \gamma_0(t) - \omega \gamma(t) \right]J_{\gamma}(r(t),\gamma(t),t) dt \\
	&\, +[J(0,\gamma(t),t)-J(r(t),\gamma(t),t)] \mathbb{E}[dN(t)] \\
	&\, + [J(r(t)+1,\gamma(t)+\alpha,t)-J(r(t),\gamma(t),t)] \mathbb{E}[dM(t)] + \ell(r(t),u(t)) \, dt \Big\}.  % _{M(t,t+dt]}
\end{align*}
Now, using $\mathbb{E}[dN(t)] = u(t)dt$ and $\mathbb{E}[dM(t)] = \gamma(t) dt$, and rearranging terms, the HJB equation follows:
\begin{align}
	0&= J_t(r(t),\gamma(t),t) +\left[\gamma_0'(t)+ \omega \gamma_0(t)- \omega \gamma(t) \right]J_{\gamma}(r(t),\gamma(t),t) \nonumber \\
	&\quad+[J(r(t)+1,\gamma(t)+\alpha,t)-J(r(t),\gamma(t),t)]\gamma(t) \nonumber \\
	&\quad +\min_{u(t,t+dt]} \ell(r(t),u(t)) + [J(0,\gamma(t),t)-J(r(t),\gamma(t),t)]u(t).  \label{eq:bellman-pde-min}
\end{align}
To be able to continue further, we need to define the loss $\ell$ and the penalty $\phi$. Following the literature on stochastic optimal
control~\cite{hanson2007}, we consider the following quadratic forms, which will turn out to be a tractable choice\footnote{\scriptsize
Considering other losses with a specific semantic meaning (\eg, $\II(r(t) \leq k)$) is a challenging direction for future work.}:
\begin{equation}
	\phi(r(t_f)) = \frac{1}{2}r^2(t_f) \quad \text{and} \quad \ell(r(t),u(t)) = \frac{1}{2} s(t) \,r^2(t) + \frac{1}{2} q\,u^2(t), \nonumber
\end{equation}
where $s(t)$ is a time significance function $s(t) \geq 0$, which favors some periods of times (\eg, times in which the follower is online\footnote{\scriptsize
Such information may be hidden but one can use the followers'{} posting activity or geographic location as a proxy~\cite{karimi2016smart}.}),
and $q$ is a given parameter, which trade-offs visibility and number of broadcasted posts.
\begin{algorithm}[t]
\small
\begin{algorithmic}[1]
\STATE \textbf{Input}: parameters $q$ and $s$ \;
\STATE \textbf{Output}: Returns time for the next post \;
\STATE $t \gets \infty$;\, $\tau \gets othersNextPost(\,)$ \;
\WHILE{$\tau < t$}
  	\STATE $\Delta \sim Sample(\sqrt{{s}/{q}}) $\;
  	\STATE $t \gets \min(t,\,\tau+\Delta)$\;
  	\STATE $\tau \gets othersNextPost(\,)$\;
\ENDWHILE
\RETURN $t$\;
\caption{\redqueen{} for fixed $s$, $q$ and one follower.}
\label{alg:sampling-redqueen}
\end{algorithmic}
\end{algorithm}

Under these definitions, we take the derivative with respect to $u(t)$ of Eq.~\ref{eq:bellman-pde-min} and
uncover the relationship between the optimal intensity and the optimal cost:
\begin{align} \label{eq:relationship-rate-j}
	u^*(t) = q^{-1}\left[J(r(t),\gamma(t),t) - J(0,\gamma(t),t)\right].
\end{align}
Finally, we substitute the above expression in Eq.~\ref{eq:bellman-pde-min} and find that the optimal cost $J$ needs to satisfy the following
nonlinear differential equation:
\begin{align}
	0&=J_t(r(t),\gamma(t),t) +\frac{1}{2}s(t)\,r^2(t)-\frac{1}{2}q^{-1}\left[J(r(t),\gamma(t),t)-J(0,\gamma(t),t)\right]^2 \nonumber \\
	&\quad + \left[\gamma_0'(t)+ \omega \gamma_0(t) - \omega \gamma(t)\right]J_{\gamma}(r(t),\gamma(t),t) +[J(r(t)+1,\gamma(t)+\alpha,t)-J(r(t),\gamma(t),t)]\gamma(t) \label{eq:bellman-pde}
\end{align}
with $J(r(t_f),\gamma(t_f),t_f)=\phi(r(t_f))$ as the terminal condition. The following technical lemma provides us with a solution to the above
equation (proven in Appendix~\ref{app:opt-con-sol}):
\begin{lemma} \label{lem:opt-con-sol}
In the space of $m$-degree polynomials, the following polynomial is the only solution to Eq.~\ref{eq:bellman-pde}:
\begin{align*}
	J(r(t),\gamma(t),t) = f(t)+ \sqrt{{s(t)}{q}} \,  r(t) + \sum_{j=1}^m g_{j}(t)\gamma^j(t),
\end{align*}
where $f(t)$ and $g_{j}(t)$ are time-varying functions which can be found by solving a linear system of differential
equations.
\end{lemma}
Given the above Lemma and Eq.~\ref{eq:relationship-rate-j}, the optimal intensity is readily given by
following theorem:
\begin{theorem}
The optimal intensity for the when-to-post problem defined by Eq.~\ref{eq:average-loss-one-follower} with quadratic loss and penalty function is given by
$u^*(t) = \sqrt{{s(t)}/{q}} \, r(t)$.
\end{theorem}
The optimal intensity only depends on the position of the most recent post by broadcaster $i$ in her follower'{}s feed and, thus, allows for a very
efficient procedure to sample posting times, which exploits the superposition theorem~\cite{Kingman1992}.
The key idea is as follows: at any given time $t$, we can view the process defined by the optimal intensity as a superposition of $r(t)$
inhomogeneous Poisson processes with intensity $\sqrt{{s(t)}/{q}} \, r(t)$ which starts at jumps of the rank $r(t)$, and find the next sample by
computing the minimum across all samples from these processes.
Algorithm~\ref{alg:sampling-redqueen} summarizes our (sampling) method, which we name \redqueen~\cite{carroll1917through}.
Within the algorithm, $\emph{Sample}(\sqrt{{s}/{q}})$ samples from a Poisson process with intensity $\sqrt{{s}/{q}}$ and
$othersNextPost(\,)$ returns the time of the next event by other broadcasters in the followers'{} feeds, once the event happens. In practice, we only need to know if the event happens before we post.
Remarkably, it only needs to sample $M(t_f)$ times from a (in)homogeneous Poisson process (if significance is time varying) and requires $O(1)$ space.

\xhdr{Optimizing for multiple followers}
Given a broadcaster $i$ with $N_i(t) = N(t)$ and $\lambda_i(t) = u(t)$ and her followers $\Ncal(i)$ with $\bm{M}_{{\scriptscriptstyle\setminus} i}(t) = \bm{M}(t)$ and
$\gammab_{{\scriptscriptstyle\setminus} i}(t) = \gammab(t)$, we can write the dynamics of $\bm{M}(t)$ and $\bm{r}(t)$, which we need to solve Eq.~\ref{eq:average-loss},
using Eqs.~\ref{eq:multi-hawkes-redqueen},~\ref{eq:hawkes-dyn} and~\ref{eq:rank-dynamics}:
\begin{align*}
	d\gammab(t) &= \left[\gammab'_0(t) + w \gammab_0(t) - w \gammab(t) \right] dt + \Db \, d\Mb(t) \\
	d\bm{r}(t) &= -\bm{r}(t)\,dN(t) + d{\bm{M}}(t).
\end{align*}
Then, consider the following quadratic forms for the loss $\ell$ and the penalty $\phi$:
\begin{align}
	\phi(\bm{r}(t_f)) &= \sum_{i=1}^n \frac{1}{2} r_i^2(t_f) \nonumber \\
	\ell(\bm{r}(t),u(t),t) &= \sum_{i=1}^n \frac{1}{2} s_i(t) r_i^2(t) + \frac{1}{2} q \,u^2(t). \nonumber
\end{align}
where $s_i(t)$ is the time significance function for follower $i$, as defined above, and $q$ is a given parameter.
Proceeding similarly as in the case of one follower, we can show that: 
\begin{align}\label{eq:optimal-rate-multi}
	u^*(t) = \sum_{i=1}^{n} \sqrt{{s_i(t)}/{q}} \, r_i(t),
\end{align}
which only depends on the position of the most recent post by broadcaster $i$ in her followers'{} feeds. Finally, we can readily adapt \redqueen (Algorithm~\ref{alg:sampling-redqueen}) to
efficiently sample the posting times using the above intensity---it only needs to sample $|\cup_{j \in \Ncal(i)} \Fcal_{j \backslash i}(t_f)|$ values from an (in)homogeneous Poisson process
(if significance is time varying) and requires $O(|\Ncal(i)|)$ space.

\vspace{2mm}
\xhdr{Optimizing with an oracle}
In this section, we consider a broadcaster $i$ with $N_i(t) = N(t)$ and $\lambda_i(t) = u(t)$, only one of her followers $j$ with $M_{j {\scriptscriptstyle\setminus} i}(t)=M(t)$,
and a constant significance $s(t) = s$. The derivation can be easily adapted to the case of multiple followers and time-varying significance.

Suppose there is an (idealized) oracle that reveals $M(t)$ from $t_0$ to $t_f$, \ie, the history $\Fcal_{j {\scriptscriptstyle\setminus} i}(t_f) = \Fcal(t_f)$ is given, and $M(t_f) = |\Fcal(t_f)| = m$.
Then, we can rewrite Eq.~\ref{eq:average-loss} as 
\begin{align}
\underset{u(t_0, t_f]}{\text{minimize}} & \quad \EE_{N(t_0, t_f]}\left[ \phi(r(t_f)) + 	\int_{t_0}^{t_f} \ell(r(s),u(s)) \, ds \right] \nonumber \\*
\text{subject to} & \quad u(t) \geq 0 \quad \forall t \in (t_0, t_f], \nonumber
\end{align}
where the expectation is only taken over all possible realizations of the counting process $N(t_0, t_f]$ since $M(t_0, t_f]$ is revealed by the oracle and, thus,
deterministic.
\begin{algorithm}[t]
\small
\begin{algorithmic}[1]
\STATE \textbf{Input}: Initial state $r_0$, interval widths $w_1,\ldots,w_{m+1}$, parameter $q$ and significance $s(t) = s$ \;
\STATE \textbf{Output}: Overall cost $J(r_0,0)$, optimal control $u_0^*,\ldots,u_m^*$ \;

\FOR{$r \gets r_0+m$ \KwTo $0$}
	\STATE $J(r,m+1) \gets \frac{1}{2}r^2$\;
\ENDFOR

\FOR{$k \gets m$ \KwTo $0$}

	\FOR{$r \gets r_0+k-1$ \KwTo $0$}
		\STATE $J(r,k)=\min\{\frac{1}{2}q + J(0,k+1)  ,\, \frac{1}{2} s w_{k+1}(r+1)^2 + J(r+1,k+1)\}$\;
	\ENDFOR
\ENDFOR

\FOR{$k \gets 0$ \KwTo $m$}
	\IF{$\frac{1}{2}q + J(0,k+1) < \frac{1}{2} s w_{k+1}(r_k+1)^2 + J(r_k+1,k+1)$}
  		\STATE $u_k^* \gets 1;\, r_{k+1} \gets 0$\;
	\ELSE
  		\STATE $u_k^* \gets 0;\, r_{k+1} \gets r_k+1$\;
	\ENDIF
\ENDFOR

\Return{$J(r_0,0),\, u_0^*,\ldots,u_m^*$}
\caption{Optimal posting times with an oracle.}
\label{alg:oracle}
\end{algorithmic}
\end{algorithm}

Similar to the previous sections, assume the loss $\ell$ and penalty $\phi$ are quadratic. The best times for user $i$ to post will always coincide with one
of the times in $\Fcal(t_f)$ since, given a posting time $\tau_i \in (t_k, t_{k+1})$, where $t_{k}, t_{k+1} \in \Fcal(t_f)$, one can reduce the cost by $(1/2)q(\tau_i-t_k)r^2(t_k)$ by
choosing instead to post at $t_k$.
As a consequence, we can discretize the  dynamics of $r(t)$ in times $\Fcal(t_f)$, and write $r_{k+1} = r_{k}+1 - (r_{k} + 1) u_k$,
where $r_k = r(t_k^{\scriptscriptstyle-})$, $u_k = u(t_k^{\scriptscriptstyle+}) \in \{0,1\}$,  $t_k \in \Fcal(t_f)$. We can easily see that $r_k$ is bounded by $0 \leq r_k < r_0+m$. Similarly, we can derive the optimal cost-to-go in
discrete-time as
\begin{align*}
	J(r_k,k) & = \min_{u_k,\ldots,u_m} \frac{1}{2}r_{m+1}^2 + \sum_{i=k}^{m} \frac{1}{2} q \, w_{i+1}\,r_{i+1}^2 + \frac{1}{2} s\,u_i^2,
\end{align*}
where $w_i = t_{i} - t_{i-1}$. 
Next, we can break the minimization and use Bellman'{}s principle of optimality,
\begin{align*}
	J(r_k,k) = \min_{u_k}  \frac{1}{2} q \, w_{k+1} \, r_{k+1}^2 + \frac{1}{2}s\,u_k^2+J(r_{k+1},k+1),
\end{align*}
and, since $u_k \in \{0, 1\}$, the above recursive equation can be written as
\begin{equation}
J(r_k,k) = \min \left\{\frac{1}{2} s + J(0,k+1)  ,\, \frac{1}{2}q\,w_{k+1}(r_{k}+1)^2 + J(r_{k}+1,k+1) \right\}. \nonumber
\end{equation}
Finally, we can find the optimal control $u^{*}_{k},\, k=0, \ldots, m$ and cost $J(r_0,0)$ by backtracking from the terminal condition $J(r_{m+1},m+1)=r_{m+1}^2/2$
to the initial state $r_0$, as summarized in Algorithm~\ref{alg:oracle}, which can be adapted to multiple followers.
Note that, in this case, the optimal strategy is not stochastic and consists of a set of optimal posting times, as one could have guessed.
However, for multiple followers, the complexity of the algorithm is $O(m^{2})$, where
$m = |\cup_{j \in \Ncal(i)} \Fcal_{j \backslash i}(t_f)|$.
%

%-------------------------------------------------------
%-------------------------------------------------------
\section{A Global Problem: Activity Maximization}
\label{sec:cheshire}
In this section, we take the perspective of an entire online social networking site and design an online algorithm to find how much should we incentivize a small number of influential users to
post more over time to increase the overall number of additional posts, shares or replies in the site.
To this aim, we first describe how to formally model such incentive mechanism in social networks. Then, we state the activity shaping problem and address the problem from the perspective of
stochastic optimal control of SDEs with jumps, similarly as in Section~\ref{sec:redqueen}.

\subsection{Triggering Additional Endogenous Actions}
Given a social network $\Gcal = (\Vcal, \Ecal)$ with $|\Vcal| = n$ users, we trigger additional endogenous user actions by directly incentivizing (\eg, paying) for $\Pb(t)$ actions, where $P_i(t)$ counts
the number of directly incentivized actions taken by user $i \in \Vcal$ before time $t$ and we can characterize $\Pb(t)$ by means of $n$ intensity functions $\ub(t)$, \ie, $\EE[d\Pb(t)] = \ub(t)dt$.
Then, if we assume the strength of influence $\Bb$ between users is the same both for organic and incentivized actions\footnote{This assumption seems reasonable in some scenarios, \eg, it is often difficult 
to notice whether an influential user is being paid for posting a message. However, one could relax this assumption by considering a different influence matrix $C \neq B$ for the additional endogenous actions 
in Eq.~\ref{eq:steering-multi-hawkes}, change the SDE, Eq.~\ref{eq:cheshire-dynamics} and HJB, Eq.~\ref{eq:hjb} accordingly, and derived the new optimal control in Theorem~\ref{thm:opt-control-cheshire}.},
as previous work~\cite{shaping14nips, farajtabar2016msc}, we can rewrite the users'{} intensities $\lambdab(t)$, given by Eq.~\ref{eq:multi-hawkes-cheshire}, as
\begin{align} \label{eq:steering-multi-hawkes}
	\lambdab(t) &= \lambdab_0(t) +  \Bb \int_0^t \kappa_{\omega}(t-s) \, d\bm{N}(s) +  \underset{\mbox{\scriptsize Additional endogenous actions}}{\underbrace{\Bb \int_0^t \kappa_{\omega}(t-s) \, d\bm{P}(s)}}.
\end{align}
Here, note that treating directly incentivized actions as a different counting process $\Pb(t)$ prevents the intensities $\lambdab(t)$ from including
their intensity but instead including only the intensity of their follow-ups (\ie, additional endogenous actions), which we aim to maximize.
Then, it is easy to derive the following alternative representation, similarly as in Proposition~\ref{prop:hawkes}, which we will use in our stochastic
optimal control algorithm:
\begin{proposition}\label{thm:hawkes-dynamics}
	Let $\bm{N}(t)$ and $\Pb(t)$ be two multidimensional counting processes with associated intensities $\bm{\lambda}(t)$, given by Eq.~\ref{eq:steering-multi-hawkes},
	and $\bm{u}(t)$, respectively. Then, the dynamics of the intensity $\bm{\lambda}(t)$ can be expressed using the following jump SDEs:
	\begin{align}\label{eq:cheshire-dynamics}
	d\bm{\lambda}(t) = \left[\lambdab'_0(t) + \omega \lambdab_0(t) - \omega \lambdab(t) \right] dt + \bm{B} \, d\bm{N}(t) + \bm{B} \, d\Pb(t)
	\end{align}
	with the initial condition $\lambdab(0)= \lambdab_0(0)$.
\end{proposition}
\vspace{-1mm}
In the remainder of the section, for ease of exposition, we assume $\lambdab_0(t) = \lambdab_0$ and thus $\lambdab_0'(t) = 0$.

\subsection{Problem Formulation}
Given a social network $\Gcal = (\Vcal, \Ecal)$ with $|\Vcal| = n$ users, our goal is to find the optimal users'{} intensities for directly incentivized actions $\ub(t)$ (for short, control intensities)
that minimize the expected value of a particular loss function $\ell(\bm{u}(t), \bm{\lambda}(t))$ of the control intensities and the users'{} intensities for not directly incentivized actions over a
time window $(t_0,t_f]$, \ie,
\begin{align}\label{eq:cheshire-prob-def}
&\underset{\bm{u}(t_0,t_f]}{\text{minimize}} \quad
\mathbb{E}_{(\bm{N},\Pb)(t_0,t_f]}\left[ \phi(\bm{\lambda}(t_f)) + \int_{t_0}^{t_f} \ell(\bm{\lambda}(s), \bm{u}(s)) \, ds \right] \nonumber \\
&\text{subject to} \quad u_i(t) \geq 0, \,\, \forall t \in (t_0,t_f], \, i=1,\ldots,n
\end{align}
where $\bm{u}(t_0,t_f]$ denotes the control intensities from $t_0$ to $t_f$, the dynamics of $\bm{N}(t)$ are given by Eq.~\ref{eq:cheshire-dynamics},
and the expectation is taken over all possible realizations of the two counting processes $\bm{N}(t)$ and $\Pb(t)$ during interval $(t_0,t_f]$.
Here, by considering a loss that is nonincreasing (nondecreasing) with respect to the intensities $\bm{\lambda}(t)$ ($\bm{u}(t)$), we will trade-off
number of directly incentivized actions and number of additional endogenous actions.
Finally, note that the optimal intensities $\bm{u}(t)$ at time $t$ may depend on the intensities $\bm{\lambda}(t)$ and thus the associated
counting process $\Pb(t)$ may be doubly stochastic.

\subsection{Stochastic Optimal Control Algorithm} \label{sec:optimal-control-cheshire}
In this section, we proceed similarly as in Section~\ref{sec:optimal-control-redqueen} and tackle the activity maximization problem defined by Eq.~\ref{eq:cheshire-prob-def} from the perspective of stochastic optimal control of SDEs
with jumps~\cite{hanson2007}.
More specifically, we first define a novel optimal cost-to-go function specially designed for this global problem and then derive and solve the corresponding HJB equation
to find the optimal control intensities.

\begin{definition}\label{thm:cost-def-cheshire}
The optimal cost-to-go $J(\bm{\lambda}(t),t)$ is defined as the minimum of the expected value of the cost of going from the state with intensity $\bm{\lambda}(t)$ at time $t$ to the final state at time $t_f$, \ie,
\begin{align}\label{eq:cost-to-go-cheshire}
J(\bm{\lambda}(t),t) = \min_{\bm{u}(t,t_f]} \mathbb{E}_{(\bm{N},\Pb)(t,t_f]}\left[ \phi(\bm{\lambda}(t_f)) + \int_t^{t_f} \ell(\bm{\lambda}(s), \bm{u}(s)) \, ds \right],
\end{align}
where the expectation is taken over all trajectories of the counting processes $\bm{N}$ and $\Pb$ in the $(t, t_f]$ interval, given the initial values of $\bm{\lambda}(t)$ and $\bm{u}(t)$.
\end{definition}
Next, we use the Bellman'{}s principle of optimality~\cite{bertsekas1995dynamic}, which the above cost-to-go $J(\bm{\lambda}(t),t)$ also allows for (proof is similar to that of Lemma~\ref{lem:bellman-opt-cond-redqueen})), to break the problem into smaller subproblems and rewrite Eq.~\ref{eq:cost-to-go-cheshire} as 
\begin{align}
J(\bm{\lambda}(t),t) &= \min_{\bm{u}(t, t+dt]} \big\{ \mathbb{E} \left[ J(\bm{\lambda}(t),t) + dJ(\bm{\lambda}(t),t) \right] + \ell(\bm{u}(t), \bm{\lambda}(t)) \, dt  \big\} \nonumber \\
0 &= \min_{\bm{u}(t, t+dt]} \big\{ \mathbb{E} \left[ dJ(\bm{\lambda}(t),t) \right] + \ell(\bm{u}(t), \bm{\lambda}(t)) \, dt  \big\},
\label{eq:bellman-opt-simplified}
\end{align}
where $dJ(\bm{\lambda}(t),t) = J(\bm{\lambda}(t+dt),t+dt) - J(\bm{\lambda}(t),t)$. Then, we explicitly differentiate $J$ with respect to time $t$ and $\lambda(t)$ using the
following Lemma (proven in the Appendix~\ref{app:diff-cost-cheshire}):
\begin{lemma}\label{lem:diff-cost-cheshire}
The differential $dJ(\bm{\lambda}(t),t)$ of the cost-to-go function $J(\bm{\lambda}(t),t)$, as defined by Eq.~\ref{eq:cost-to-go-cheshire}, is given by
\begin{align*}
	J_t(\bm{\lambda}(t),t) \, dt &+ [w \lambdab_0 - w \bm{\lambda}(t)]^T \nabla_{{\bm{\lambda}}}J(\bm{\lambda}(t),t) \, dt
	+ \sum_{i} \left[J(\bm{\lambda}(t)+\bm{b}_i, t)-J(\bm{\lambda}(t), t) \right] dN_i(t) \\ % \vspace*{-5mm}
	&+ \sum_{i} \left[J(\bm{\lambda}(t)+\bm{b}_i, t)-J(\bm{\lambda}(t), t) \right] dP_i(t),
\end{align*}
where $\bm{b}_i$ is the $i$'{}th column of $\bm{B}$, $J_t$ is derivative of $J$ with respect to $t$ and $\nabla_{\bm{\lambda}} J$ is the gradient of $J$ with respect
to $\bm{\lambda}(t)$.
\end{lemma}
Next, if we plug the above equation in Eq.~\ref{eq:bellman-opt-simplified} and use that $\mathbb{E}[N_i(t)]=\lambda_i(t)\,dt$ and $\mathbb{E}[P_i(t)]=u_i(t)\,dt$, the HJB
equation follows:
\begin{align}
	0 &= J_t(\bm{\lambda}(t),t)
	+ [w \lambdab_0 - w \bm{\lambda}(t)]^T \nabla_{{\bm{\lambda}}}  J(\bm{\lambda}(t),t)+ \bm{\lambda}^T(t) \, \Delta_B J
	+ \min_{\bm{u}(t)}  \ell(\bm{u}(t), \bm{\lambda}(t)) +  \bm{u}^T(t) \, \Delta_B J, \label{eq:hjb}
\end{align}
where $\Delta_{B}J$ denotes a vector whose $i$'{}th element is given by $(\Delta_{B}J)_i = J(\bm{\lambda}(t)+\bm{b}_i, t)-J(\bm{\lambda}(t), t)$.
\begin{algorithm}[!t]
\small
\begin{algorithmic}[1]
\STATE\textbf{Initialization: } \\
\STATE Compute $\bm{H}(t)$ and $\mathbf{g}(t)$ \;
\STATE $\mathbf{u}(t) \leftarrow -\bm{S}^{-1} \big[ \Bb^T (\bm{g}(t) + \bm{H}(t) \lambdab_0) + \frac{1}{2} \diag(\Bb^T \bm{H}(t) \Bb) \big]$ \;
\vspace{1mm}

\STATE\textbf{General subroutine: } \\
\STATE $(i, \tau) \gets Sample(\mathbf{u}(t))$ \;
\STATE $(j, s) \gets NextAction(\,)$ \;
 \WHILE{$s < \tau$}
 \STATE $\lambdab_N(t) \leftarrow \Bb \mathbf{e}_j \kappa_{\omega}(t - s)$ \;
 \STATE $\mathbf{u}_N(t) \leftarrow -\bm{S}^{-1} \Bb^T \bm{H}(t) \lambdab_N(t)$ \;
 \STATE $(k, r) \leftarrow Sample(\mathbf{u}_N(t))$ \;
 \IF{$r < \tau$}
 	\STATE $\tau \leftarrow r$ \;
	\STATE $i \leftarrow k$ \;
 \ENDIF
 \STATE $\mathbf{u}(t) \leftarrow \mathbf{u}(t) + \mathbf{u}_N(t)$ \;
 \STATE $(j, s) \leftarrow NextAction(\,)$ \;
 \ENDWHILE
 \STATE $\lambdab_P(t) \leftarrow \Bb \mathbf{e}_i \kappa_{\omega}(t - \tau)$ \;
 \STATE $\mathbf{u}_P(t) \leftarrow -\bm{S}^{-1} \Bb^T \bm{H}(t) \lambdab_P(t)$ \;
 \STATE $\mathbf{u}(t) \leftarrow \mathbf{u}(t) + \mathbf{u}_P(t)$ \;
 \STATE {\bf return} $(i, \tau)$
 %
% \vspace{-7mm}
\caption{\mbox{\cheshire: it returns user $i$ and time $\tau$ for the next incentivized action}} \label{alg:sampling-cheshire}
\end{algorithmic}
%\end{multicols}
\end{algorithm}

To solve the above equation, we need to define the loss and penalty functions, $\ell$ and $\phi$. Similarly as in Section~\ref{sec:redqueen}, we consider the following quadratic
forms, which will turn out to be a tractable choice:
\begin{equation*}% \label{eq:loss-func}
\ell(\bm{u}(t), \bm{\lambda}(t)) = -\frac{1}{2} \bm{\lambda}^T(t) \,\bm{Q}\, \bm{\lambda}(t) + \frac{1}{2} \bm{u}^T(t) \,\bm{S}\, \bm{u}(t)
\quad \mbox{and} \quad \phi(\bm{\lambda}(t_f)) = -\frac{1}{2} \bm{\lambda}^T(t_f) \,\bm{F}\, \bm{\lambda}(t_f),
\end{equation*}
where $\bm{Q}$, $\bm{F}$ and $\bm{S}$ are given symmetric matrices\footnote{\scriptsize In practice, we will consider diagonal matrices.} with
$q_{ij} \geq 0$, $f_{ij} \geq 0$ and $s_{ij} \geq 0$ for all $i, j \in [n]$.
These matrices allow us to trade-off the number of directly incentivized actions over time and the number of additional endogenous actions, both
over time and at time $t_f$.
Under these definitions, we can find the relationship between the optimal intensity and the optimal cost by solving the
minimization in the HJB equation:
\begin{align}
&\underset{\bm{u}(t)}{\text{minimize}} \quad
\bm{u}^T(t) \, \Delta_B J + \frac{1}{2} \bm{u}^T(t) \,\bm{S}\, \bm{u}(t) \quad \text{subject to} \quad  u_i(t) \geq 0, \,\, i=1,\ldots,n. \nonumber
\end{align}
By taking the differentiation with respect to $\bm{u}(t)$, the solution of the unconstrained minimization is given by 
\begin{align} \label{eq:opt-ctrl}
	\bm{u}^*(t) = - \bm{S}^{-1} \Delta_{B} J,
\end{align}
which is the same as the solution to the constrained problem given that $s_{ij} \geq 0$, by definition, and $(\Delta_B J)_i \leq 0$, as proved in the Appendix~\ref{app:positive-delta}.
Then, we substitute Eq.~\ref{eq:opt-ctrl} into Eq.~\ref{eq:hjb} and find that the optimal cost $J$ needs to satisfy the following partial
differential equation:
\begin{align}
	0 &= J_t(\bm{\lambda}(t),t)
	+ [w \lambdab_0 - w \bm{\lambda}(t)]^T \nabla_{{\bm{\lambda}}}J(\bm{\lambda}(t),t) + \bm{\lambda}^T(t) \, \Delta_B J
	-\frac{1}{2} \bm{\lambda}^T(t) \,\bm{Q}\, \bm{\lambda}(t)
	- \frac{1}{2} \Delta_B J^T \bm{S}^{-1} \Delta_B J,  \label{eq:hjb2}
\end{align}
with $J(\bm{\lambda}(t_f), t_f) = \phi(\bm{\lambda}(t_f))$ as the terminal condition. The following lemma provides us with a solution to the above equation
(proven in Appendix~\ref{sec:quad-proposal-cheshire}):
\begin{lemma} \label{lem:quad-proposal-cheshire}
In the space of $m$-degree degree polynomials, the following quadratic form is the only solution to Eq.~\ref{eq:hjb2}:
\begin{equation}
J(\bm{\lambda}(t), t) = f(t) + \bm{g}(t)^T \bm{\lambda}(t) + \frac{1}{2} \bm{\lambda}(t)^T \bm{H}(t) \bm{\lambda}(t). \nonumber
\end{equation}
where $\bm{g}(t)$ and $\bm{H}(t)$ can be found by solving the following differential equations:
\begin{align}
	\dot{\bm{H}}(t) &= (\omega \Ib-\Bb)^T \bm{H}(t) + \bm{H}(t) (\omega \Ib-\Bb) \nonumber
	+ \bm{H}(t) \Bb \bm{S}^{-1} \Bb^T \bm{H}(t) + \bm{Q} \nonumber  \\ % \label{equ:ode-h} \\
	\dot{\bm{g}}(t) &= [\omega \Ib-\Bb^T+\bm{H}(t) \Bb \bm{S}^{-1} \Bb^T] \bm{g}(t) - \omega  \bm{H}(t) \lambdab_0 \nonumber
	 + \frac{1}{2} \left[ \bm{H}(t) \Bb \bm{S}^{-1} - \Ib \right] \diag(\Bb^T \bm{H}(t) \Bb). \nonumber
\end{align}
\end{lemma}

In the above lemma, note that the first differential equation is a matrix Riccati differential equation, which can be solved using many well-known efficient numerical
solvers~\cite{garrett2013}, and the second one is a first order differential equation which has closed form solution. Both equations are solved backward in time with
final conditions $\bm{g}(t_f)=\bm{0}$ and $\bm{H}(t_f)=-\bm{F}$.

Finally, given the above cost-to-go function, the optimal intensity is given by the follo\-wing theorem:
\begin{theorem}\label{thm:opt-control-cheshire}
The optimal intensity for the activity maximization problem defined by Eq.~\ref{eq:cheshire-prob-def} with quadratic loss and penalty function is given by
\begin{align*}
\bm{u}^*(t) = -\bm{S}^{-1} \big[ \Bb^T \bm{g}(t) + \Bb^T \bm{H}(t) \bm{\lambda}(t)+\frac{1}{2} \diag(\Bb^T \bm{H}(t) \Bb) \big].
\end{align*}
\end{theorem}
Since the optimal intensity is linear in $\lambdab(t)$, it allows for an efficient procedure to sample the times of the users'{} directly
incentivized actions:
at any given time $t$, we can view the multidimensional control signal $\bm{u}(t)$ as a superposition of inhomogeneous multidimensional
Poisson processes, one per non incentivized action, which start when the actions take place.
Algorithm~\ref{alg:sampling-cheshire} summarizes our method, which we name \cheshire~\cite{carroll1917through}. 

Within the algorithm,
\emph{NextAction}() returns the time of the next (non directly incentivized) action in the network as well as the identity of the user who
takes the action, once the action happens,
$\bm{e}_j$ is an indicator vector where the entry corresponding to user $j$ is $1$,
and $\emph{Sample}(\bm{u}(t))$ samples from a multidimensional inhomogeneous Poisson process with intensity $\bm{u}(t)$
and it returns both the sampled time and dimension (\ie, user).
Moreover, note that the algorithm initially plans a user $i$ and time $\tau$ for the next directly incentivized action, however, if before $\tau$, a new organic
action takes place at $s < \tau$ and the intensity $\lambdab(t)$ changes, then the algorithm updates the user and time for the next directly incentivized action
using the superposition principle.
To sample from a multidimensional inhomogeneous Poisson process, there exist multiple methods \eg, refer to~\cite{lewis1979simulation}.
Finally, note that one can precompute most of the quantities the algorithm needs, \eg, lines 2-3, $\Bb \bm{e}_j$ in line 8, and $\bm{S}^{-1} \Bb^T \bm{H}(t)$
in line 9.
Given these precomputations, the algorithm only needs to perform $O(n)$ operations and sample $\bm{1}^{T} \Nb(t_f)$ times from an inhomogeneous Poisson
process.

%-------------------------------------------------------
%-------------------------------------------------------
\section{Experiments}
\label{sec:experiments}
In this section, we validate \redqueen (Algorithm~\ref{alg:sampling-redqueen}) and \cheshire (Algorithm~\ref{alg:sampling-cheshire}) using both synthetic and real data gathered from Twitter and
compare their performance with several state of the art methods and competitive baselines 
\cite{He2012}.
\subsection{When-to-post} \label{sec:experiments-redqueen}
\begin{figure}[t]
	\centering
	\adjustbox{trim={0.25\width} 0 {0.25\width} 0,clip}{\includegraphics[width=1.2\textwidth]{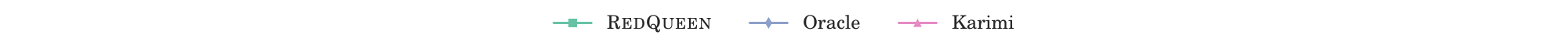}} \\
	\subfloat[\normalsize Position over time]{ % (lower is better)]{
	\includegraphics[width=.3\textwidth]{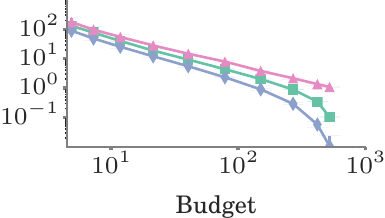}} \hspace{1mm}
	\subfloat[\normalsize Time at the top]{ % (higher is better)]{
	\includegraphics[width=.3\textwidth,]{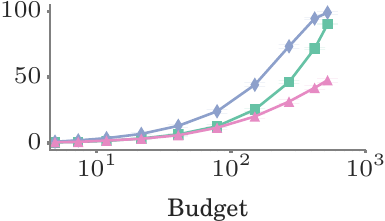}}
	\caption{Optimizing for one follower. Performance of \redqueen in comparison with the oracle and the method by~\cite{karimi2016smart}
	against number of broadcasted events. The feeds counting processes $M(t)$ due to other broadcasters are Hawkes processes with $\gamma_0=10$,
	$\alpha=1$ and $w=10$. 	In all cases, the time horizon $t_f - t_0$ is chosen such that the number of stories posted by other broadcasters is $\sim$$1000$.
	Error bars are too small to be seen.}
	\label{fig:performance-synthetic-one-follower}
\end{figure}
\begin{figure}[t]
      \centering
      \adjustbox{trim={0.25\width} 0 {0.25\width} 0,clip}{\includegraphics[width=1.2\textwidth]{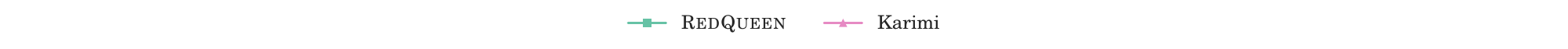}} \\
      \begin{tabular}{cc}
      \includegraphics[width=.3\textwidth]{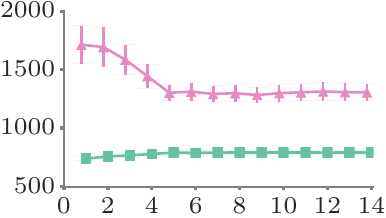} &
      {\includegraphics[width=.3\textwidth]{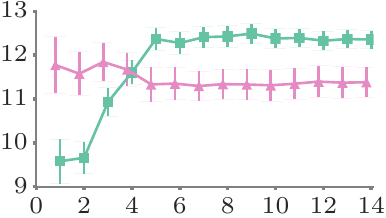}} \\
      {(a) Position over time} & {(b) Time at the top}
      \end{tabular} \label{fig:performance-synthetic-multiple-followers-number}
      \caption{Optimizing for multiple followers. Performance of \redqueen in comparison with the method by~\cite{karimi2016smart} against number of followers.
      The feeds counting processes $M(t)$ due to other broadcasters follow piecewise constant intensities, where the intensity of each follower remains constant within each piece,
      it varies as a half-sinusoid across pieces and it starts with a random initial phase. The performance of both methods stays constant upon addition of more followers.
      }
	\label{fig:performance-synthetic-multiple-followers}
\end{figure}

\subsubsection{Experiments on synthetic data}

\xhdr{Experimental setup}
We evaluate the performance via two quality measures: position over time, ${\scriptstyle \int_{0}^{T} r(t) dt}$, 
and time at the top, ${\scriptstyle \int_{0}^{T} \II(r(t) < 1) dt}$ and compare the performance of \redqueen against the oracle, described in Section~\ref{sec:optimal-control-redqueen},
and the method by~\cite{karimi2016smart}, which, to the best of our knowledge, is the state of the art.
Unless otherwise stated, we set the significance $s_i(t) = 1,\, \forall\, t, i$ and use the parameter $q$ to control the number of posts by \redqueen\footnote{\scriptsize The expected number
of posts by \redqueen{} are a decreasing function of $q$. Hence, we can use binary search to guess $q$ and then use averaging over multiple simulation runs to estimate the number of
posts made.}.

\xhdr{Optimizing for one follower}
We first experiment with one broadcaster and one follower against an increasing number of events (or budget).
We generate the counting processes $M(t)$ due to other broadcasters using Hawkes processes, as defined by Eq.~\ref{eq:multi-hawkes-redqueen}. We perform {$10$ independent simulation
runs} and compute the average and standard error (or standard deviation) of the quality measures.
Fig.~\ref{fig:performance-synthetic-one-follower} summarizes the results, which show that our method: (i) consistently outperforms the method by Karimi et al. by large margins;
(ii) achieves at most $3$$\times$ higher position over time than the oracle as long as the budget is $<$$30$\% of the posted events by all other broadcasters; and, (iii)
achieves $>$$40$\% of the value of time at the top that the oracle achieves.

\xhdr{Optimizing for multiple followers}
Next, we ex\-pe\-ri\-ment with one broadcaster and multiple followers.
In this case, we generate the counting processes $M(t)$ due to other broadcasters using piece-wise constant intensity functions. More specifically,
we simulate the feeds of each follower for $1$ day, using $24$ $1$-hour long segments, where the rate of posts remains constant per follower
in each segment and the rate itself varies as a half-sinusoid (\ie, from $\sin{0}$ to $\sin{\pi}$), with each follower starting with a random initial phase.
This experimental setup reproduces volume changes throughout the day across followers'{} feeds in different time-zones and closely resembles the
settings in previous work~\cite{karimi2016smart}.
The total number of posts by the \redqueen broadcaster is kept nearly constant and is used as the budget for the other baselines. Additionally, for Karimi'{}s
method, we provide as input the true empirical rate of tweets per hour for each user. Here, we do not compare with the oracle since, due to its quadratic
complexity, it does not scale.

Figure~\ref{fig:performance-synthetic-multiple-followers} summarizes the results. In terms of position over time, \redqueen outperforms Karimi'{}s method by a factor of
$2$.
In terms of time at the top, \redqueen achieves $\sim$18\% lower values than Karimi'{}s method for $1$-$4$ followers but $\sim$10\% higher values for $>$$5$ followers.
A potential reason Karimi'{}s method performs best in terms of time at the top for a low number of followers and piecewise constant intensities is that, while the number of
followers is low, there are segments which are clearly favorable and thus Karimi'{}s method concentrates posts on those. However, as the number of followers increases, there
are no clear favorable segments and advance planning does not give Karimi'{}s method any advantage.
On the other hand, \redqueen, due to its online nature, is able to adapt to transient variations in the feeds.
\begin{figure}[t]
	\centering
        \begin{tabular}{cc}
        {\includegraphics[width=.3\textwidth]{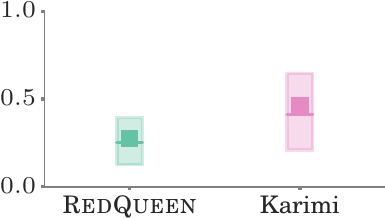}} & {\includegraphics[width=.3\textwidth]{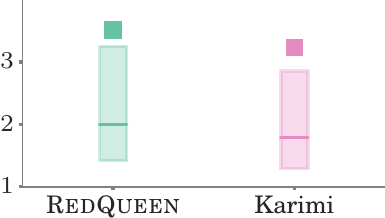}}\\
        {(a) Position over time} & {(b) Time at the top}
        \end{tabular}
	\caption{Performance of \redqueen and the method by~\cite{karimi2016smart} for $2000$ Twitter users, picked
	at random. The solid horizontal line (square) shows the median (mean) quality measure, normalized with respect to the value achieved by the users'{} actual
	\emph{true} posts, and the box limits correspond to the 25\%-75\% percentiles.}
	\label{fig:performance-real}
\end{figure}

\subsubsection{Experiments on real data}
\xhdr{Dataset description and experimental setup}
We use data gathered from Twitter as reported in previous work~\cite{cha2010measuring}, which comprises profiles of $52$ million
users, $1.9$ billion directed follow links among these users, and $1.7$ billion public tweets posted by the collected users. The follow
link information is based on a snapshot taken at the time of data collection, in September 2009.
Here, we focus on the tweets published during a two month period, from July 1, 2009 to September 1, 2009, in order to be able to consider
the social graph to be approximately static, and sample $2000$ users uniformly at random as broadcasters and record all the tweets
they posted.
Then, for each of these broadcasters, we track down their followers and record all the (re)tweets they posted as well as reconstruct
their timelines by collecting all the (re)tweets published by the people they follow.
We assign equal significance to each follower but filter out those who follow more than $500$ people since, otherwise, they would 
dominate the optimal strategy.
Finally, we tune $q$ such that the total number of tweets posted by our method is equal to the number of tweets the broadcasters
tweeted during the two month period (with a tolerance of $10\%$).
\begin{figure*}[t!]
	\centering
	\begin{tabular}{cc}
        \includegraphics[width=.3\textwidth]{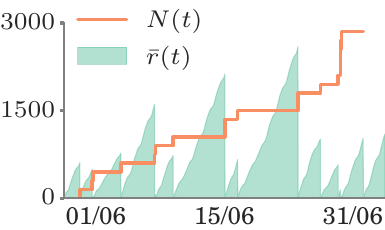} &
	\includegraphics[width=.3\textwidth]{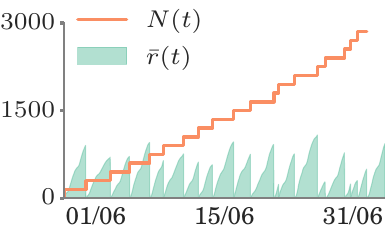} \\
	 {(a) True posts} & {(b) \redqueen (without significance)} \\
	 {$\frac{1}{T} \int_{0}^{T} \bar{r}(t) dt = 698.04$} & {$\frac{1}{T} \int_{0}^{T} \bar{r}(t) dt = 389.45$} \\
	 \vspace{2mm} {} \\
	\includegraphics[width=.3\textwidth]{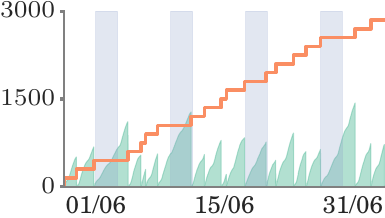} &
	\includegraphics[width=.3\textwidth]{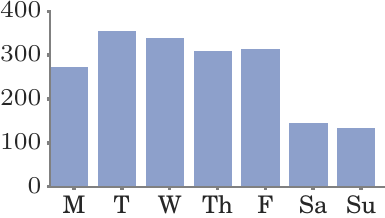} \\
	{(c) \redqueen (with significance)}  & {(d) Followers' (re)tweets per weekday}\\
  	 {$\frac{1}{T} \int_{0}^{T} \bar{r}(t) dt = 425.25$} & {}
	 \end{tabular}
	\caption{A broadcaster chosen from real data. Panels compares the position over time $\bar{r}(t) = \sum_{i=0}^{\Ncal} r(t) / \Ncal$ (in green; lower is better) for the most recent tweet posted by a real user against the most
	recent one \emph{posted} by a simulation run of \redqueen without and with significance. Here, the orange staircases represent the counts $N(t)$ of the tweets posted by the real user and \redqueen over time. The shaded area in panel (c) highlights weekends. We can see that \redqueen avoided tweeting on weekends, when the followers are less likely to be active/logged-in, as seen in panel (d).}
	\label{fig:individual-user-real}
\end{figure*}

\xhdr{Solution quality}
We only compare the performance of our method against the method by~\cite{karimi2016smart} since the oracle does not scale to the size of real data.
Moreover, for the method by Karimi et al., we divide the two month period into ten segments of approximately one week to fit the piecewise constant intensities of the
followers'{} timelines, which the method requires.
Fig.~\ref{fig:performance-real} summarizes the results by means of box plots, where position over time and time at the top are normalized with respect to the value
achieved by the broadcasters'{} actual \emph{true} posts during the two month period.
That means, if $y=1$, the optimized intensity achieves the same position over time or time at the top as the broadcaster's{} true posts.
In terms of position over time and time at the top, \redqueen consistently outperforms competing methods by large margins and achieves $0.28$$\times$ lower average position
and $3.5$$\times$ higher time at the top, in average, than the broadcasters'{} true posts---in fact, it achieves lower position over time (higher time at the top) for
$100$\% ($99.1$\%) of the users.

\xhdr{Time significance} We look at the actual broadcasting strategies for one real user and investigate the effect of a time varying
significance. We define $s_i(t)$ to be the probability that follower $i$ is online on that weekday, estimated empirically using the (re)tweets the follower posted as in~\cite{karimi2016smart}.
Fig.~\ref{fig:individual-user-real} compares the position over time for the most recent tweet posted by a real user against the most recent one \emph{posted} by a
simulation run of \redqueen with and without time varying significance. We can see that without significance information, \redqueen posts at nearly an even pace.
However, when we supply empirically estimated significance, \redqueen avoids tweeting at times the followers are unlikely to be active, \ie, the weekends, denoted by the shaded areas in panel (c) of Fig.~\ref{fig:individual-user-real}.
Due to this, the average position (maximum position) falls from $389{.}45$ ($1085{.}17$) to $425{.}25$ ($1431{.}0$), but is still lower than $698{.}04$ ($2597{.}9$)
obtained by the user's original posting schedule.

\subsection{Activity Maximization}
\subsubsection{Experiments on Synthetic Data}

In this section, we first shed light on \cheshire'{}s sampling strategy in two small Kronecker networks~\cite{LeskovecCKFG10} by recording, on the one hand, the number of directly
incentivized actions per node and, on the other hand, the number of not directly incentivized actions per node in comparison with an uncontrolled setup.
Then, we compare the performance of our method against several baselines and state of the art methods~\cite{shaping14nips, farajtabar2016msc} on a variety of large Kronecker networks
and provides a scalability analysis.

\begin{figure*}[t]
	\captionsetup[subfigure]{labelformat=empty}
	\centering
	% core-periphery
	\begin{tabular}{cccc}
	\subfloat[Uncontrolled, $\Nb(t_f)$]{ \includegraphics[width=0.20\textwidth]{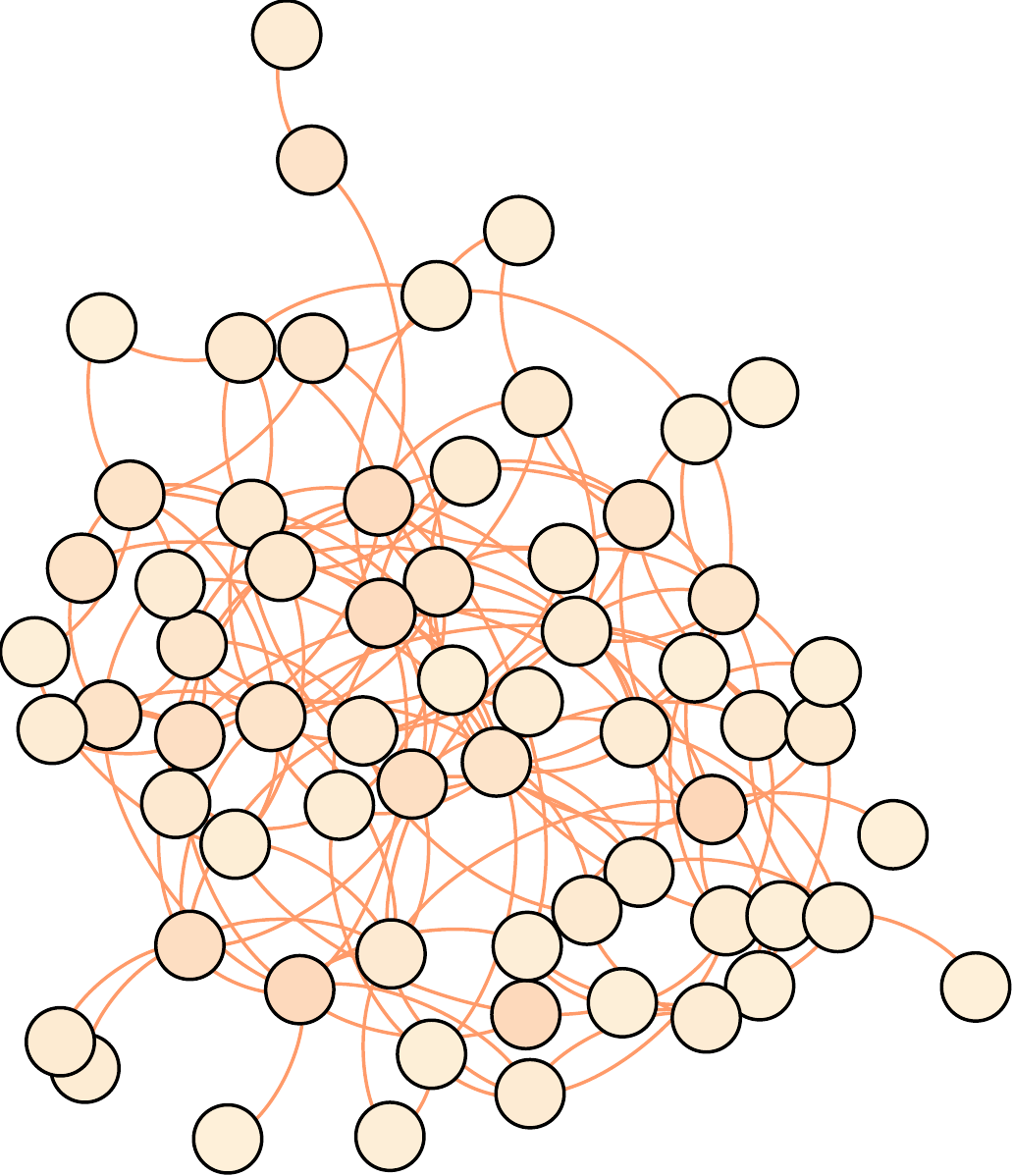} \label{fig:1_1}} &
	\subfloat[\cheshire, $\Nb(t_f)$]{ \includegraphics[width=0.20\textwidth]{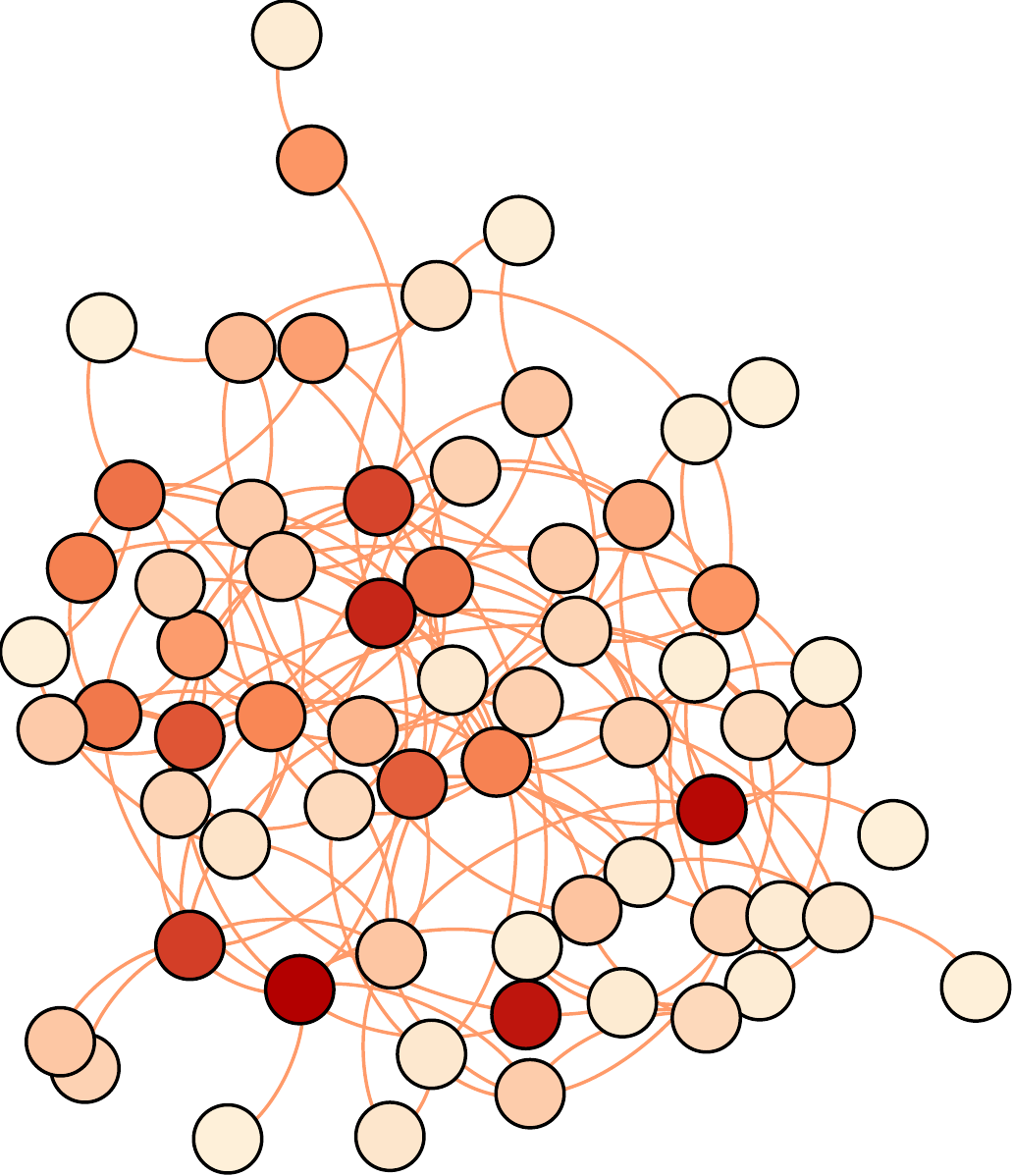} \label{fig:1_2}} &
	\subfloat[\cheshire, $\Pb(t_f)$]{ \includegraphics[width=0.20\textwidth]{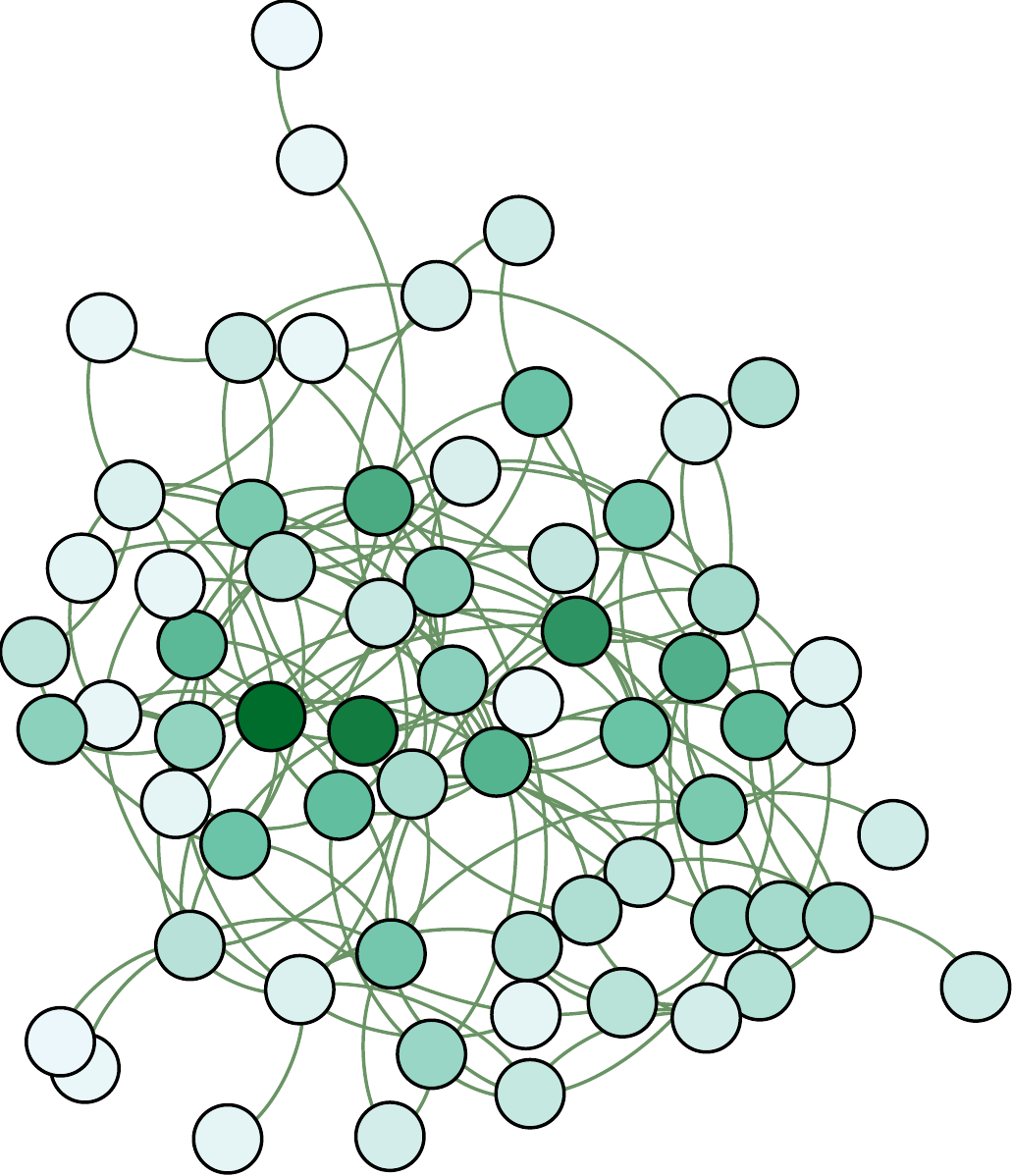} \label{fig:1_3}} &
 	\subfloat[Total number of actions]{\includegraphics[width=0.20\textwidth]{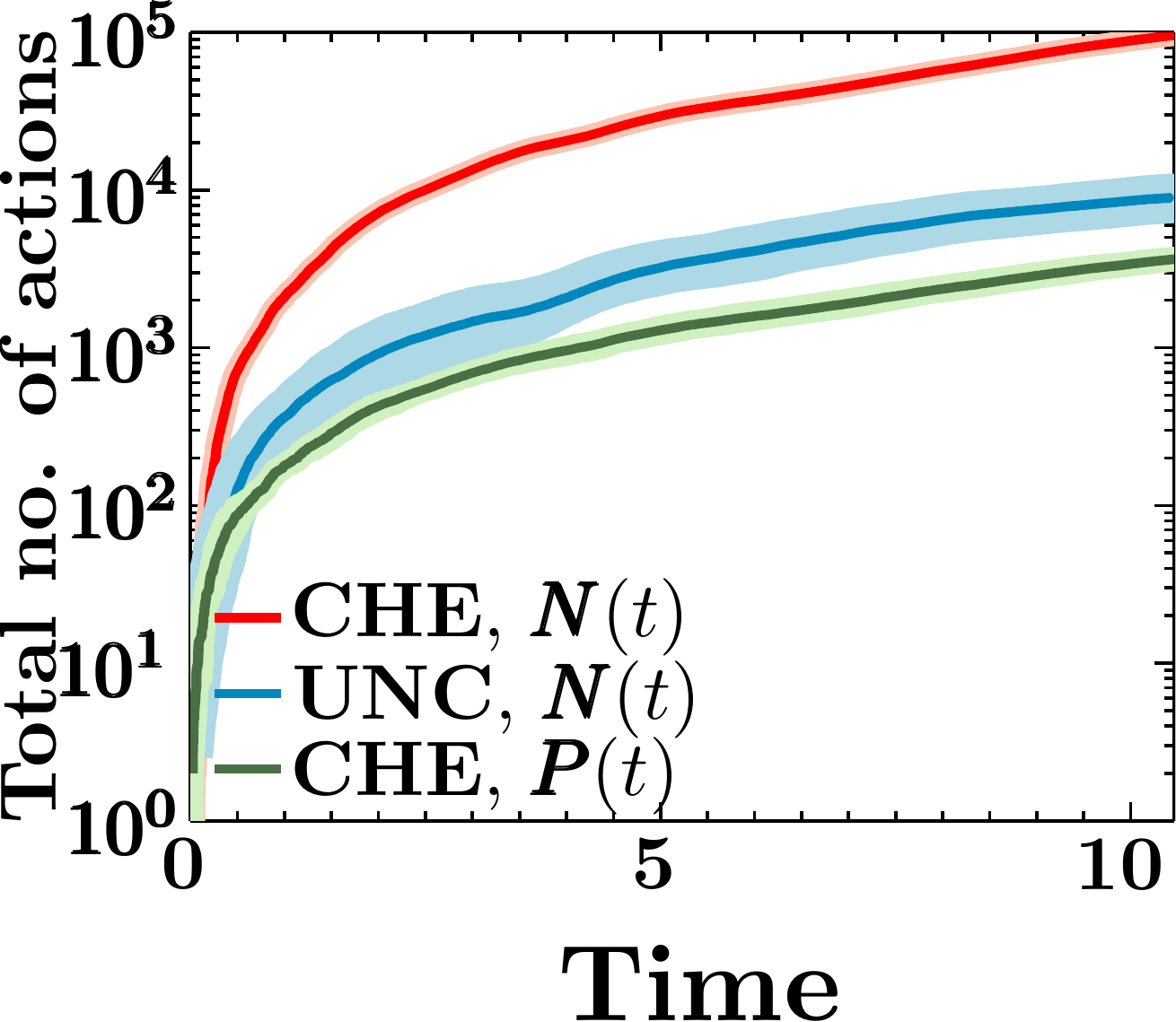} \label{fig:1_4}} \vspace*{2mm} \\
         \multicolumn{4}{c}{} \hfill (a) Core-Periphery network \hfill \\
         % dissortative
	\subfloat[Uncontrolled, $\Nb(t_f)$]{ \includegraphics[width=0.20\textwidth]{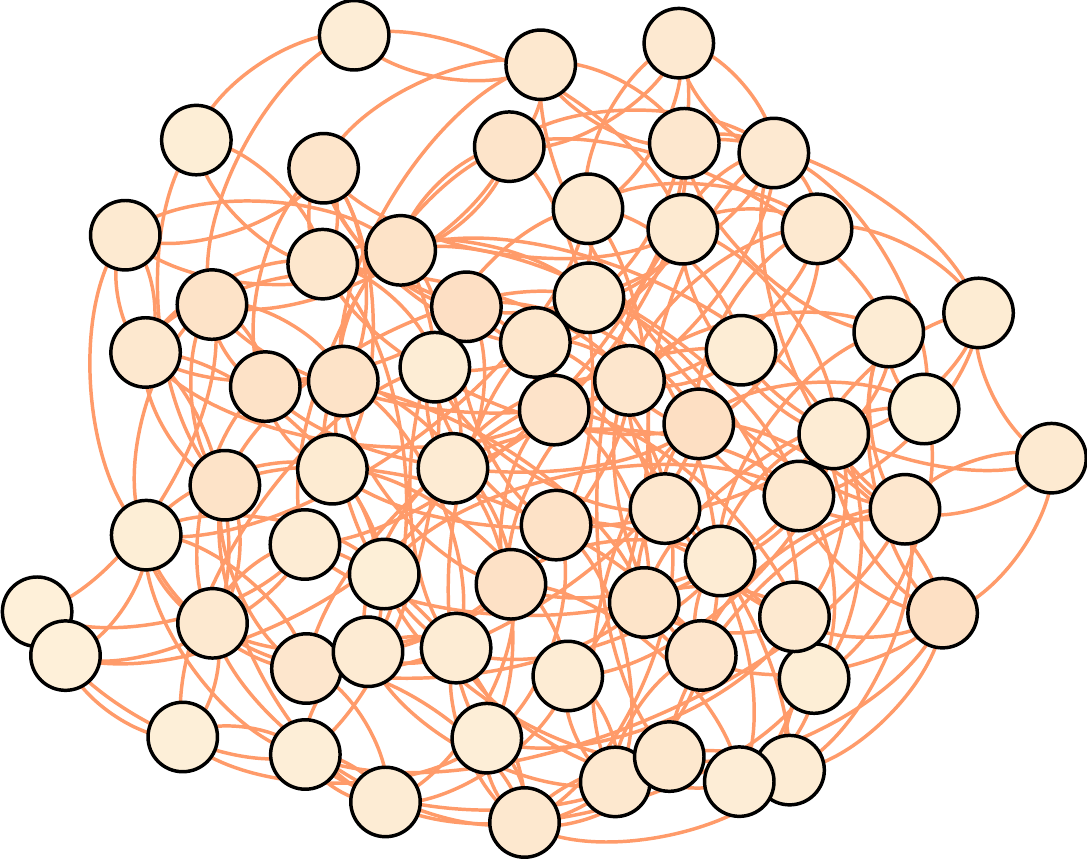} \label{fig:1_5}}  &
	\subfloat[\cheshire, $\Nb(t_f)$]{ \includegraphics[width=0.20\textwidth]{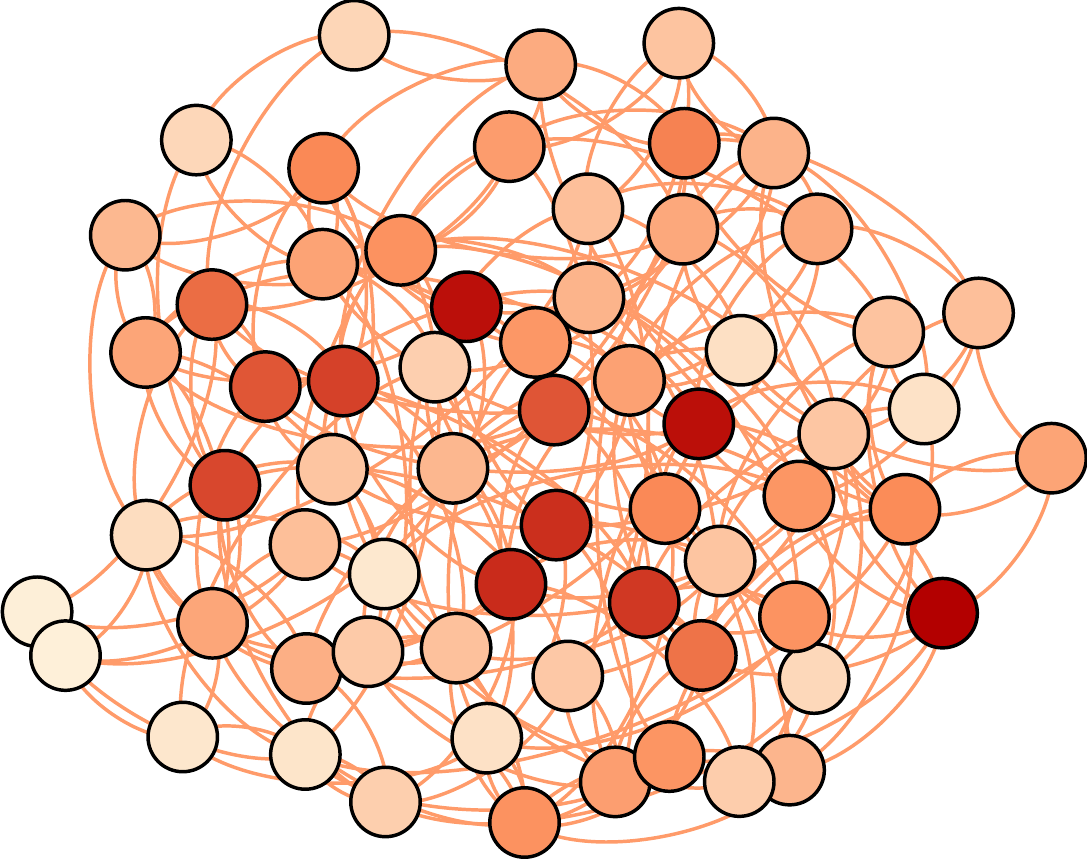} \label{fig:1_6}} &
	\subfloat[\cheshire, $\Pb(t_f)$]{ \includegraphics[width=0.20\textwidth]{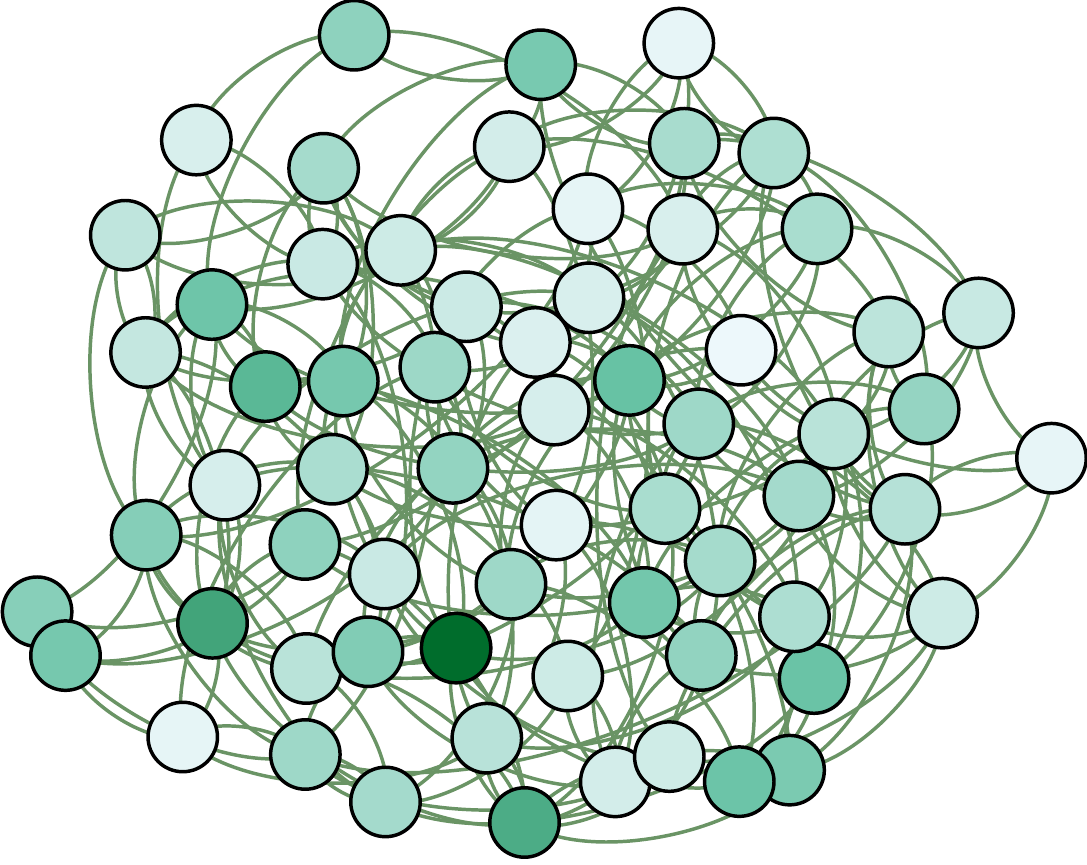} \label{fig:1_7}}  &
	\subfloat[Total number of actions]{\includegraphics[width=0.20\textwidth]{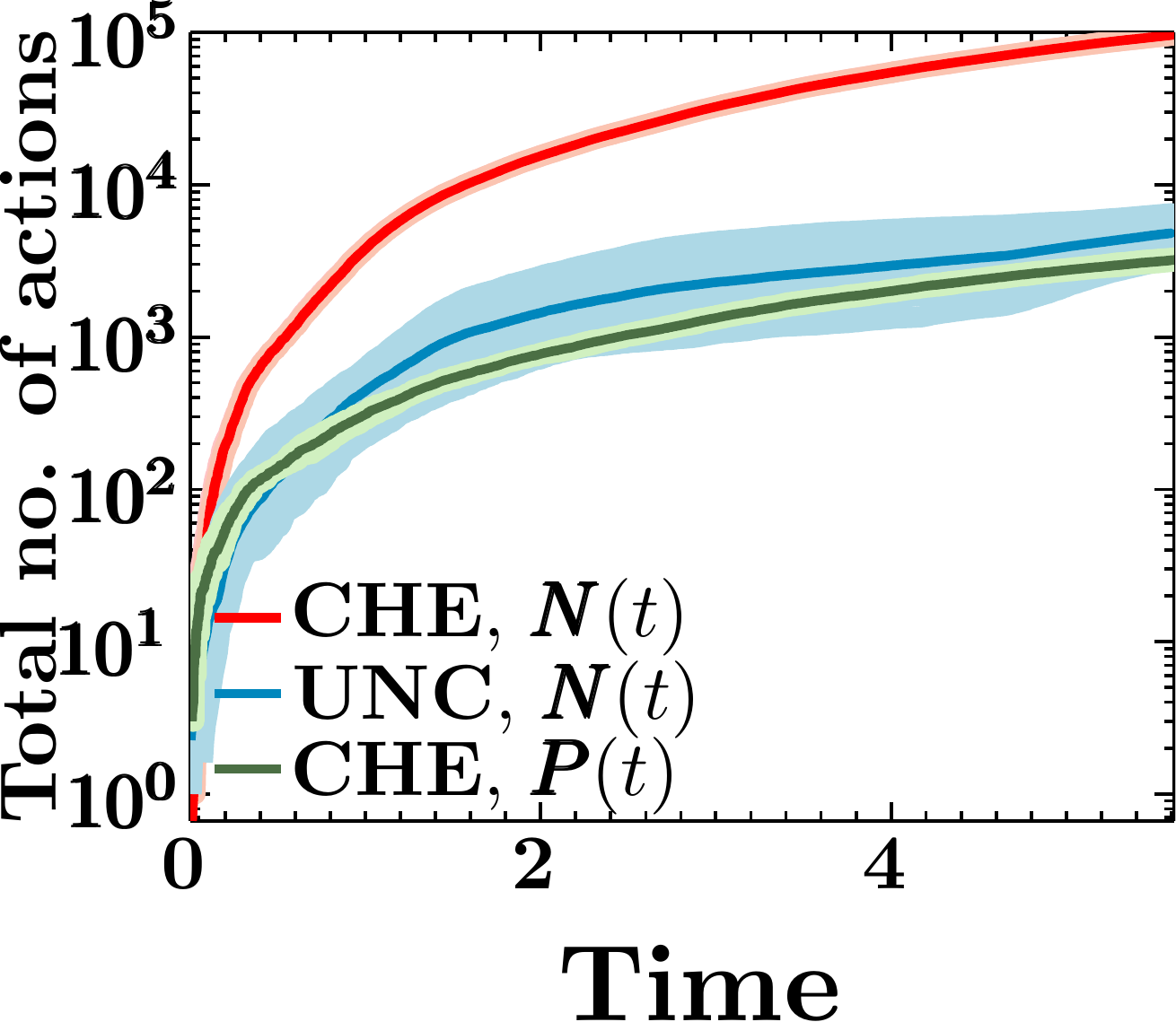} \label{fig:1_8}} \vspace*{2mm} \\
	\multicolumn{4}{c}{} \hfill (b) Dissortative network \hfill \vspace{-0.1cm} \\
	\end{tabular}

 	\caption{Activity on two 64-node networks, $G_1$ (top) and $G_2$ (bottom) under uncontrolled (Eq.~\ref{eq:multi-hawkes-cheshire}; without \cheshire) and
		      controlled (Eq.~\ref{eq:steering-multi-hawkes}; with \cheshire) dynamics.
		      The first and second columns visualize the final number of non incentivized actions $\bm{N}(t_f)$ under uncontrolled and controlled
		      dynamics, where darker red corresponds to higher number of actions.
		      The third column visualizes the final number of incentivized actions $\Pb(t_f)$ under controlled dynamics, where darker green corresponds
		      to higher number of actions.
		      The fourth column shows the temporal evolution of the number of incentivized and non incentivized actions across the whole networks for controlled
		      and uncontrolled dynamics, where the solid line is the average across simulation runs and the shadowed region represents the standard error.
		      By incentivizing a relatively small number of actions ($\sim$3{,}600 actions), \cheshire is able to increase the overall number of (non incentivized) actions dramatically
		      ($\sim$$96{,}000$ vs $\sim$$4{,}800$ actions).
	}
	\label{fig:thdem}
	\vspace{-3mm}
\end{figure*}
\xhdr{Sampling strategy}
We generate two small Kronecker networks with 64 nodes, a small core-periphery (parameter matrix $[0.96 , 0.3; 0.3, 0.96]$) and a dissortative network (param. matrix $[0.3, 0.96 ; 0.96,0.3]$), shown in Figure~\ref{fig:thdem}.
For each network, we draw $\Bb$ from a uniform distribution $U(0,10)$, $\lambdab_0$ also from a uniform distribution $U(0, 10)$ for $20$\% of the nodes and $\lambdab_0 = 0$ for the remaining $80$\%, and set $\omega=16$, $t_0 = 0$
and $t_f = 5.5$.
Then, we compare the number of actions $\Nb(t)$ over time under uncontrolled dynamics (Eq.~\ref{eq:multi-hawkes-cheshire}; without \cheshire) and controlled dynamics (Eq.~\ref{eq:steering-multi-hawkes}; with \cheshire).
In both cases, we perform $20$ simulation runs and sample non directly incentivized actions using Ogata'{}s thinning algorithm~\cite{Ogata1981}.
In the case of controlled dynamics, we sample directly incentivized actions using Algorithm~\ref{alg:sampling-cheshire}.

Figure~\ref{fig:thdem} summarizes the results in terms of the number of non directly incentivized and incentivized actions, which show that:
(i) a relatively small number of incentivized actions (fourth column; $\sim$3{,}600 actions) result in a dramatic increase in the overall number of (non directly
incentivized) actions with respect the uncontrolled setup (fourth column; $\sim$$96{,}000$ vs $\sim$$4{,}800$ actions);
(ii) the majority of the incentivized actions concentrate in a small set of influential nodes (third column; nodes in dark green);
and, (iii) the variance of the overall number of actions (fourth column; shadowed regions) is consistently reduced when using \cheshire, in other words, the networks
become more robust.
\begin{figure*}[t]
\centering
  { \includegraphics[width=0.956\textwidth]{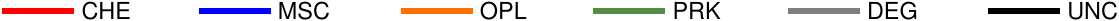}}\\ \vspace*{-3mm}
 \hspace*{-0.6cm}\subfloat[Core-periphery]{\setcounter{subfigure}{1} \includegraphics[height=0.18\textwidth]{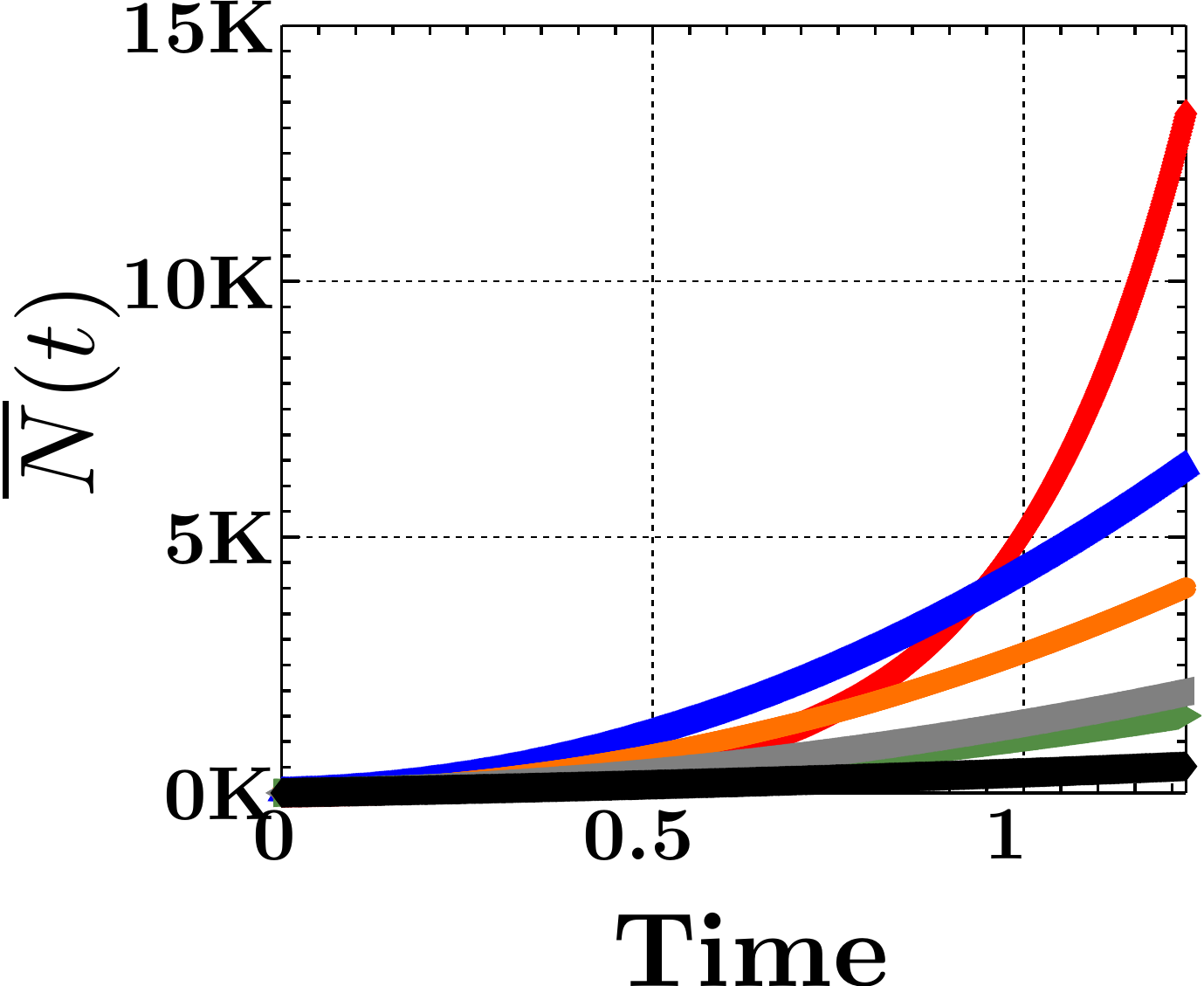}}
\hspace*{0.0cm}\subfloat[Hierarchical]{\includegraphics[height=0.18\textwidth]{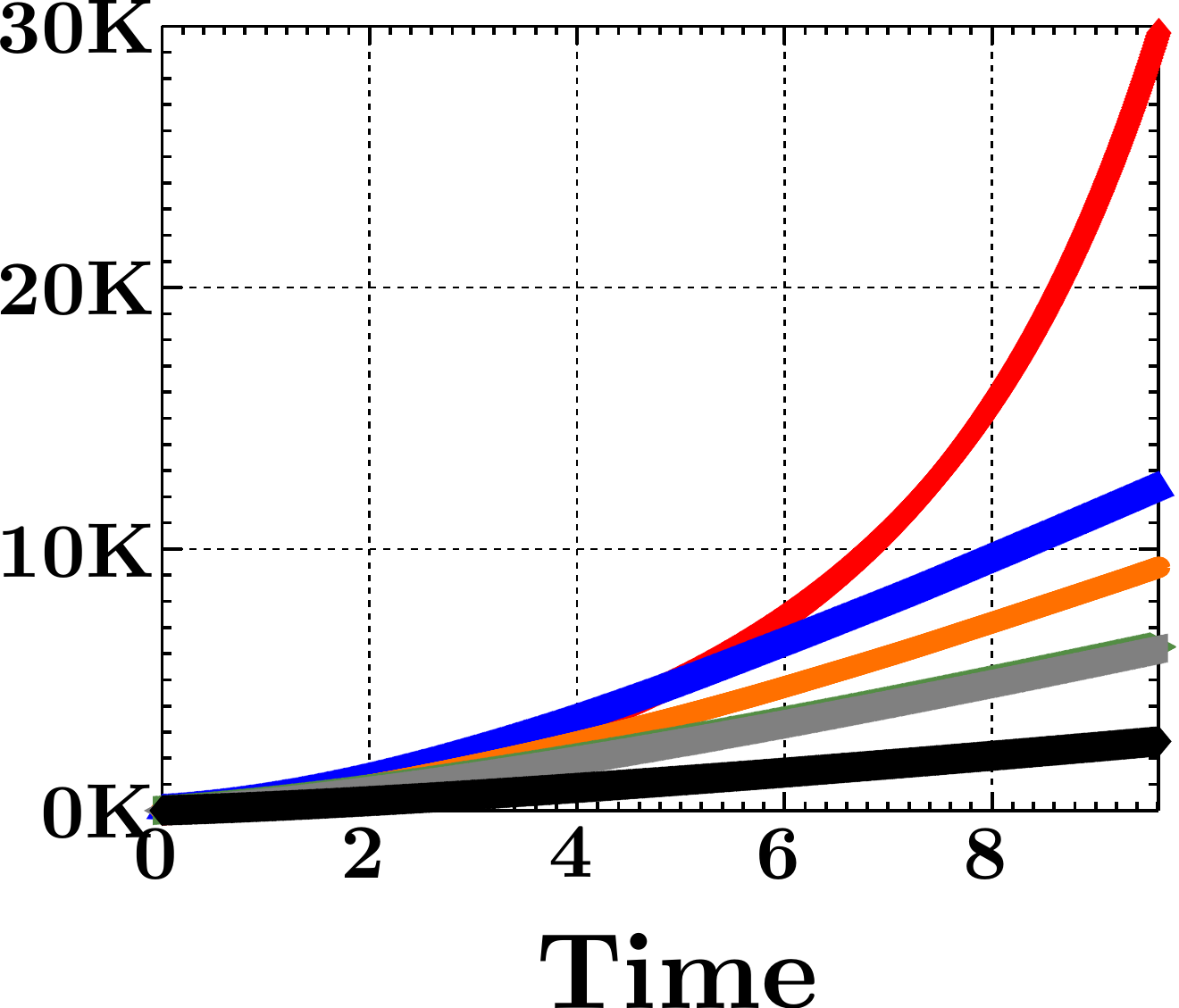}}
\hspace*{0.0cm}\subfloat[Homophily]{\includegraphics[height=0.18\textwidth]{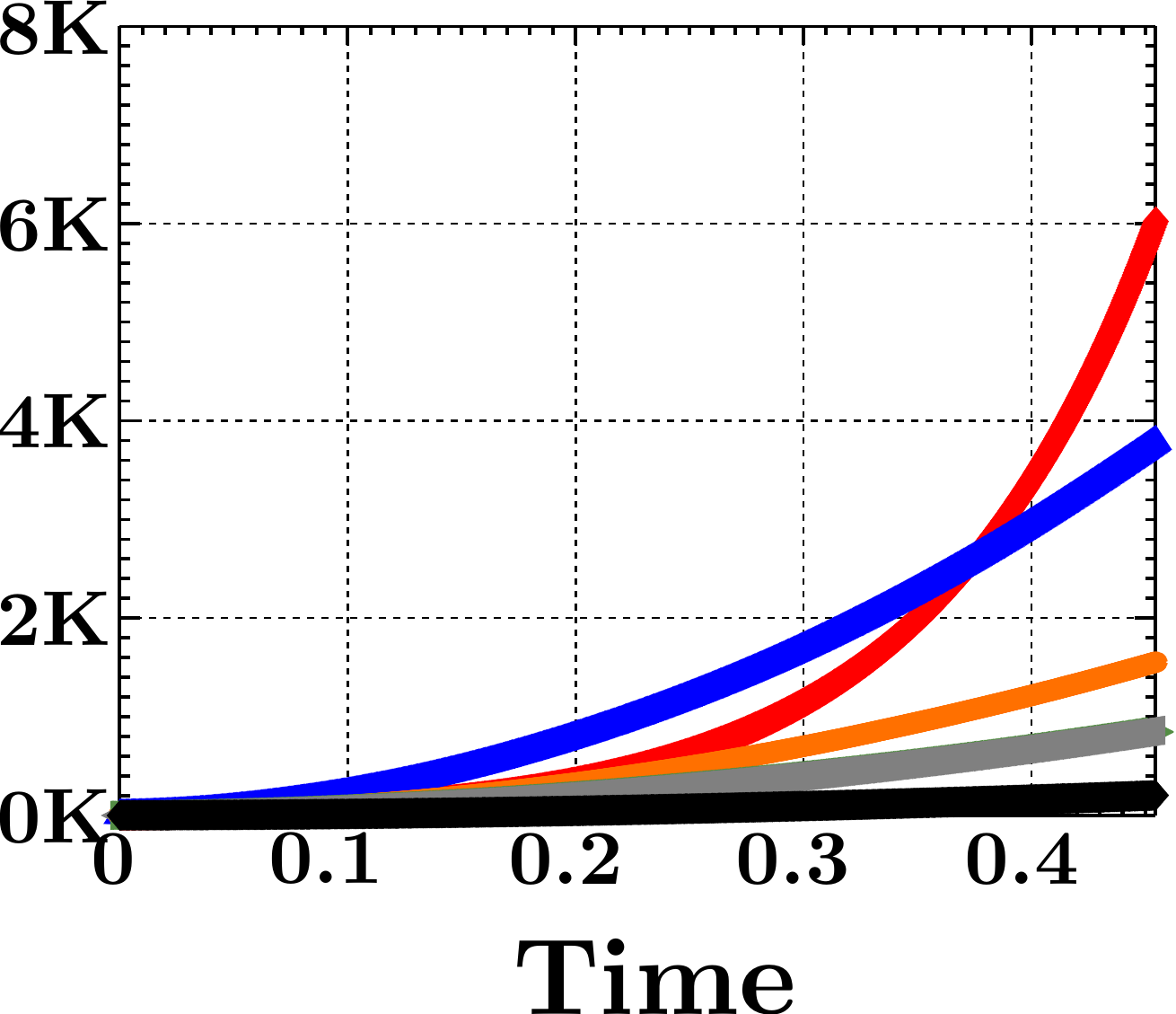}}
\hspace*{0.0cm}\subfloat[Heterophily]{ \includegraphics[height=0.18\textwidth]{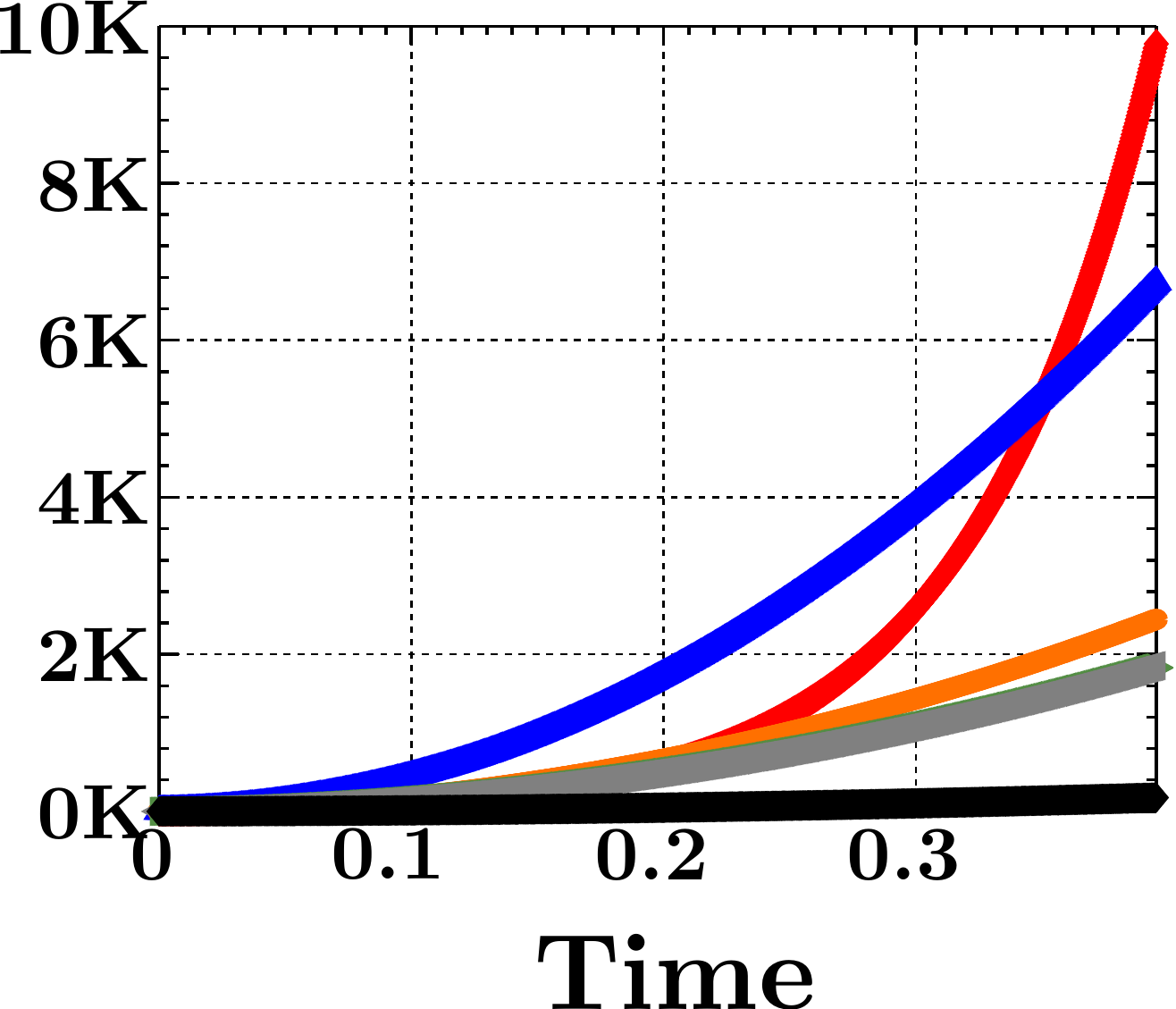}}
\hspace*{0.0cm}\subfloat[Random]{ \includegraphics[height=0.18\textwidth]{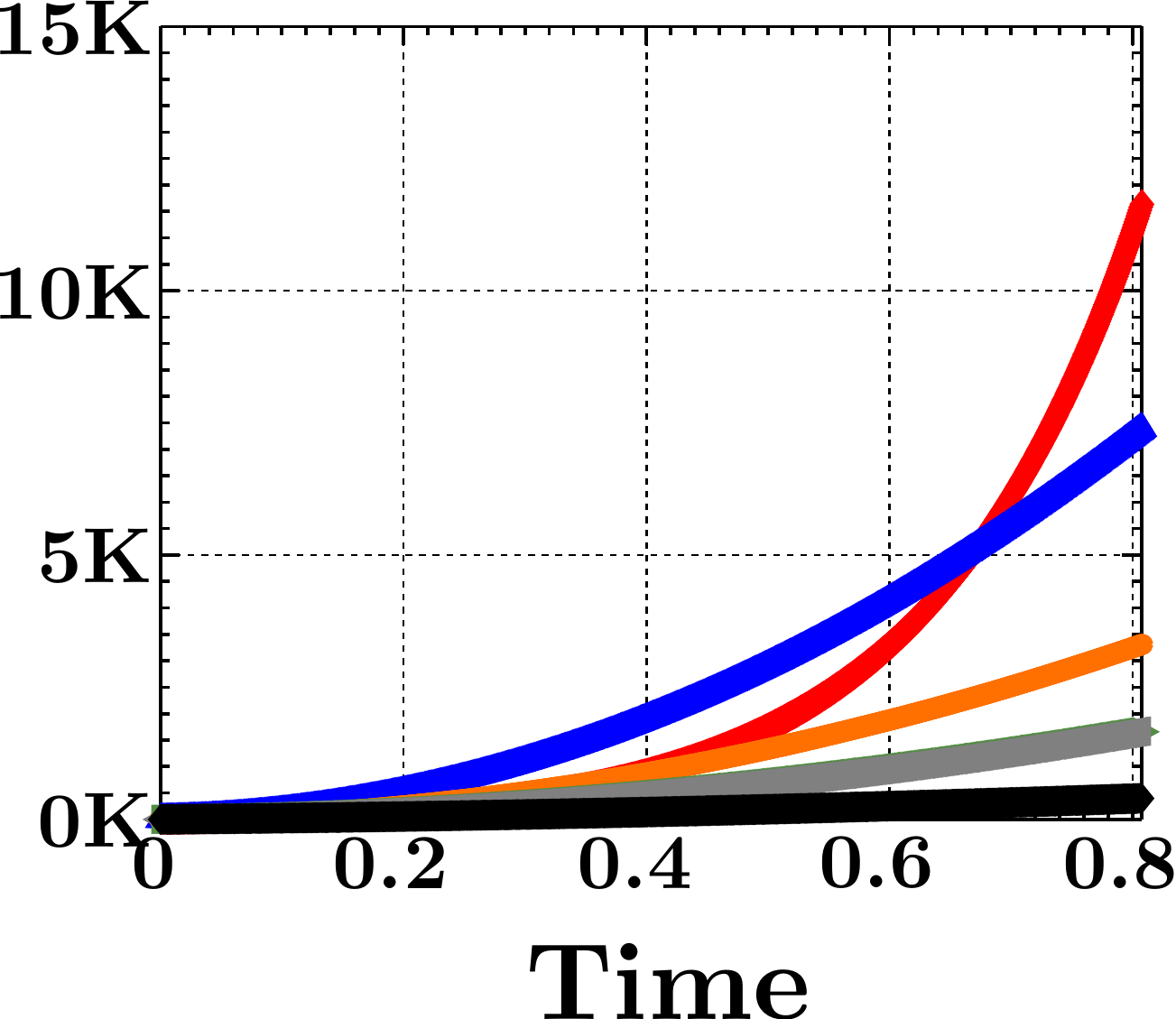}}
\vspace{-1mm}
\caption{Performance over time of \cheshire against several competitors for several types of Kronecker networks. Performance is measured in terms of overall
number of tweets $\bar{N}(t) = \sum_{u \in \Vcal} \EE[N_u(t)]$. In all cases, we tune the parameters $Q$, $S$ and $F$ such that the total number of incentivized
tweets \emph{posted} by our method is equal to the budget used in the competing methods and baselines.}
\label{fig:NwTSyn}
\vspace{-2mm}
\end{figure*}
\begin{figure*}[t]
\centering
 \hspace*{-0.6cm}\subfloat[Running time vs $t_f$]{\setcounter{subfigure}{1} \includegraphics[height=0.18\textwidth]{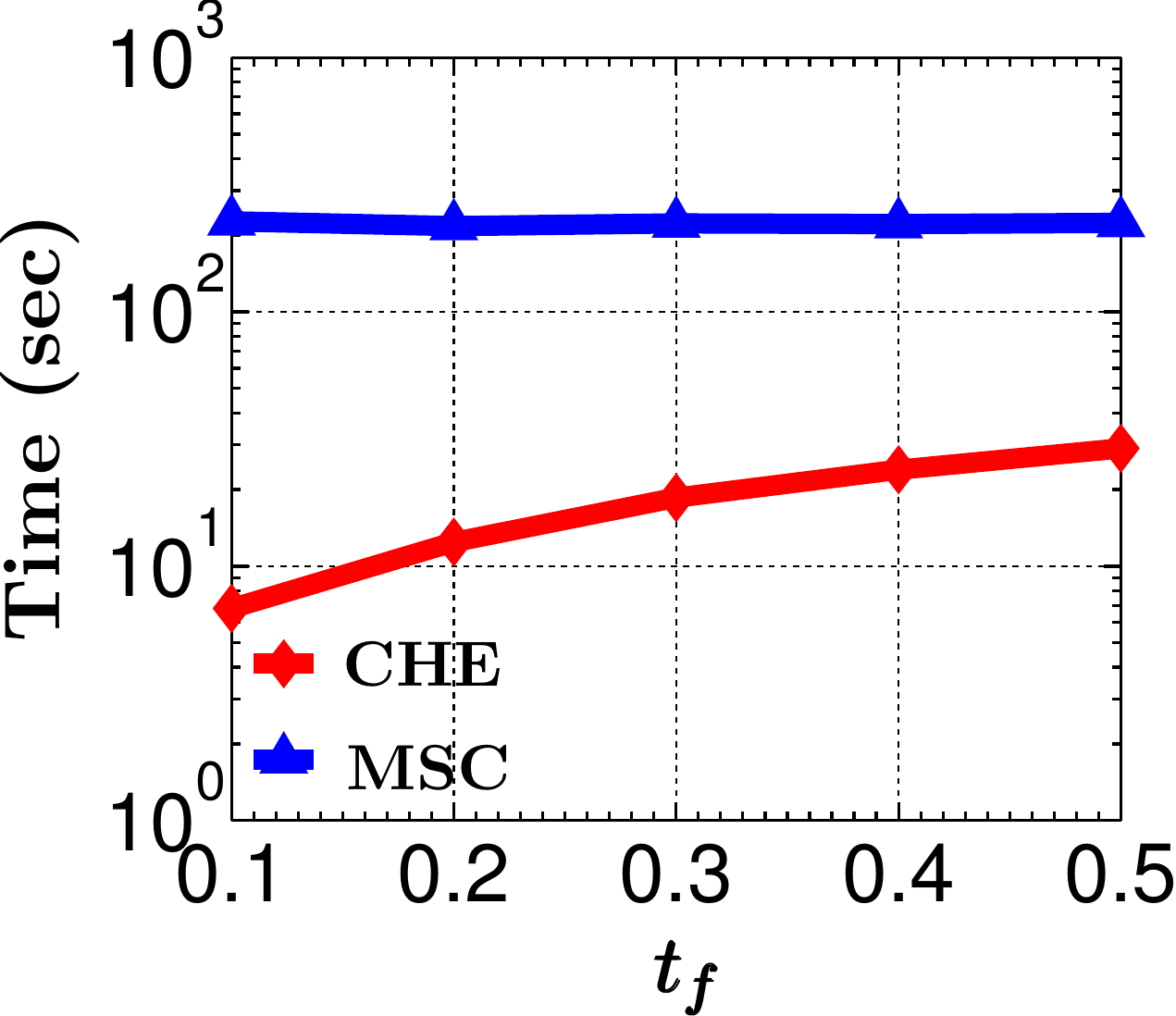}}
\hspace*{1cm}\subfloat[Running time vs \#nodes]{\includegraphics[height=0.18\textwidth]{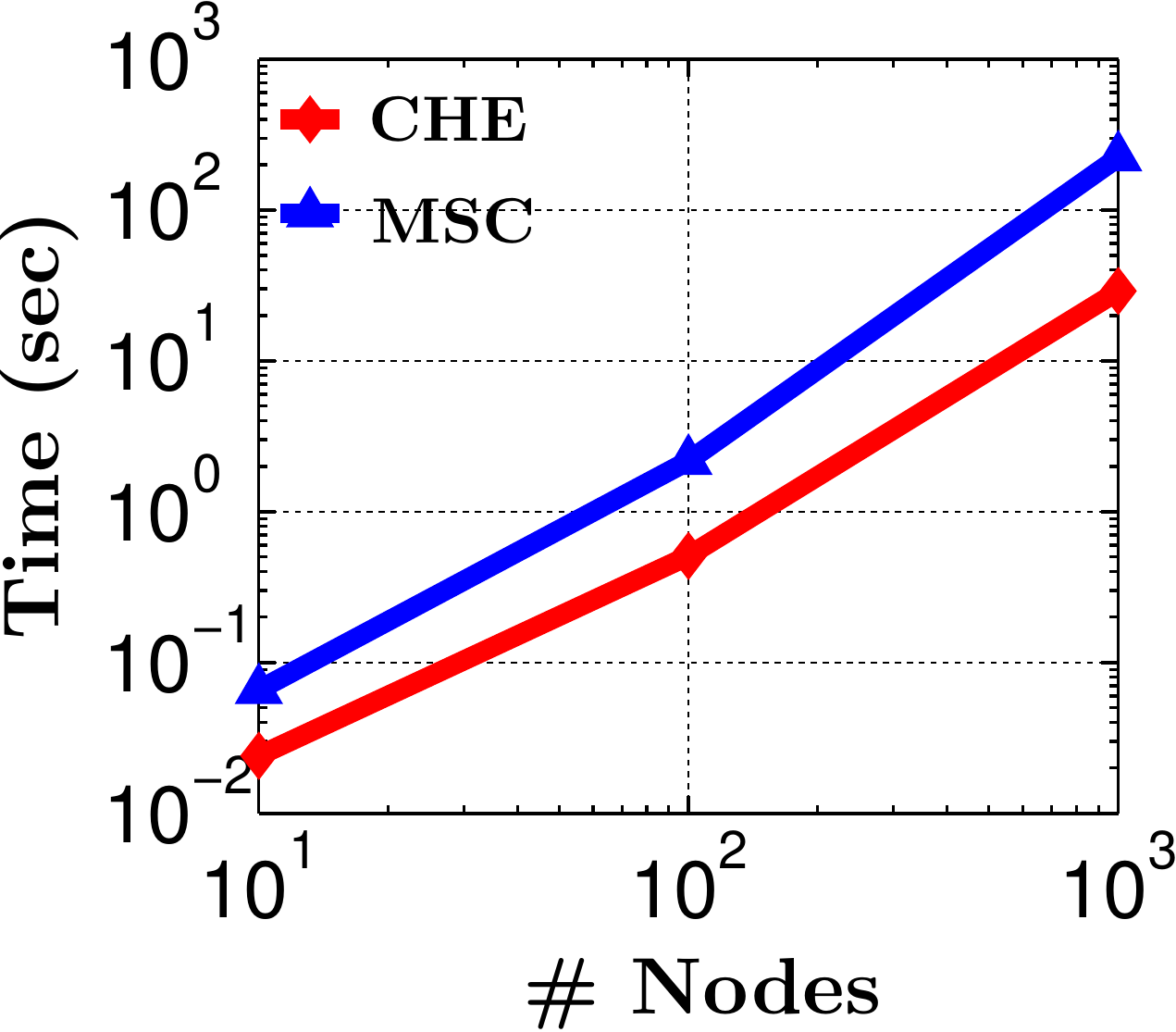}}
 \vspace{-1mm}
\caption{Scalability of \cheshire against several competitors. Panel (a) shows the running time against the cut-off time $t_f$ for a $1{,}000$
node Kronecker network. Panel (b) shows the running time for several Kronecker networks of increasing size with $t_f = 0.5$. In both panels,
the average degree per node is $10$. The experiments are carried out in a single machine with 24 cores and 64 GB of main memory.}
\label{fig:NwTSynS}
\vspace{-2mm}
\end{figure*}

\xhdr{Performance}
We experiment with five different types of large Kronecker networks~\cite{LeskovecCKFG10} with $512$ nodes:
(i) homophily (assortative) networks (parameter matrix $[0.96 , 0.3; 0.3, 0.96]$); 
(ii) heterophily (dissortative) networks ($[0.3, 0.96 ; 0.96,0.3]$); 
(iii) random networks ($[0.7, 0.7; 0.7,0.7]$);
(iv) hierarchical networks ($[0.9, 0.1 ; 0.1, 0.9]$); and,
(v) core-periphery networks ($[0.9, 0.5 ; 0.5, 0.3]$).
For each network, we draw $\lambdab_0$ and $\Bb$ from a uniform distribution $U(0,1)$ and set $\omega=100$.
We compare the performance of our algorithm, \cheshire, with two state of the art methods, which we denote as
OPL~\cite{shaping14nips} and MSC~\cite{farajtabar2016msc}, and three baselines, which we denote as PRK, DEG, and UNC.
OPL is a provably optimal algorithm for the offline setting and MSC is a suboptimal algorithm that discretizes the time window of interest in several rounds and, at 
any given round, it computes the control signal using the feedback from previous rounds to maximize the activity.
PRK and DEG distribute the users'{} control intensities $\mathbf{u}(t)$ proportionally to the user'{}s page rank and outgoing degree in the network,
respectively.
UNC simply considers all control intensities to be zero, \ie, it represents the uncontrolled dynamics.

For the large Kronecker networks, Figure~\ref{fig:NwTSyn} compares the performance of our algorithm against others in terms of overall average number of tweets $\bar{N}(t) = \sum_{u \in \Vcal} \EE[N_u(t)]$ for a
fixed budget $\bar{P}(t_f) = \sum_{u \in \Vcal} \EE[P_u(t_f)] \approx 6.1$K. We find that: (i) our algorithm consistently outperforms competing methods by large margins at time $t_f$; (ii) it triggers up
to $50$\%--$100$\% more posts than the second best performer by time $t_f$; (iii) MSC tends to use the budget too early, as a consequence, although it initially beats our method,
it eventually gets outperformed by time $t_f$; and, (iv) the baselines PRK and DEG have an underwhelming performance, suggesting that the network structure alone is not an
accurate measure of influence.

\xhdr{Scalability}
Figure~\ref{fig:NwTSynS} shows that our algorithm scales to large networks and is almost an order of magnitude faster than the second best
performer, MSC~\cite{farajtabar2016msc}. For example, our algorithm takes $\sim$$30$ seconds to steer a network with $1{,}000$ nodes and
average degree of $10$ while MSC takes  $\sim$$4$ minutes.

\subsubsection{Experiments on Real Data}
In this section, we experiment with data gathered from Twitter and show that our model can maximize the number of online actions
more effectively than several baselines and state of the art methods~\cite{shaping14nips, farajtabar2016msc}.
\begin{table}[t]
\small
{\centering\begin{tabular}{|p{2.4cm}|c|c|c|c|}
\hline%
 \textbf{Dataset}&\textbf{  $\mathbf{|\Vcal|}$} & \textbf{$\mathbf{|\Ecal|}$} & \textbf{$|\Hcal(T_{\text{Data}})|$} &\textbf{$T=T_\text{simulation}$} \\ \hline \hline
{Elections}  &   231 &1108  	& 1584 	& 120.2\\ \hline
{Verdict}    &  1059  & 10691  & 17452	& 22.11 \\ \hline
{Club}       &  703  & 4154  	&  9409	& 19.23\\ \hline
{Sports}      &  703  & 4154    &  7431	& 21.53\\ \hline%
{TV Show}      &  947 & 10253 	& 13203	& 12.11\\ \hline%
\end{tabular}
\caption{Real datasets statistics}
\label{tab:datasets}}
\end{table}

\xhdr{Experimental setup}
We experiment with five Twitter data sets about current real-world events, where actions are tweets (and retweets).
To create each data set, we used the Twitter search API\footnote{\scriptsize \url{https://dev.twitter.com/rest/public/search}} to collect all the tweets (corresponding 
to a 2-3 weeks period around the event date) that contain hashtags related to:
\begin{itemize}[noitemsep,nolistsep]
\item \textbf{Elections: } British election, from May 7 to May 15, 2015.
\item \textbf{Verdict:} Verdict for the corruption-case against Jayalalitha, an Indian politician, from May 6 to May 17, 2015.
\item \textbf{Club:} Barcelona getting the first place in La-liga, from May 8 to May 16, 2016.
\item \textbf{Sports: } Champions League final in 2015, between Juventus and Real Madrid, from May 8 to May 16, 2015.
\item \textbf{TV Show:} The promotion on the TV show ``Games of Thrones'', from May 4 to May 12, 2015.
\end{itemize}
We then built the follower-followee network for the users that posted the collected tweets using the Twitter rest API\footnote{\scriptsize \url{https://dev.twitter.com/rest/public}}.
Finally, we filtered out users that posted less than 200 tweets during the account lifetime, follow less than 100 users, or have less than 50 followers. An account of the data set statistics 
is given in Table~\ref{tab:datasets}.\\
\begin{figure*}[t]
  \captionsetup[subfigure]{justification=centering}
  \centering
  { \includegraphics[width=0.956\textwidth]{legend_IwB_new}}\\ \vspace*{-2mm}
 \subfloat[\small Elections,  $\bar{P}(t_f)\approx 3K$]{\setcounter{subfigure}{1} \includegraphics[height=0.16\textwidth]{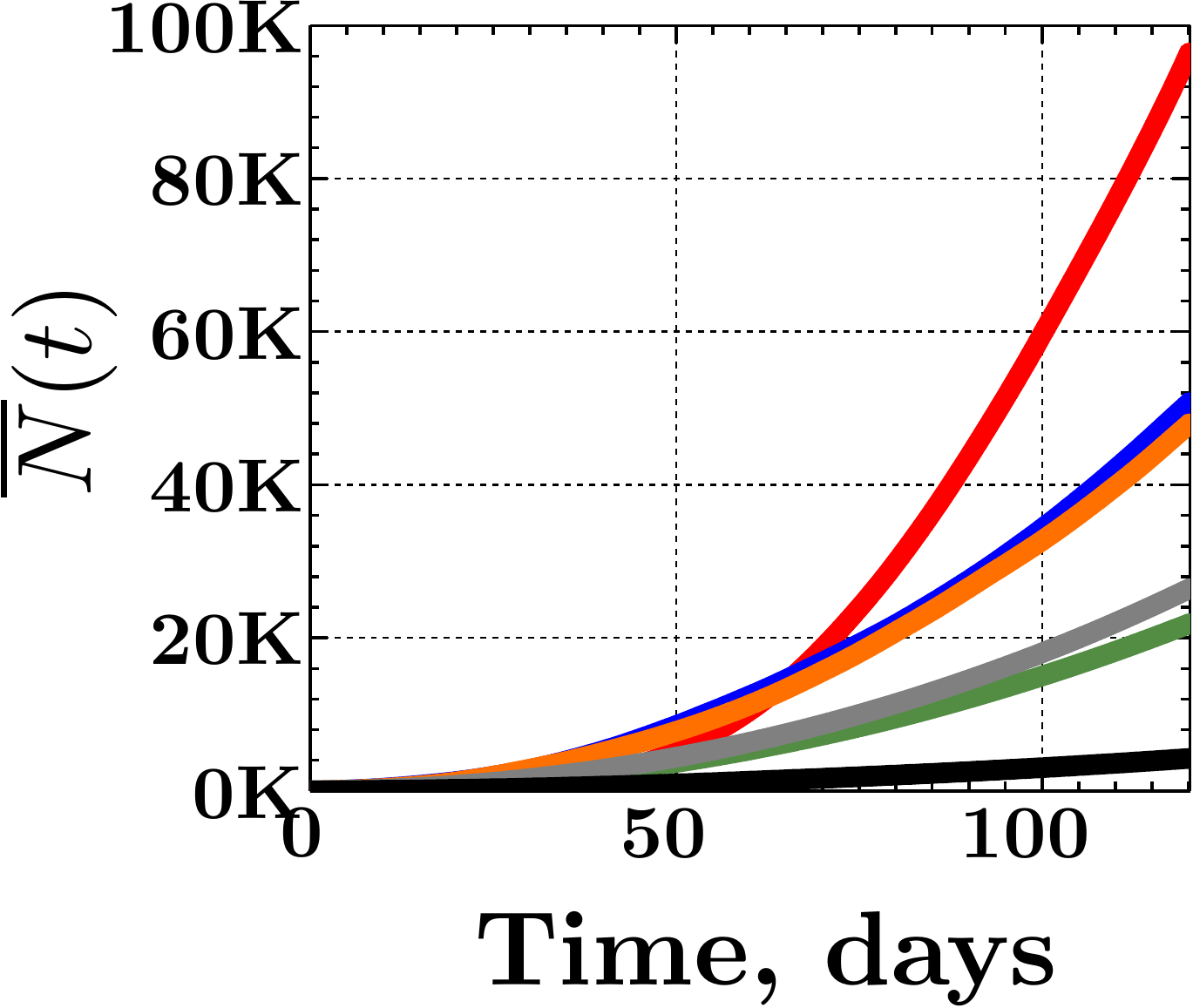}}
\hspace*{2mm}\subfloat[\small  Verdict, $\bar{P}(t_f)\approx14K$]{\includegraphics[height=0.16\textwidth]{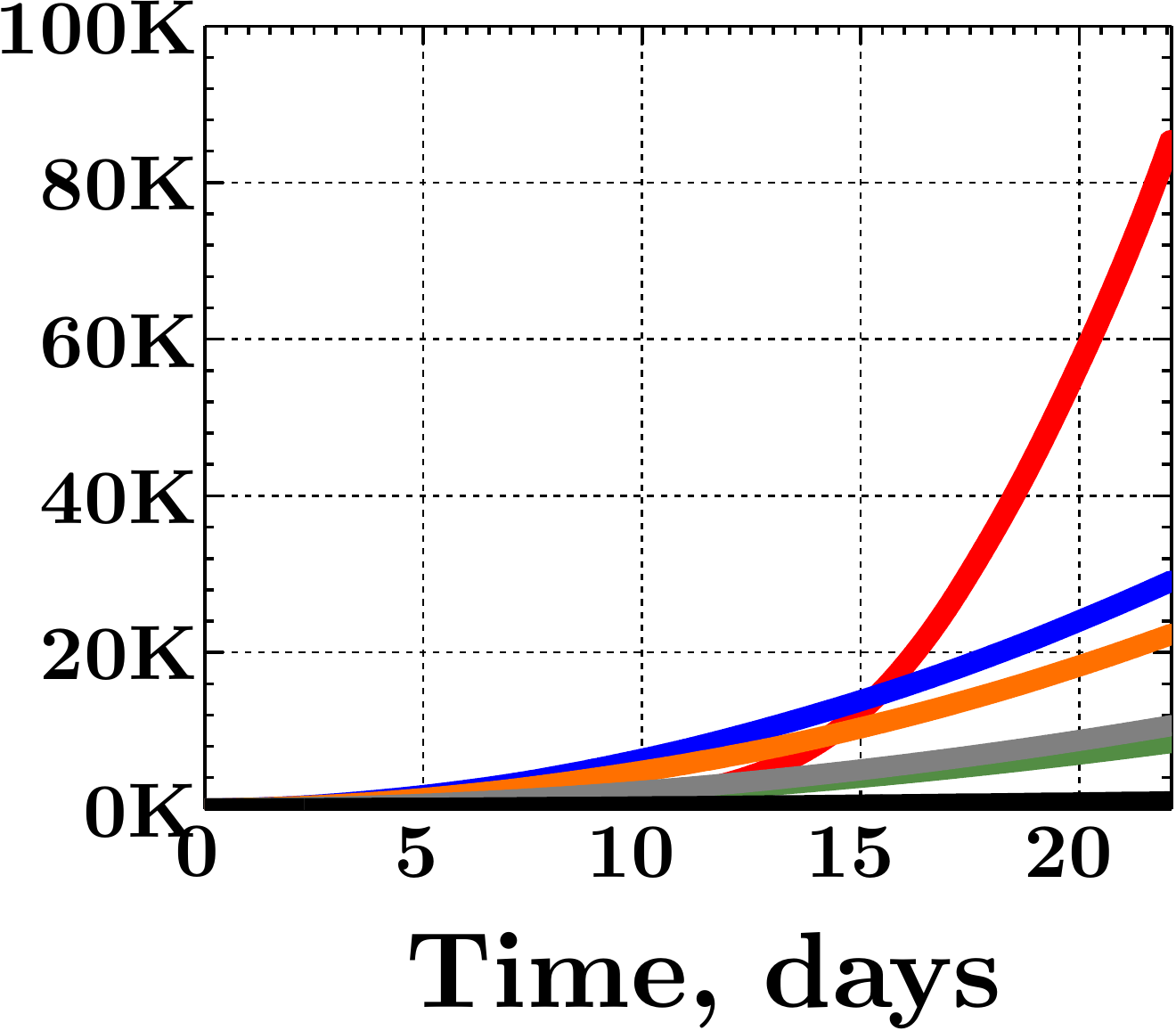}}
\hspace*{2mm}\subfloat[\small  Club, $\bar{P}(t_f)\approx 5K$ ]{\includegraphics[height=0.16\textwidth]{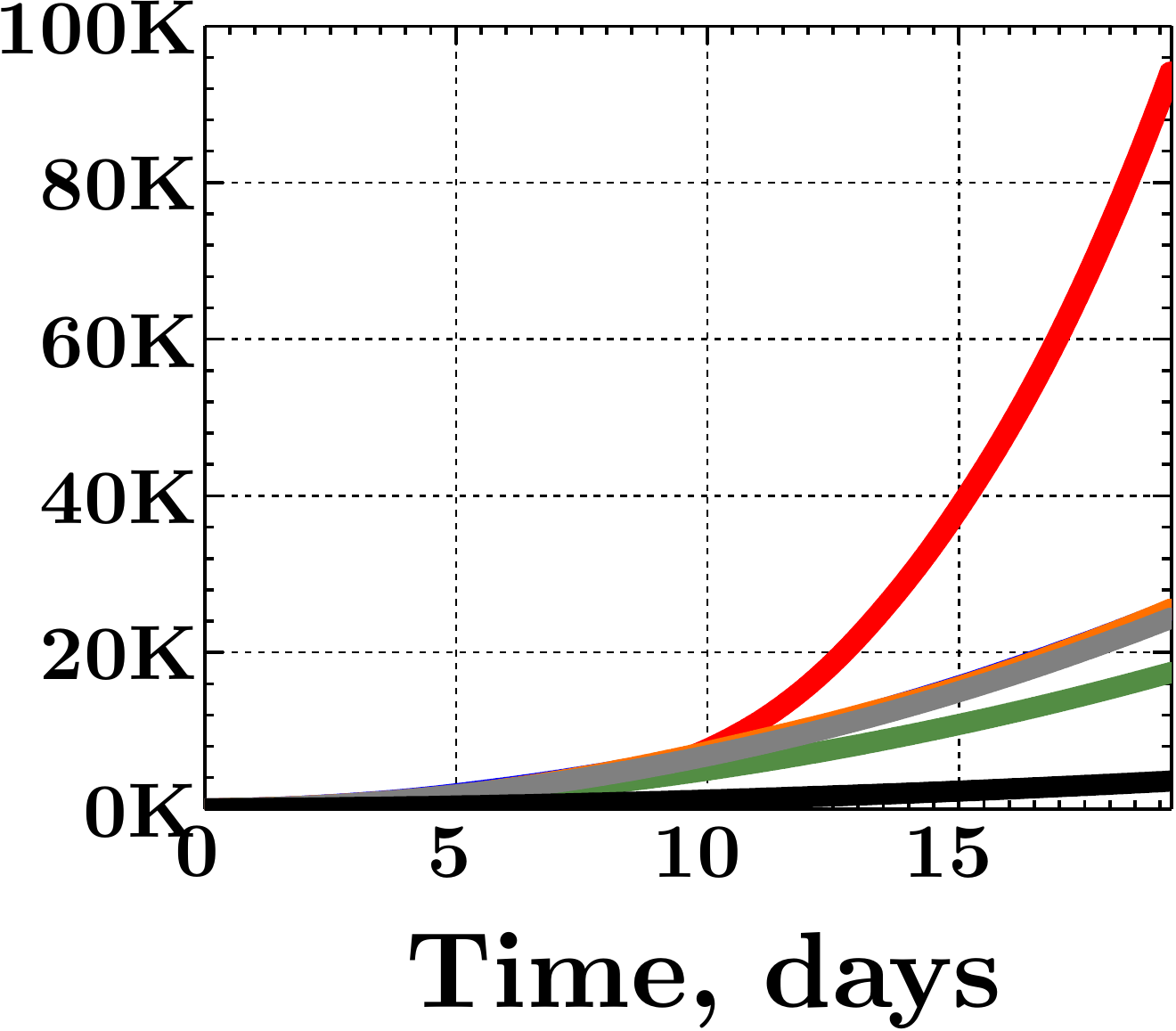}}
\hspace*{2mm}\subfloat[\small  Sports, $\bar{P}(t_f)\approx 5K$]{ \includegraphics[height=0.16\textwidth]{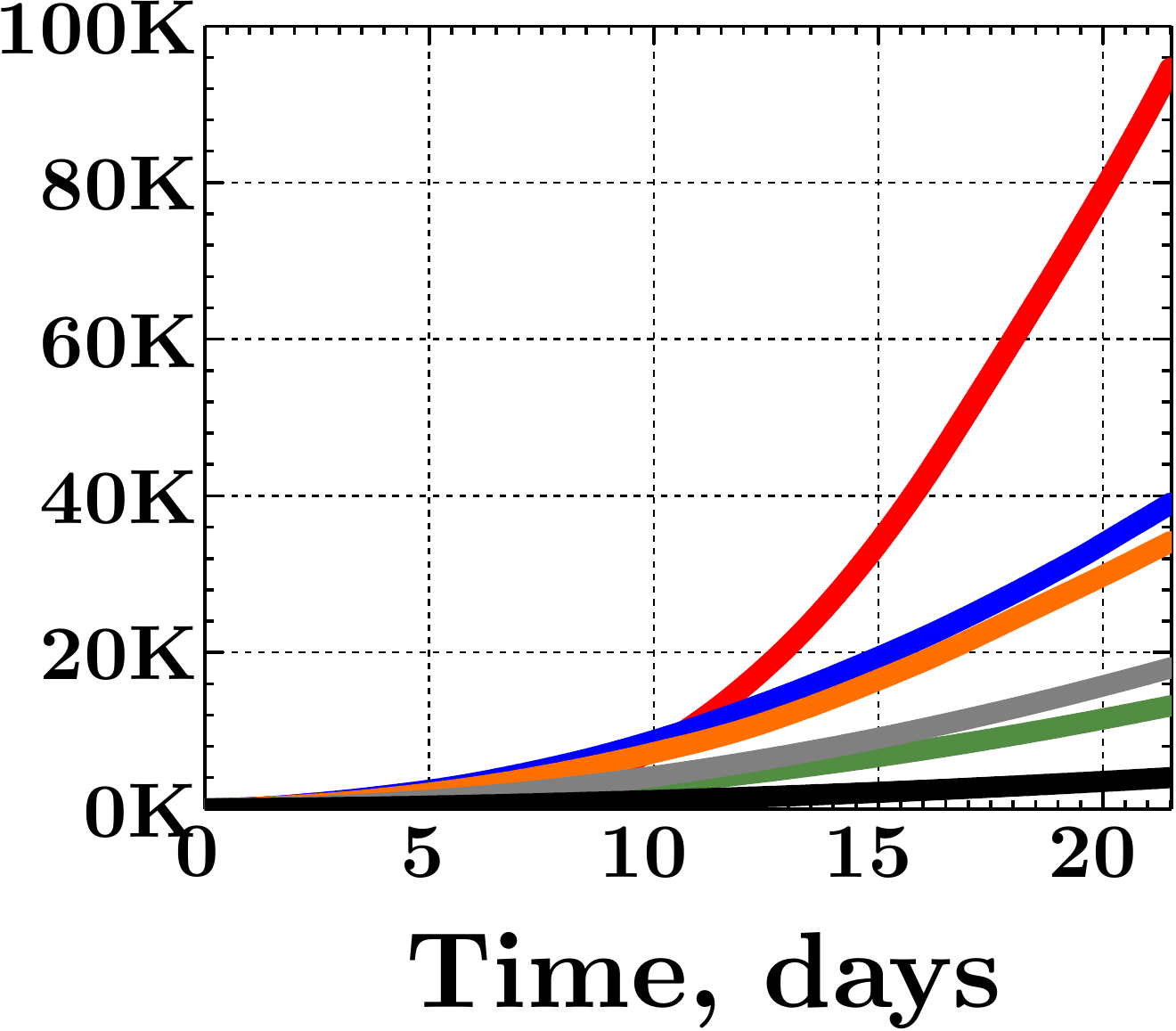}}
\hspace*{2mm}\subfloat[\small  Series, $\bar{P}(t_f)\approx 13K$]{ \includegraphics[height=0.16\textwidth]{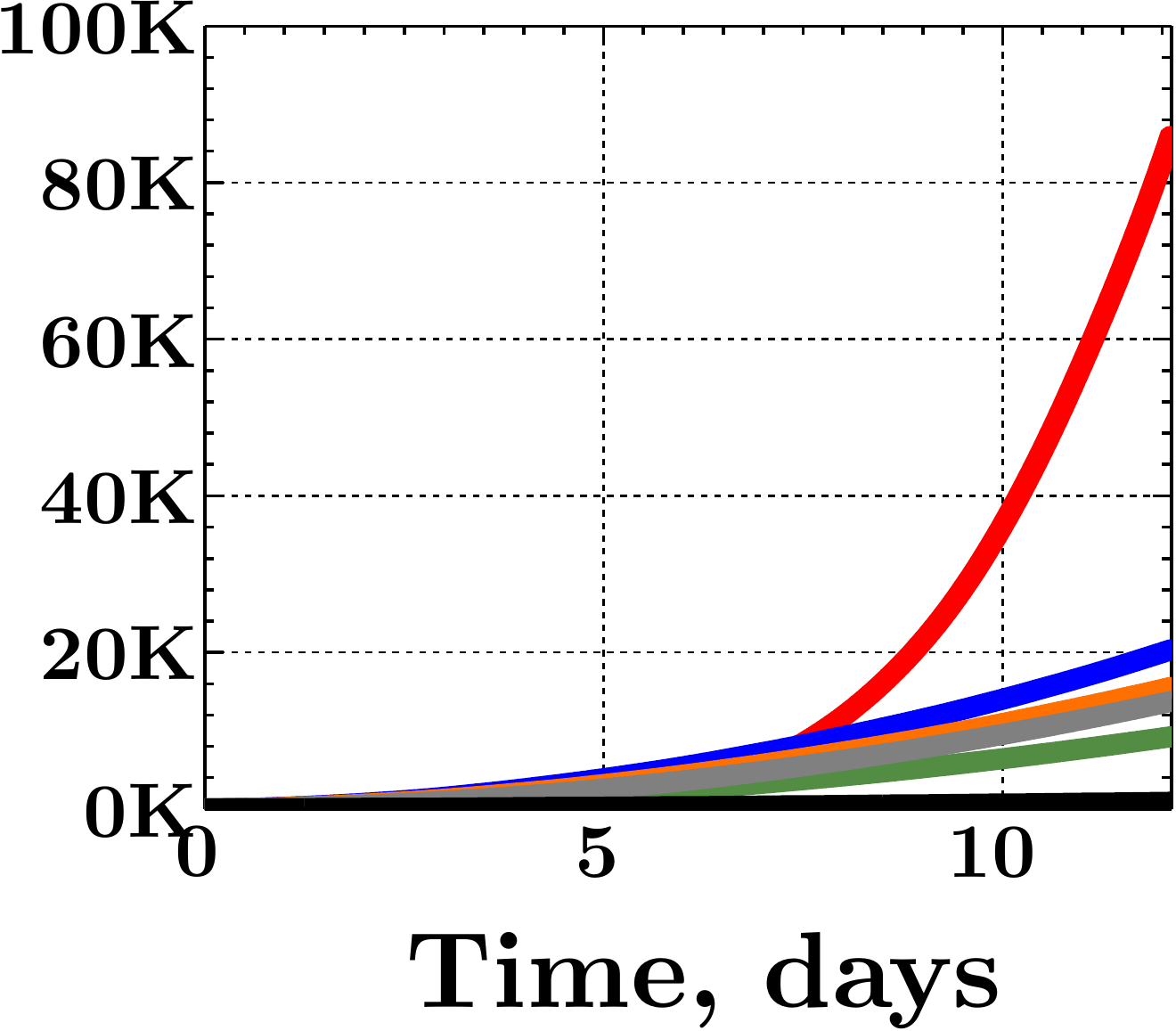}}
\caption{Performance over time of \cheshire against several competitors for each Twitter data set. Performance is measured in terms of overall number of tweets
$\bar{N}(t) = \sum_{u \in \Vcal} \EE[N_u(t)]$. In all cases, we tune the parameters $\bm{Q}$, $\bm{S}$ and $\bm{F}$ to be diagonal
matrices such that the total number of incentivized tweets \emph{posted} by our method is equal to the budget used in the competing methods and baselines.
\cheshire (in red) consistently outperforms competing methods over time and it triggers up to $100$\%--$400$\% more posts than the second best performer (in blue) as time
goes by.}
\label{fig:NwT}
\end{figure*}

Similarly as in the synthetic experiments, we compare the performance of our algorithm with two state of the art methods, OPL~\cite{shaping14nips} and
MSC~\cite{farajtabar2016msc}, and three baselines, PRK, DEG and UNC.
More in detail, we proceed as follows.

For each data set, we estimate the influence matrix $\Bb$ of the multidimensional Hawkes process defined by Eq.~\ref{eq:multi-hawkes-cheshire} using maximum likelihood,
as elsewhere~\cite{shaping14nips,Valera2015}. Moreover, we set the decay parameter $\omega$ of the corresponding exponential triggering kernel $\kappa(t)$ by
cross-validation.
Then, we perform $20$ simulation runs for each method and baseline, where we sample non directly incentivized actions from the multidimensional Hawkes process learned
from the corresponding Twitter data set using Ogata'{}s thinning algorithm~\cite{Ogata1981}.
For the competing methods and baselines, the control intensities $\bm{u}(t)$ are deterministic and thus we only need to sample incentivized actions from inhomogeneous
Poisson processes~\cite{lewis1979simulation}. For our method, the control intensities are stochastic and thus we sample incentivized actions using Algorithm~\ref{alg:sampling-cheshire}.
Here, we compare their performance in terms of the (empirical) average number of tweets $\EE[\Nb(t)]$.
In the above procedure, for a fair comparison, we tune the parameters $\bm{Q}$, $\bm{S}$ and $\bm{F}$ to be diagonal matrices such that the total number of incentivized tweets \emph{posted} by our method
is equal to the budget used in the state of the art methods and baselines.

\xhdr{Results}
We first compare the performance of our algorithm against others in terms of overall average number of tweets $\bar{N}(t) = \sum_{u \in \Vcal} \EE[N_u(t)]$ for a fixed budget
$\bar{P}(t_f) = \sum_{u \in \Vcal} P_u(t_f)$. Figure~\ref{fig:NwT} summarizes the results, which show that: (i) our algorithm consistently outperforms competing methods
by large margins; (ii) it triggers up to $100$\%--$400$\% more posts than the second best performer as time goes by; and, (iii) the baselines PRK and DEG have
an underwhelming performance, suggesting that the network structure alone is not an accurate measure of influence.

\begin{figure*}[t]
\captionsetup[subfigure]{labelformat=empty}
\centering
 { \includegraphics[width=0.956\textwidth]{legend_IwB_new}}\\ \vspace*{-2mm}
 \hspace*{-0.6cm}\subfloat[Elections]{\setcounter{subfigure}{1} \includegraphics[height=0.18\textwidth]{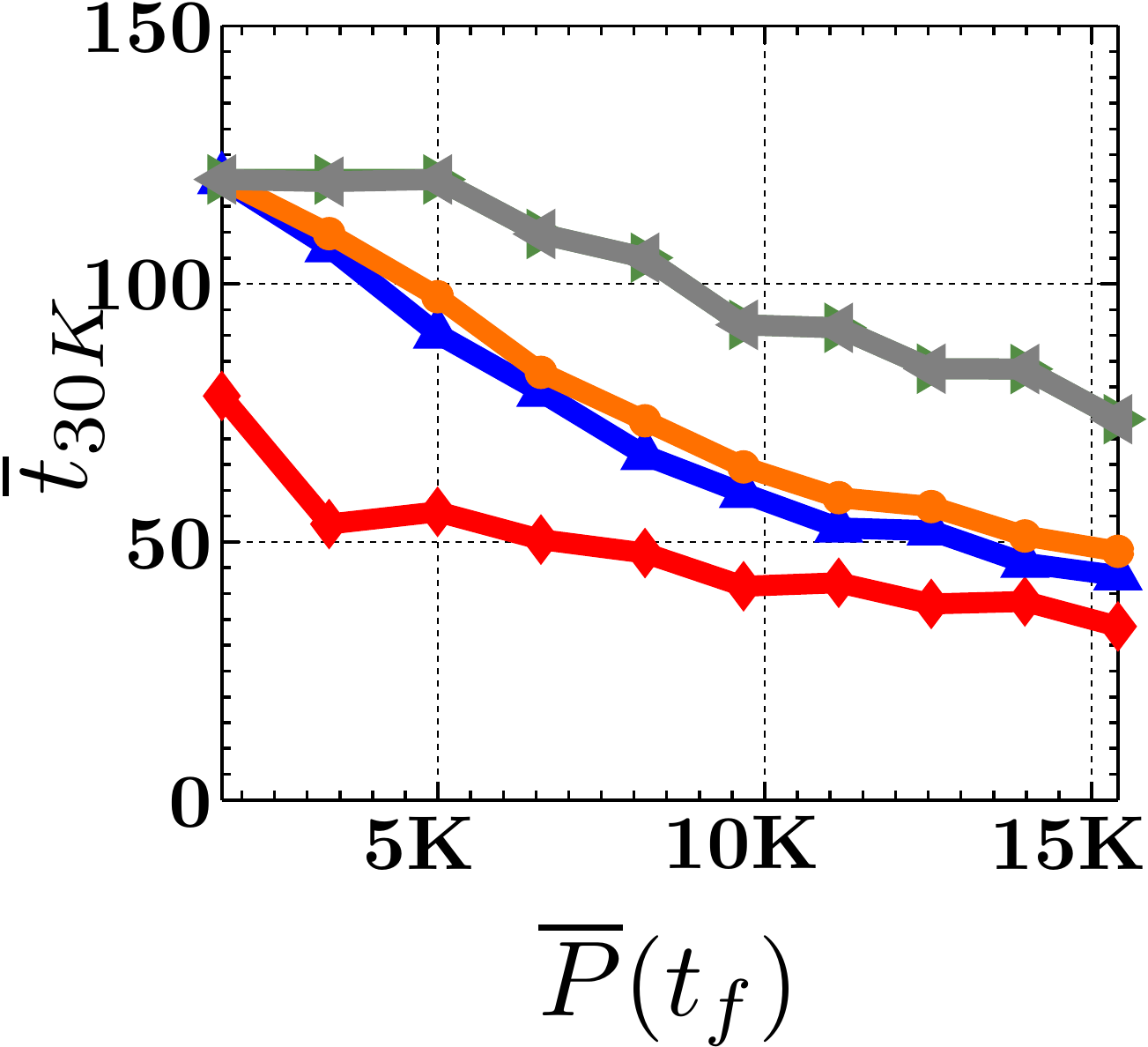}}
\hspace*{0.1cm}\subfloat[Verdict]{\includegraphics[height=0.18\textwidth]{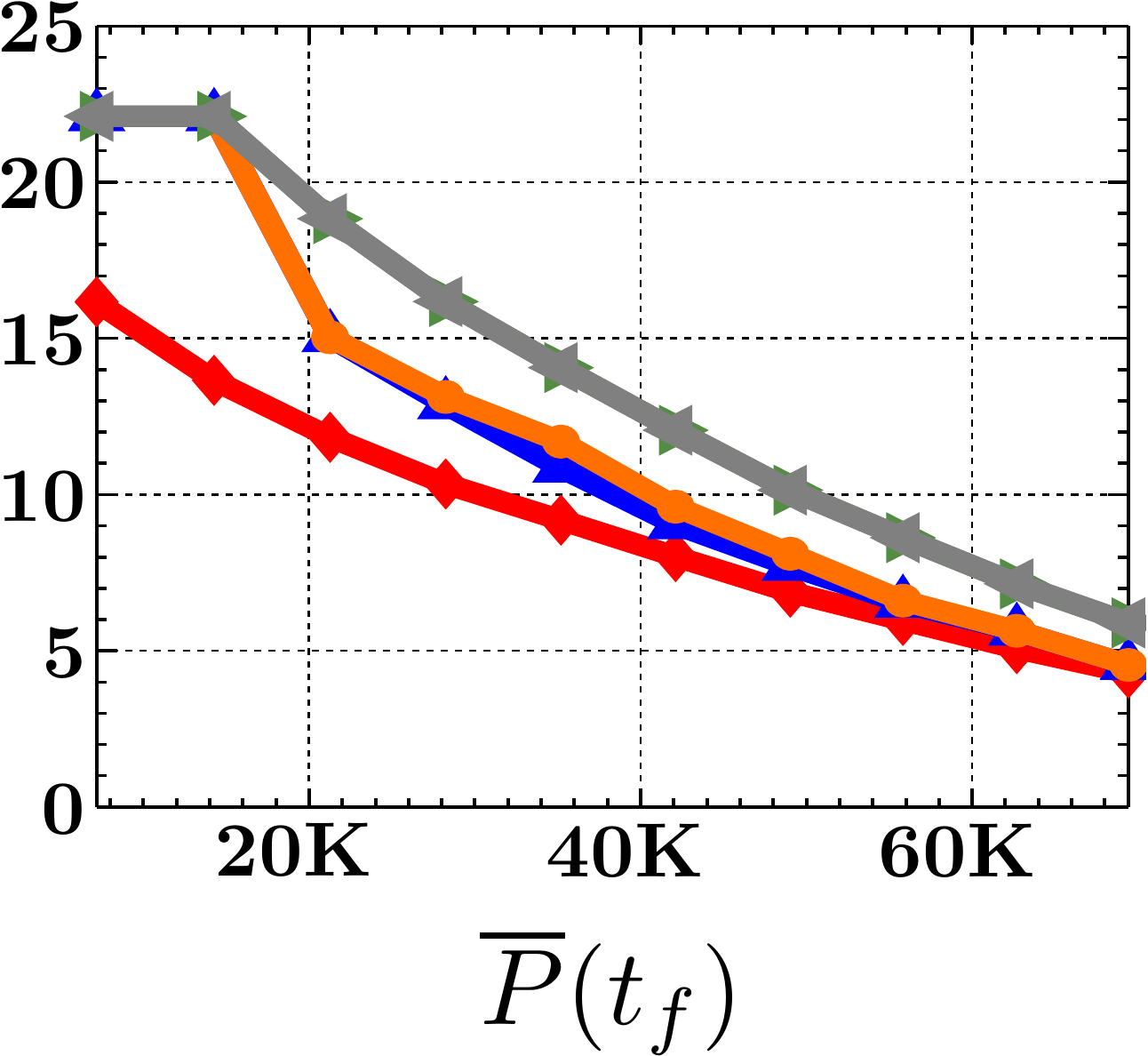}}
\hspace*{0.1cm}\subfloat[Club]{\includegraphics[height=0.18\textwidth]{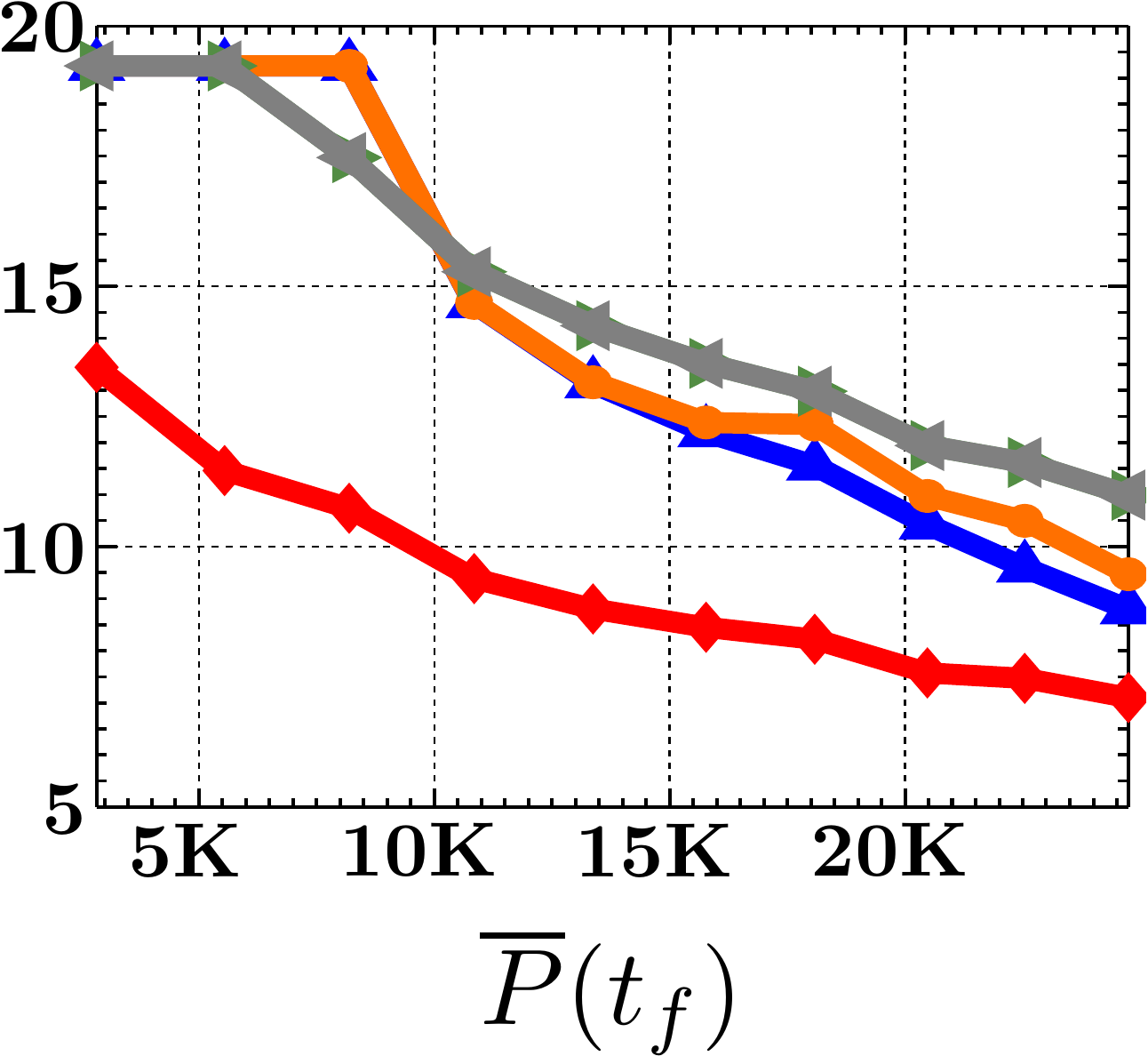}}
\hspace*{0.1cm}\subfloat[Sports]{ \includegraphics[height=0.18\textwidth]{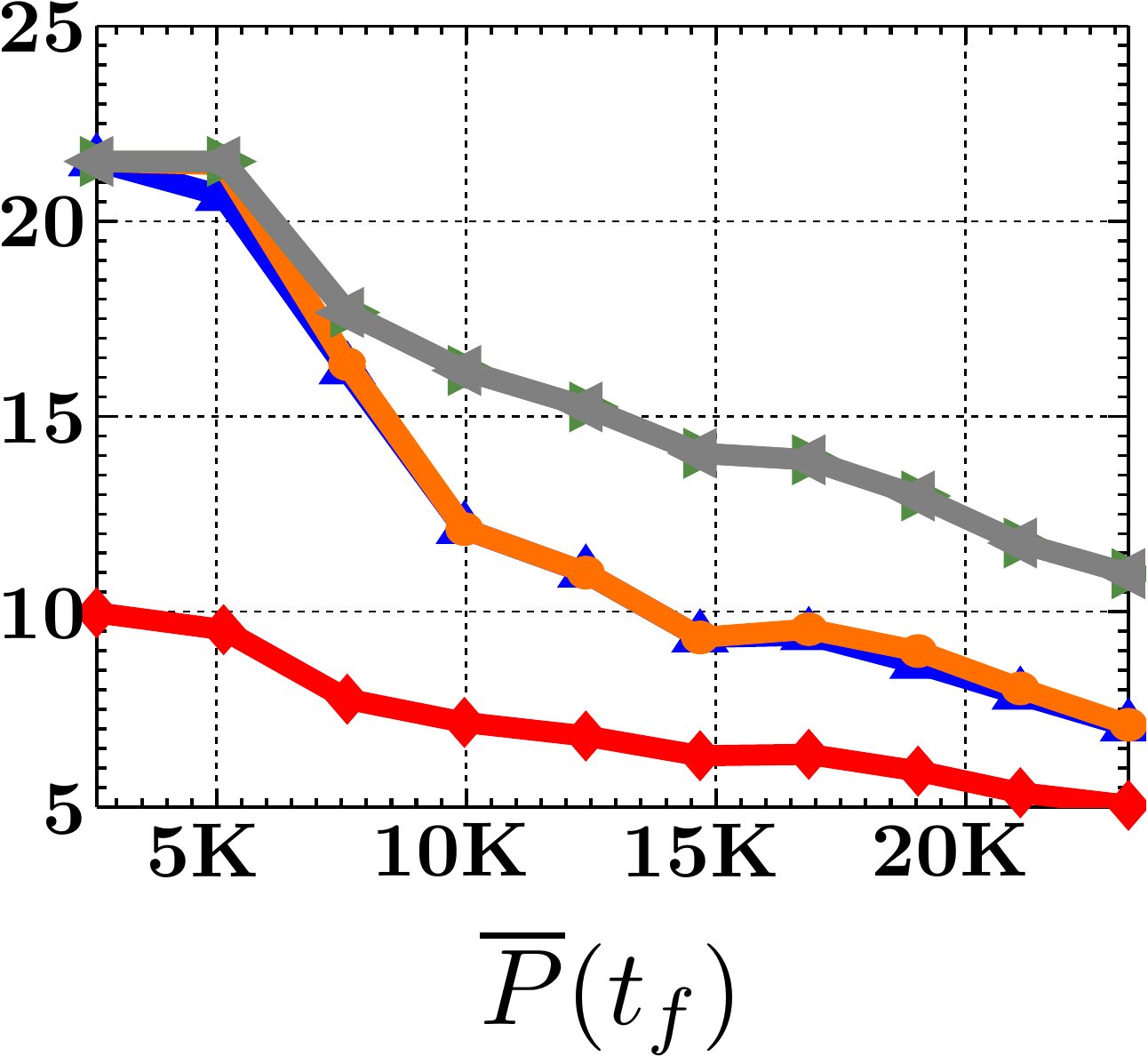}}
 \hspace*{0.1cm}\subfloat[Series]{ \includegraphics[height=0.18\textwidth]{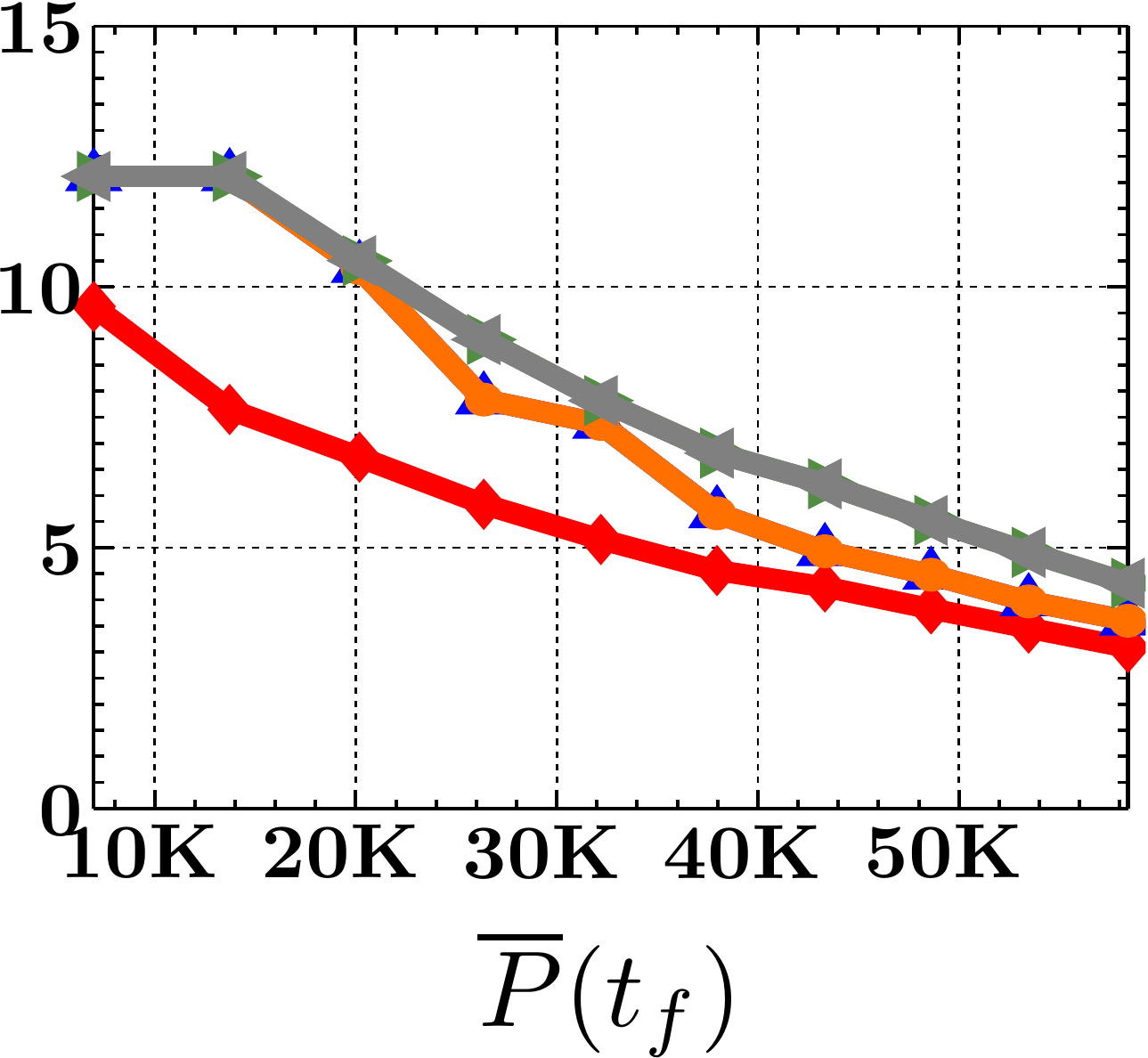}}
\caption{Performance vs. number of directly incentivized tweets for each Twitter data set. Performance is measured in terms of the average time $\bar{t}_{30K}$ required by each
method to reach a milestone of $30{,}000$ tweets.
\cheshire (in red) consistently reaches the milestone faster than the competing methods, \eg, $20$\%--$50$\% faster than the second best performer (in blue) for low
budgets.}
\label{fig:NuT}
\end{figure*}

Next, we evaluate the performance of our algorithm against others with respect to the available budget. To this aim, we compute the average time $\bar{t}_{30K}$ required by
each method to reach a milestone of $30{,}000$ tweets against the number of directly incentivized tweets $\bar{P}(t_f)$ (\ie, the budget). Here, we do not report the results for the
uncontrolled dynamics (UNC) since it did not reach the milestone after $10$$\times$ the time the slowest competitor took to reach it.
Figure~\ref{fig:NuT} summarizes the results, which show that: (i) our algorithm consistently reaches the milestone faster than the competing methods; (ii) it exhibits
a greater competitive advantage when the budget is low; and, (iii) it reaches the milestone $20$\%--$50$\% faster than
the second best performer for low budgets.
 
%-------------------------------------------------------
%-------------------------------------------------------
\section{Conclusions}
\label{sec:conclusions}
In this paper, we developed efficient online algorithms for steering social activity, both at a user (local) and network (global) level, based on stochastic
optimal control of stochastic differential equations (SDEs) with jumps.
In doing so, we established a previously unexplored connection between optimal control of jump SDEs and doubly stochastic marked temporal
point processes, which is of independent interest.
We experimented with synthetic and real-world data gathered from Twitter and showed that our algorithms can consistently steer social
activity more effectively than the state of the art.

Our work also opens many venues for future work. For example, from the perspective of an individual user, we considered social networks
that sort stories in the users'{} feeds in reverse chronological order (\eg, Twitter, Weibo). Extending our methodology to social networks that
sort stories algorithmically (\eg, Facebook) is a natural next step.
Moreover, we assume that only one broadcaster is using \redqueen. A very interesting follow-up would be augmenting our framework to consider
multiple broadcasters under cooperative, competitive and adversarial environments.
From the perspective of an entire social net\-wor\-king site, our experimental evaluation is based on simulation, using models whose parameters ($\Bb$,
$\omega$) are learned from data. It would be very interesting to evaluate our method using actual interventions in a social network.
From both perspectives, our algorithms optimize quadratic losses and assume the model parameters do not change over time, however, it
would be useful to derive optimal broadcasting intensities for other losses with well-defined semantics and time-varying model parameters.
Finally, optimal control of jump SDEs with doubly stochastic temporal point processes can be potentially applied to design online algorithms for a
wide variety of control problems in social and information systems, such as human learning~\cite{reddy2016unbounded} or rumor control~\cite{friggeri2014rumor}.

%-------------------------------------------------------
%-------------------------------------------------------
\appendix
\section{Proof of Proposition~\ref{prop:hawkes}}
\label{app:hawkes-dynamics}
Using the left continuity of Poisson processes and the definition of derivative $d\lambdab(t)=\lambdab(t+dt) - \lambdab(t)$ we can find the dynamics of the process using
Ito'{}s calculus~\cite{hanson2007} as follows:
\begin{align*}
	d\lambdab(t)
	&=\lambdab'_0(t) \, dt + \Cb \int_0^{t+dt} \kappa_{\omega}(t+dt-s)\,d\bm{N}(s) - \Cb \int_0^t \kappa_{\omega}(t-s)\,d\bm{N}(s) \\
	&=\lambdab'_0(t) \, dt + \Cb \int_0^{t+dt} (\kappa_{\omega}(t-s)+\kappa_{\omega}'(t-s) \, dt) \, d\bm{N}(s) - \Cb \int_0^t \kappa_{\omega}(t-s)\,d\bm{N}(s) \\
	&=\lambdab'_0(t) \, dt + \Cb \int_t^{t+dt} \kappa_{\omega}(t-s)\,d\bm{N}(s) + dt \, \Cb \int_0^{t+dt} \kappa_{\omega}'(t-s)\,d\bm{N}(s) \\
	&=\lambdab'_0(t) \, dt + \Cb\,\kappa_{\omega}(0) \, d\bm{N}(t) - \omega \, dt \, \Cb \int_0^{t+dt} \kappa_{\omega}(t-s)\,d\bm{N}(s)	 \\
	&=\lambdab'_0(t) \, dt + \Cb\,d\bm{N}(t)  - \omega \, dt \, \Cb \int_0^{t} \kappa_{\omega}(t-s)\,d\bm{N}(s)		 \\
	&=\left[\lambdab'_0(t) + \omega \lambdab_0(t) - \omega \lambdab(t) \right] dt + \Cb\,d\bm{N}(t).
\end{align*}

%%%%
%%%%

\section{Proof of Lemma~\ref{lem:bellman-opt-cond-redqueen}} \label{app:bellman-opt-cond-redqueen} 
\begin{align*}
	&J(\gamma(t),r(t),t) = \min_{u(t,t_f]} \mathbb{E}_{(N,M)(t,t_f]} \left[ \phi(r(t_f)) + \int_t^{t_f} \ell(r(\tau),u(\tau)) \, d\tau \right ] \\
	&= \min_{u(t,t_f]} \mathbb{E}_{(N,M)(t,t_f]} \bigg[ \phi(r(t_f)) + \int_t^{t+dt} \ell(r(\tau),u(\tau)) \, d\tau \nonumber + \int_{t+dt}^{t_f} \ell(r(\tau),u(\tau)) \, d\tau \bigg ] \nonumber \\
	&= \min_{u(t,t_f]} \mathbb{E}_{(N,M)(t,t+dt]}\bigg[\mathbb{E}_{(N,M)(t+dt,t_f]} \Big[ \phi(r(t_f)) + \ell(t,r,u) \, dt  \nonumber + \int_{t+dt}^{t_f} \ell(r(\tau),u(\tau)) \, d\tau \Big ] \bigg] \\
	&= \min_{u(t,t+dt]} \min_{u(t+dt,t_f]} \mathbb{E}_{(N,M)(t,t+dt]}\bigg[\ell(r(t),\gamma(t),t) \, dt \nonumber +\mathbb{E}_{(N,M)(t+dt,t_f]} \Big[ \phi(r(t_f)) + \int_{t+dt}^{t_f} \ell(r(\tau),u(\tau)) \, d\tau \Big ] \bigg]  \\
	&= \min_{u(t,t+dt]} \mathbb{E}_{(N,M)(t,t+dt]} \left[ J(\gamma(t+dt),r(t+dt),t+dt)\right ] + \ell(r(t),u(t)) \, dt.
\end{align*}

\section{Proof of Lemma~\ref{lem:diff-cost-redqueen}}
\label{app:diff-cost-redqueen}
According to the definition of differential,
\begin{align*}
	dJ(r(t),\gamma(t),t) &:= J(r(t+dt),\gamma(t+dt),t+dt) - J(r(t),\gamma(t),t) \\
	&\,\,= J(r(t)+dr(t),\gamma(t)+d\gamma(t),t+dt) - J(r(t),\gamma(t),t).
\end{align*}
To evaluate the first term in the right hand side of the above equality we substitute $dr(t)$ and $d\gamma(t)$ using Eq.~\ref{eq:dynamics-one-follower}. Then, using
the zero-one jump law~\cite{Kingman1992} we can write:
\begin{align*}
	dJ &= J\big(0,\gamma(t)+m(t)dt,t+dt\big)\,dN(t) + J\big(r(t)+g(t),\gamma(t)+m(t)dt+\alpha,t+dt\big)\,dP(t) \\
	&+ J\big(r,\gamma(t)+m(t) dt,t+dt\big)\,(1-dN(t))(1-dP(t)) - J\big(r(t),\gamma(t),t\big),
\end{align*}
where $m(t) = \gamma_0'(t) + \omega \gamma_0(t) - \omega \gamma(t)$.
Then, we can expand the first three terms in the right hand sides:
\begin{align*}
	J\big(0,\gamma+m(t)dt,t+dt\big) &=J(0,\gamma(t),t)+J_\gamma(0,\gamma(t),t) m(t) dt +J_t(0,\gamma(t) ,t)dt \\
	J\big(r(t)+1,\gamma(t)+m(t)dt+\alpha,t+dt\big) &= J(r(t)+1,\gamma(t)+\alpha,t) + J_\gamma(r+1,\gamma(t)+\alpha,t) m(t) dt \\
	J\big(r(t),\gamma(t)+m(t)dt,t+dt\big) &= J(r(t),\gamma(t),t) + J_\gamma(r(t),\gamma(t),t) m(t) dt+J_t(r(t), \gamma(t), t)dt,
\end{align*}
using that the bilinear differential form $dt \, dN(t)=0$~\cite{hanson2007} and $dN(t)dP(t)=0$ by the zero-one jump law. Finally,
\begin{align*}
dJ(r(t),\gamma(t),t) &= J_t(r(t),\gamma(t),t)dt  + \left[\gamma_0'(t) + \omega \gamma_0(t) - \omega \gamma(t) \right]J_{\gamma}(r(t),\gamma(t),t)dt \\
	&\quad +[J(0,\gamma(t),t)-J(r(t),\gamma(t),t)] dN(t) \\
	&\quad + [J(r(t)+1,\gamma(t)+\alpha,t)-J(r(t),\gamma(t),t)] dP(t),
\end{align*}
which concludes the proof.

\section{Proof of Lemma~\ref{lem:opt-con-sol}} \label{app:opt-con-sol} 
Consider the following proposal for the cost-to-go:
\begin{align*}
	J(r(t),\gamma(t),t) = \sum_{i=0}^n \sum_{j=0}^m f_{ij}(t)r^i(t)\gamma^j(t),
\end{align*}
where $m$ and $n$ are arbitrary large numbers and $f_{ij}(t)$ are time-varying functions. Now, substitute this proposal in to Eq. \ref{eq:bellman-pde}:
\begin{align*}
	0&=\sum_{i=1}^n f'_{i0} \, r^i + f_{i0}\,(r+1)^i \gamma - f_{i0} \, r^i \gamma
	+\sum_{j=1}^m f'_{0j} \gamma^j + j(\gamma'_0+w\gamma_0-w\gamma)f_{0j} \gamma^{j-1} +f_{0j} (\gamma+\alpha)^j \gamma - f_{0j} \gamma^{j+1} \\
	&+\sum_{i=1}^n \sum_{j=1}^m j(\gamma'_0+w\gamma_0-w\gamma)f_{ij} r^i \gamma^{j-1}
	+\sum_{i=1}^n \sum_{j=1}^m f_{ij} (r+1)^i (\gamma+\alpha)^j\gamma - f_{ij} r^i \gamma^{j+1} \\
	&-\frac{1}{2}q^{-1}\bigg[\sum_{i=1}^n f_{i0} r^i + \sum_{i=1}^n \sum_{j=1}^m  f_{ij} r^i \gamma^j \bigg]^2 +\frac{1}{2}s\,r^2 +f_{00}'
\end{align*}
where for notational simplicity we omitted the time argument of functions. To find the unknown functions $f_{ij}(t)$,  we equate the coefficient of different variables.
If we consider the coefficient of $r^{2n}$, we have $f_{n0}(t)=0$. We can continue this argument for $n-1,n-2,\cdots,2$ to show that $\forall i\geq 2; \, f_{i0}(t)=0$.
Similar reasoning for coefficients of $r^{2i}\gamma^{2j}$ shows that $\forall j, i\geq 2; f_{ij}(t)=0$. Finally, the coefficient of $r^2$ is $s(t)/2 - q^{-1} f^2_{10}(t)/2=0$,
so $f_{10}(t)=\sqrt{s(t)q}$. If we rename $f_{0j}(t)$ to $g_j(t)$ and $f_{00}(t)$ to $f(t)$, then it follows that
\begin{align*}
	J(r(t),\gamma(t),t) = f(t)+ \sqrt{s(t) q} \, r(t) + \sum_{j=1}^m g_{j}(t)\gamma^j(t).
\end{align*}
We can continue the previous method to find the remaining coefficients and completely define the cost-to-go function. If we equate the coefficient of $\gamma^j$ to zero
we would have a system of first oder differential equation which its $j$'th row for $j > 1$ is
\begin{align*}
	g_j'(t) -g_{j-1}(t) + j(\alpha- w) g_j(t) + (j+1)\big(\gamma_0'(t)+w\gamma_0(t)+\frac{j}{2}\alpha^2\big)g_{j+1}(t)
	+ \sum_{k=2}^{m-j} \binom{j+k}{k+1} \alpha^{k+1} g_{j+k}(t) = 0
\end{align*}
We can also evaluate the corresponding differential equation for $j=1$. When $\gamma_0(t)=\gamma_0$, we can express these differential equations using a matrix differential
equation $\bm{g}'(t) = A\bm{g}(t)$.
and its solution is $\bm{g}(t) = c_1 e^{\zeta_1 t} \bm{u}_1 + c_2 e^{\zeta_2 t} \bm{u}_2 + \cdots + c_n e^{\zeta_n t} \bm{u}_n	$
where $\zeta_i$ and $\bm{u}_i$ are eigenvalue and eigenvector of matrix $A$ and $c_i$ is a constant found using the terminal conditions. Since in triangular matrices diagonal entries
are eigenvalues, we have
	$ \bm{g}(t) = \sum_{j=1}^m c_i e^{j(w-\alpha)} \bm{u}_i $.
We can approximate general time varying $\gamma_0(t)$ using piecewise function and repeat the above procedure for each piece.

\section{Proof of Lemma~\ref{lem:diff-cost-cheshire}}
\label{app:diff-cost-cheshire}
According to the definition of differential,
\begin{align*}
	dJ(\lambdab(t),t) := J(\lambdab(t+dt),t+dt) - J(\lambdab(t),t) = J(\lambdab(t)+d\lambdab(t),t+dt) - J(\lambdab(t),t).
\end{align*}
To evaluate the first term in the right hand side of the above equality we substitute $d\lambdab(t)$ using Eq.~\ref{eq:cheshire-dynamics}. Then, using the zero-one jump
law~\cite{Kingman1992} we can write:
\begin{align*}
	&J(\lambdab(t)+d\lambdab(t),t+dt) = J(\lambdab(t)+ m(t)\,dt + \Bb \, d\bm{N}(t) + \Bb \, d\Pb(t), t+dt) \\
	&= \sum_i J(\lambdab(t)+ m(t)\,dt + \bm{b}_i , t+dt) dN_i(t) + \sum_i J(\lambdab(t)+ m(t)\,dt + \bm{b}_i, t+dt) dP_i(t) \\
	&\quad+ J(\lambdab(t)+ m(t)\,dt, t+dt) \prod_i [1-dN_i(t)][1-dP_i(t)] \\
	&=J(\lambdab(t)+ m(t)\,dt, t+dt) [1-\sum_{i} dN_i(t)+dP_i(t)] + \sum_i J(\lambdab(t)+ m(t)\,dt + \bm{b}_i , t+dt) [dN_i(t)+dP_i(t)]
	 \\
	&=J(\lambdab(t)+ m(t)\,dt, t+dt) + \sum_i [J(\lambdab(t)+ m(t)\,dt + \bm{b}_i , t+dt) - J(\lambdab(t)+ m(t)\,dt, t+dt)] [dN_i(t)+dP_i(t)]
\end{align*}
where $m(t) := \omega \lambdab_0 - \omega \lambdab(t)$ and we used that the bilinear differential form $dt \, dN(t)=0$~\cite{hanson2007}.
By total derivative rule, it follows that
\begin{align*}
	J(\lambdab(t)+ m(t)\,dt + \bm{b}_i , t+dt) &=
	J(\lambdab(t)+\bm{b}_i,t) + \nabla_{\lambdab}J(\lambdab(t)+\bm{b}_i,t) m(t)\,dt + J_t(\lambdab(t)+\bm{b}_i,t)\,dt \\
	J(\lambdab(t)+ m(t)\,dt, t+dt) &=
	J(\lambdab(t),t) + \nabla_{\lambdab}J(\lambdab(t),t) m(t)\,dt + J_t(\lambdab(t),t)\,dt.
\end{align*}
Then, the differential is given by:
\begin{align*}
	dJ(\lambdab(t)+d\lambdab(t),t+dt) &=  J(\lambdab(t),t) + (\omega \lambdab_0 - \omega \lambdab(t))^{T} \nabla_{\lambdab}J(\lambdab(t),t) \,dt + J_t(\lambdab(t),t)\,dt \\
	&+ \sum_{i} \left[J(\bm{\lambda}(t)+\bm{b}_i, t)-J(\bm{\lambda}(t), t) \right] [dN_i(t)+dP_i(t)],
\end{align*}
which completes the proof.

\section{Proof of $(\Delta_B J)_i \leq 0$}
\label{app:positive-delta}
Lets $t < s$, then according to the definition we can write,
\begin{align*}
	\lambdab(s) = \lambdab_0 + \int_0^s \kappa_{\omega}(s-\tau) \, \Bb \, d\bm{N}(\tau) = \lambdab_0 + \int_0^t \kappa_{\omega}(s-\tau) \, \Bb \, d\bm{N}(\tau) + \int_t^s \kappa_{\omega}(s-\tau) \, \Bb \, d\bm{N}(\tau).
\end{align*}
For the exponential kernel $\kappa_{\omega}(t)=e^{-wt}$ we have,
\begin{align*}
	\int_0^t \kappa_{\omega}(s-\tau) \, \Bb \, d\bm{N}(\tau) = \int_0^t e^{-w(s-\tau)} \, \Bb \, d\bm{N}(\tau) = e^{-w(s-t)} \int_0^t e^{-w(t-\tau)} \, \Bb \, d\bm{N}(\tau) = e^{-w(s-t)} (\lambdab(t) - \lambdab_0 )
\end{align*}
so given the value of $\lambdab(t)$ at time $t$ then we can write $\lambdab(s)$ for later times as
\begin{align*}
	\lambdab(s) &= \lambdab_0 + e^{-w(s-t)} (\lambdab(t) - \lambdab_0 )
	 + \int_t^s \kappa_{\omega}(s-\tau) \, \Bb \, d\bm{N}(\tau)
\end{align*}
Lets consider a process $\bm{\xi}(s)$ with intensity value at time $t$ equal to $\lambdab(t)+\bm{b}_i$ as,
\begin{align*}
	\bm{\xi}(s) &= \lambdab_0 + e^{-w(s-t)} (\lambdab(t) + \bm{b}_i - \lambdab_0 )
	 + \int_t^s \kappa_{\omega}(s-\tau) \, \Bb \, d\bm{N}(\tau).
\end{align*}
Since $\bm{b}_i \succeq 0$, then given the same history in interval $(t,s)$, we have $\bm{\xi}(s) \succeq \lambdab(s)$. 
Then, we have:
\begin{align*}
	\ell(\bm{\xi}(s), \bm{u}(s)) \leq \ell(\lambdab(s), \bm{u}(s)).
\end{align*}

Now, taking the integration, then expectation (over all histories) and finally the minimization from the above inequality does not change the direction of inequality. So it readily follows the required result.
\begin{align*}
	J(\lambdab(t)+\bm{b}_i, t) \leq J(\lambdab(t), t).
\end{align*}

\section{Proof of Lemma \ref{lem:quad-proposal-cheshire}}\label{sec:quad-proposal-cheshire}
Consider the following proposal of degree three for the cost-to-go function:
\begin{align*}
	J(\lambdab(t), t) = f(t) + \sum_i g_i(t)  \lambda_i(t) + \sum_i \sum_j \lambda_i(t) \lambda_j(t) H_{ij}(t) + \sum_i\sum_j\sum_k \lambda_i(t) \lambda_j(t) \lambda_k(t) H_{ijk}(t)
\end{align*}
If we plug this proposal in Eq.~\ref{eq:hjb}, and evaluate the coefficient of fourth degree terms like $ \lambda_i^2(t)\lambda_j^2(t)$ and equate them to zero, then we can find the unknown coefficients $H_{ijk}(t)$'s as follows:
\begin{align*}
	\forall i,j,t: \,\,	\sum_k \big(\sum_\ell B_{\ell k}^2 \big) H^2_{ijk}(t) = 0
\end{align*}
Since the sum of positives terms is zero if and only if they all be zero, then $H_{ijk}(t)$ and consequently the terms with degree three in the proposal are all zero. So the proposal reduces to a quadratic proposal.

It is quite straightforward to extend this argument for proposals with order $m > 3$  and by equating the degree $2m-1$ terms, similarly conclude that coefficients of degree $m$ terms in the proposal are zero. If we repeat this argument for $m-1,\ldots,3$, we deduce that any proposal with arbitrary degree $m \geq 2$, would result in a quadratic optimal cost.

%-------------------------------------------------------
%-------------------------------------------------------

{
\bibliographystyle{abbrv}
\bibliography{refs}
}

\end{document}